\documentclass[aps,prb,showkeys,showpacs,twocolumn,superscriptaddress,letter]{revtex4}

\usepackage{amsmath}
\usepackage{amsfonts, amssymb,amsxtra}
\usepackage[]{graphicx}

\usepackage{color}                                      
\definecolor{d_red}{cmyk}{0.00, 0.81, 1.00, 0.27}
\definecolor{d_orange}{cmyk}{0.00, 0.33, 1.00, 0.00}
\definecolor{d_blue}{cmyk}{0.78, 0.47, 0.00, 0.20}
\definecolor{d_lgreen}{cmyk}{0.07, 0.00, 0.79, 0.29}
\definecolor{d_green}{cmyk}{0.66, 0.00, 0.71, 0.56}
\definecolor{d_blue}{cmyk}{0.78, 0.47, 0.00, 0.20}
\definecolor{d_dblue}{cmyk}{0.91, 0.79, 0.00, 0.22}
\definecolor{d_pink}{cmyk}{0.0, 0.79, 0.37, 0.29}
\definecolor{d_purple}{cmyk}{0.16, 0.54, 0.00, 0.70}
\definecolor{d_paleblue}{cmyk}{0.669, 0.338, 0.00, 0.373}
\definecolor{d_dpaleblue}{cmyk}{0.441, 0.290, 0.00, 0.580}
\definecolor{d_brown}{cmyk}{0.0, 0.490, 0.930, 0.350}
\definecolor{d_turquoise}{cmyk}{0.630, 0.04, 0.0, 0.440}

\newcommand{\bra}[1]{\langle~\!\!#1~\!\!|}
\newcommand{\ket}[1]{|~\!\!#1~\!\!\rangle}
\newcommand{\braket}[2]{\langle#1|#2\rangle}
\newcommand{\braopket}[3]{\langle#1|#2|#3\rangle}
\newcommand{\av}[1]{\langle #1 \rangle}

\usepackage{hyperref}
\usepackage[figure,table]{hypcap}               
\hypersetup{
pdftitle = {},
pdfsubject = {},
pdfauthor = {},
pdfkeywords = {},
pdfcreator = {},
pdfproducer = {LaTeX with hyperref},
colorlinks = true,
linkcolor = red, 
anchorcolor = red,
citecolor = red, 
filecolor = red, 
menucolor = red, 
pagecolor = red, 
urlcolor  = red, 
breaklinks = true,
pdfstartview = FitV,
pdfhighlight = /I,
pdfpagelayout = OneColumn,
hypertexnames=true 
}

\begin{document}

\title{Dynamics, synchronization, and quantum phase transitions of two dissipative spins}


\author{Peter P. Orth}
\affiliation{Department of Physics, Yale University, New Haven, Connecticut 06520, USA}
\author{David Roosen}
\affiliation{Institut f\"ur Theoretische Physik, Johann Wolfgang Goethe--Universit\"at, 60438 Frankfurt/Main, Germany }
\author{Walter Hofstetter}
\affiliation{Institut f\"ur Theoretische Physik, Johann Wolfgang Goethe--Universit\"at, 60438 Frankfurt/Main, Germany }
\author{Karyn Le Hur}
\affiliation{Department of Physics, Yale University, New Haven, Connecticut 06520, USA}

\date{\today}

\begin{abstract}
We analyze the static and dynamical properties of two Ising-coupled quantum spins embedded in a common bosonic bath as an archetype of dissipative quantum mechanics. First, we elucidate the ground state phase diagram for an ohmic and a subohmic bath using a combination of bosonic numerical renormalization group (NRG), analytical techniques and intuitive arguments. Second, employing the time-dependent NRG we investigate the system's rich dynamical behavior arising from the complex interplay between spin-spin and spin-bath interactions. Interestingly, spin oscillations can synchronize due to the proximity of the common non-Markovian bath and the system displays highly entangled steady states for certain nonequilibrium initial preparations. We complement our non-perturbative numerical results by exact analytical solutions when available and provide quantitative limits on the applicability of the perturbative Bloch-Redfield approach at weak coupling. 
\end{abstract}



\maketitle

\section{Introduction}
\label{sec:motiv-intr}
A quantum system is never completely isolated from its environment which results in noticeable effects such as decoherence, dissipation and entanglement.~\cite{lehur_entanglement_spinboson} One prominent example embodies a two-level (spin-$1/2$) system interacting with a collection of harmonic oscillators, the so-called spin-boson model.~\cite{RevModPhys.59.1,weissdissipation,KarynLeHur-UnderstandingQPT-Article} The latter displays a rich behavior ranging from damped Rabi oscillations to localization in one of the two states, and has been widely studied as a paradigm of quantum dissipation and quantum-to-classical transitions.~\cite{RevModPhys.75.715} As it constitutes the elementary unit of a quantum computer (qubit), much work was recently directed toward understanding and controlling the dissipative spin-boson dynamics in nonequilibrium situations such as time-dependent external fields.~\cite{PhysRevA.82.032118,PhysRevB.79.115137,nalbach:220401,roosen_hofstetter_lehur_unpublished,PhysRevLett.98.100504} 
The model is of particular importance because it may be implemented in a variety of different experimental contexts, for example, the tunneling of defects in solid-state systems,~\cite{GrabertWipf-SpinBosonTunnelingOfDefects1990} electron transfer in chemical reactions~\cite{Marcus1956, Marcus1985} or qubit designs based on the Josephson effect.~\cite{RevModPhys.73.357,J.E.Mooij08131999,Manucharyan_coherent_osc_in_sc_loop} Other systems that are described by the spin-boson Hamiltonian are trapped ions,~\cite{porras:010101} quantum emitters coupled to surface plasmons,~\cite{PhysRevB.82.075427} and the cold-atom quantum dot setup.~\cite{recati:040404,orth:051601,PhysRevLett.104.200402,PhysRevLett.105.045303} Further variants of spin-boson models involve two-level atoms interacting with a single quantized mode of an electromagnetic cavity.~\cite{RevModPhys.73.565,Schoelkopf_Nature_2008, koch:023811,Hartmann_Plenio_LPRev_2008} 

The environmental influence on the phase coherence between the two spin states is of crucial importance in the field of quantum computing, as it sets a limit to the timescale where coherent quantum logical operations can be performed. In this context, it is essential to extend the system to multiple two-level systems (or qubits), as operations involving two-qubits, {\it e.g.}, the CNOT gate, are required to obtain a complete set of quantum logical operations. In addition, the presence of a second spin allows to address the competition between spin-spin and spin-bath interactions and the resulting interplay between quantum control and dissipation. 

\begin{figure}[t]
  \centering
  \includegraphics[width=.7\linewidth]{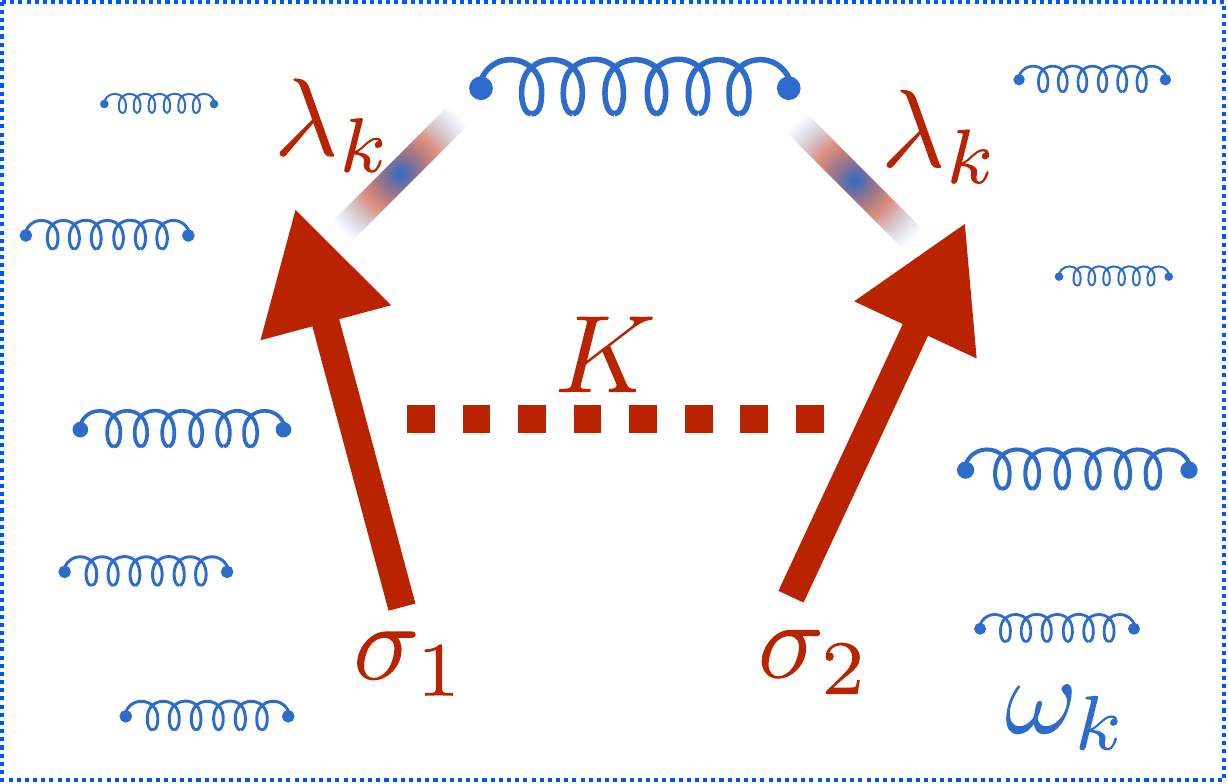}
\caption{Two quantum spins-$\frac12$, $\boldsymbol{\sigma}_{1}$ and $\boldsymbol{\sigma}_2$, coupled through an Ising interaction $K$. The spins are also entangled, via their $\sigma^z$-components, to a common reservoir of bosonic oscillator modes with frequencies $\omega_k$. The bath is characterized by the spectral density $J(\omega) = 2 \pi \alpha \omega^s \omega_c^{1-s} \theta(\omega_c - \omega)\theta(\omega)$, where $s=1$ $(s<1)$ refers to an ohmic (subohmic) bosonic environment.}
    \label{fig:1}
\end{figure}

In the present article, we investigate such a generalization of the single spin-boson model and consider two quantum spins $\{\boldsymbol \sigma_{1}$, $\boldsymbol \sigma_2\}$ that are coupled to each other via an Ising-type coupling and interact with a common bath of harmonic oscillator modes, as described by the Hamiltonian (see also Fig.~\ref{fig:1})
\begin{align}
  \label{eq:1}
  H &=  \sum_{j=1}^2 \Bigl[ \frac{\Delta_j}{2} \sigma^x_j + \frac{\epsilon_j}{2} \sigma^z_j + \frac{\sigma^z_j}{2} \sum_{k>0} \lambda_k (b^\dag_k + b_k) \Bigr] + \frac{K}{4} \sigma^z_1 \sigma^z_2 \nonumber \\ & \qquad + \sum_{k>0} \omega_k b_k^\dag b_k\,.
\end{align}
We set the reduced Planck constant $\hbar=1$.
Here, $\sigma^{x,y,z}_{1,2}$ are the usual Pauli matrices describing the two spins and $b_k$ is the bosonic annihilation operator of the bath mode with frequency $\omega_k$. The free spin part of the Hamiltonian contains the tunneling amplitudes $\Delta_{1,2}$, bias fields $\epsilon_{1,2}$ and the bare Ising interaction constant $K$. The effects of the bosonic environment on the spins are fully captured by the bath spectral density~\cite{RevModPhys.59.1,weissdissipation,KarynLeHur-UnderstandingQPT-Article} 
\begin{equation}
  \label{eq:2}
   J(\omega) = \pi \sum_k \lambda_k^2 \delta(\omega - \omega_k) = 2 \pi \alpha \omega^s \omega_c^{1-s} \theta(\omega_c - \omega) \theta(\omega) \,,
\end{equation}
which we assume to behave as a power law $\omega^s$ ($s>0$) up to the cutoff frequency $\omega_c$. Hereafter, we will be studying exponents in the range $\frac12 \leq s \leq 1$, where the case $s=1$ $(s<1)$ refers to an ohmic (subohmic) bosonic bath. The strength of the coupling to the bath is characterized by the dimensionless dissipation constant $\alpha \geq 0$. 

For the spin-bath interaction, for simplicity, we use identical coupling constants $\lambda_k$ for both spins. This corresponds to the case where the spins are spatially close to each other. Specifically, we assume their separation $d_{12}$ to be smaller than the shortest wavelength of the bath excitations: $d_{12} \lesssim \lambda_c = v_s/\omega_c$, where $v_s$ is the sound velocity in the bath.~\cite{orth:051601,PhysRevLett.102.160501}

There are several reasons for considering an Ising-like coupling $\frac{K}{4} \sigma^z_1 \sigma^z_2$ between the two spins. First, there are experimental situations where such an $\text{SU}(2)$-broken coupling is realized, for instance in capacitively coupled quantum dots where the operators $\sigma^z_j$ describe charge states on the dot.~\cite{PhysRevLett.75.705,PhysRevB.53.1034,PhysRevB.53.3893} Other examples are the cold-atom quantum dot setting, trapped ions and superconducting qubits. Second, since the bath couples to the $\sigma^z$ component of the spins, it automatically induces an indirect (ferromagnetic) Ising interaction between the spins which is mediated by a coherent exchange of phonons. This results in a renormalization of $K$ to $K_r = K - 4 \alpha \omega_c /s$. Therefore, even for zero $K$, the spins are Ising-coupled. We note that, in general, the bath induced interaction decays with the spatial distance between the spins $d_{12}$ on a lengthscale given by $\lambda_c \sim \omega_c^{-1}$.~\cite{orth:051601,PhysRevLett.102.160501} 

The two-spin boson model allows to address the competition between spin-spin entanglement, characterized for instance by the concurrence, and spin-bath entanglement, characterized for instance by the entanglement entropy.~\cite{PhysRevB.77.155420, Campagnano2010,PhysRevB.75.035134,PhysRevLett.89.277901,PhysRevLett.102.160501,PhysRevLett.91.070402} The entanglement entropy also contains information about the coherence between different spin states.~\cite{lehur_entanglement_spinboson}
We will show below that for a particular initial preparation, the system exhibits a non-trivial steady-state, where the spins are strongly entangled with the bath while maintaining coherence between different spin configurations. 

Whereas for some experimental realizations the description of independent bosonic reservoirs is appropriate, {\it e.g.}, in the case of  quantum dots coupled to independent leads,~\cite{PhysRevLett.75.705} there are others, where the spins couple to a common bath, {\it e.g.}, the cold-atom~\cite{recati:040404,orth:051601,PhysRevLett.104.200402,PhysRevLett.105.045303} and trapped ion setup.~\cite{porras:010101} Here, we assume a common bath because we are mostly interested in studying the competition between the coherent and dissipative parts of the interaction induced by the bath, leading to dynamical spin synchronization and highly entangled steady states. The other situation has been addressed for instance in Refs.~\onlinecite{PhysRevB.69.214413, Naegele2010622, 1367-2630-10-11-115010}.

In the following, we aim to investigate not only the static properties of the ground state but also the nonequilibrium dynamics of the system, both for an ohmic and a subohmic boson bath. In the subohmic case, we mainly consider the experimentally relevant situation of $s=1/2$.~\cite{PhysRevLett.97.016802,PhysRevLett.103.206401} We apply the powerful non-perturbative numerical renormalization group (NRG).\cite{Wilson1975, PhysRevB.71.045122, RevModPhys.80.395,PhysRevLett.95.086406} To solve for the dynamics of the system, we employ the recently developed time-dependent NRG (TD-NRG),~\cite{PhysRevLett.95.196801,PhysRevB.74.245113} that we compare to exact solutions, available at special points in the parameter space, and to the Bloch-Redfield master equation approach~\cite{PhysRevA.67.042319} at weak dissipation.

The paper is outlined as follows. 
In Sec.~\ref{sec:model-ground-state}, we calculate the zero temperature phase diagram as a function of dissipation strength $\alpha$ and Ising coupling $K$, both for $s=1/2$ and $s=1$. As a reminiscence of the single spin-boson model, it contains a delocalized phase ($\av{\sigma^z_{1,2}} = 0$) for small dissipation and a localized phase ($\av{\sigma^z_{1,2}} \neq 0$ for $\epsilon_{1,2} =0^+$) for large dissipation. We give a physically intuitive explanation for the asymmetry between the ferromagnetic ($K<0$) and antiferromagnetic ($K>0$) regions of the phase diagram. 

In Sec.~\ref{sec:phase-trans-scal}, we investigate the critical properties at the phase transitions such as the behavior of the entanglement entropy across the transition, or the scaling of spin expectation values, that occurs for a subohmic bath. 

In Sec.~\ref{sec:non-equil-dynam}, we explore the nonequilibrium dynamics of the two spins after a quantum quench of parameters. We typically polarize the spins initially by applying large bias fields along the $z$ or $x$-direction that we switch off at time $t=0$. 
We begin our analysis in Sec.~\ref{sec:exact-decoh-with} with the exactly solvable case of zero transverse fields $\Delta_{1,2} = 0$, where we show that our TD-NRG results perfectly agree with the exact analytical solution. In Sec.~\ref{sec:beat-decoh-breakd} we investigate the regime of weak spin-bath coupling, and compare TD-NRG to the commonly employed perturbative Bloch-Redfield approach. We give quantitative limits on the applicability of the Redfield method. 
In Sec.~\ref{sec:synchr-spin-dynam}, we find that, interestingly, the bath is able to synchronize spin oscillations via a coherent exchange of phonons, even at weak spin-bath coupling. This phenomenon is not captured in the Bloch-Redfield master equation approach, where the backaction of the bath on the spins is neglected. This method thus fails to correctly describe the spin dynamics even in the perturbative regime.
In Sec.~\ref{sec:case-k_r-=}, we investigate the spin dynamics for vanishing (renormalized) Ising interaction $K_r=0$ and highlight similarities and differences to the single spin-boson model. We elaborate on the case of weak-dissipation in Sec.~\ref{sec:weak-dissipation}, where we compute the quality factor of the damped oscillations. In Sec.~\ref{sec:toulouse-point-1}, we discuss the dynamics at the generalized Toulouse point $\alpha = 1/2$. In Sec.~\ref{sec:large-spin-bath}, we examine the crossover to the regime of strong spin-bath coupling for general Ising coupling, and point out differences between the case of an ohmic and a subohmic bath.
In Sec.~\ref{sec:gener-highly-entangl}, we describe that a highly entangled steady state can emerge from the dynamics if the system is prepared far from equilibrium. 
 We finally conclude in Sec.~\ref{sec:conclusions}, and leave the details of some of our calculations to the Appendix. 

\section{Ground State Phases}
\label{sec:model-ground-state}

In this Section, we employ the bosonic NRG~\cite{PhysRevB.71.045122,PhysRevLett.95.086406} to calculate the ground state phase diagram corresponding to the Hamiltonian in Eq. (\ref{eq:1}) as a function of dissipation strength $\alpha$ and Ising coupling $K$. We present the results for the ohmic, $s=1$, as well as the subohmic case of $s=1/2$. We point out similarities and differences to the situation of the single spin-boson model and to that of a two-spin model with two separate baths. 

Throughout this study we use the following parameters for our NRG calculations (we use the common notation): a discretization parameter of $\Lambda = 1.4$, a total of $N_{b,0} = 599$ bosonic modes in the first iteration and $N_{b,N} = 
6$ in the following ones, while keeping $N_{\text{Lev}} = 200$ low-energy levels in each NRG 
iteration. 

We obtain a qualitative understanding of the phase diagram by using the fact that the fast bath modes follow the spin dynamics adiabatically in the sense known from the famous Born-Oppenheimer approximation.~\cite{RevModPhys.59.1,weissdissipation} The spins are dressed by the bath phonons, and as a result the energy separation of the two lowest-energy spin states becomes renormalized. 
This situation is reminiscent of the single spin-boson model. There, the tunneling splitting $\Delta$ also becomes renormalized by the bath, and in the ohmic case, one finds a renormalized value of $\Delta_r = \Delta (\frac{\Delta}{\omega_c})^{\alpha/(1-\alpha)}$ for $\alpha < 1$ and a complete quench of the tunneling for $\alpha> 1$, where the system is thus localized.~\cite{RevModPhys.59.1,weissdissipation}
\begin{figure}[t]
  \centering
  \includegraphics[width=1.0\linewidth]{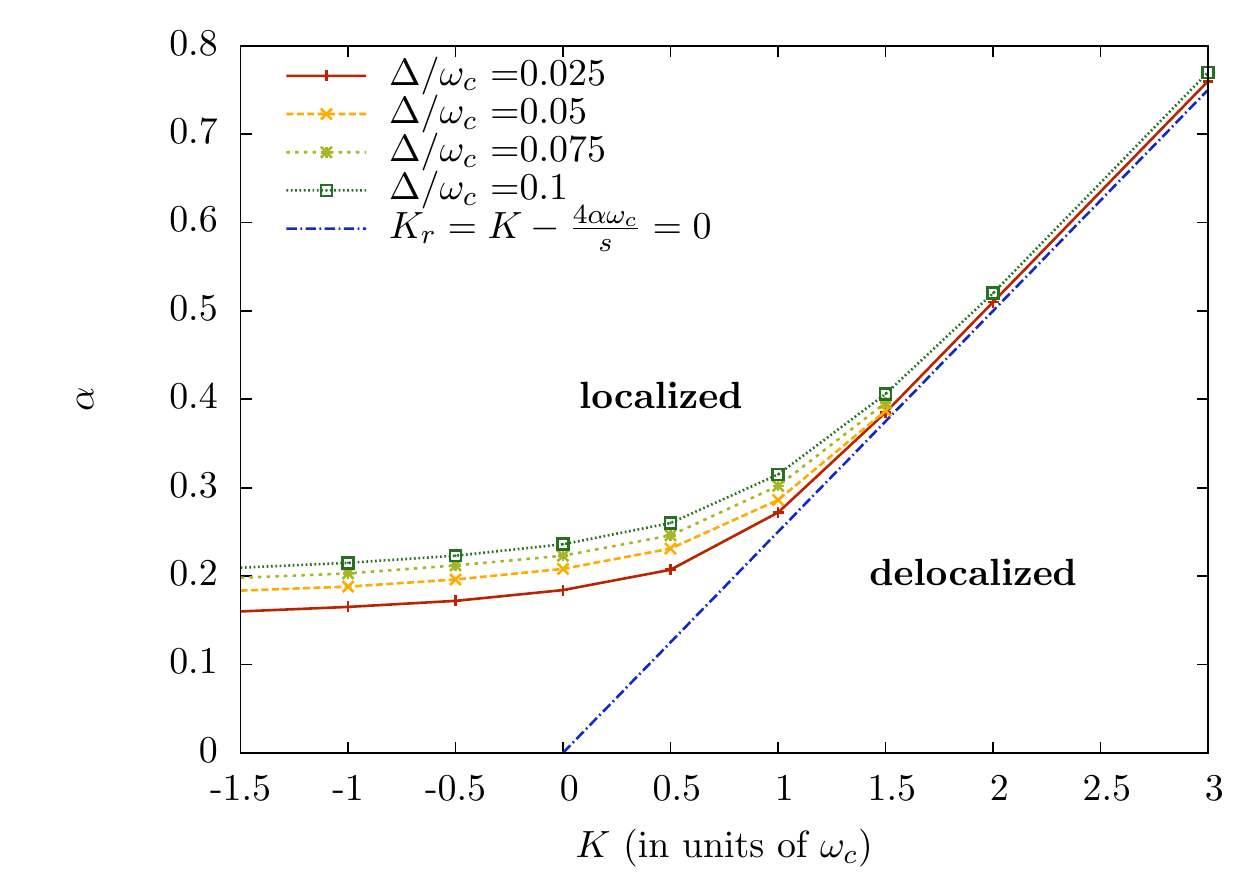}
  \caption{Phase diagram of the ohmic two-spin boson model as a function of dissipation strength $\alpha$ and Ising coupling $K$. Different curves correspond to different values of tunneling amplitudes $\Delta_1 = \Delta_2 \equiv \Delta$. For infinitesimal bias fields $\epsilon_{1,2} = -10^{-8} \omega_c$ the ground state of the system in the localized region is given by $\ket{\!\!\uparrow \uparrow} \otimes \ket{\Omega}$, where $\ket{\Omega}$ is a shifted bath vacuum [see Eq.~\eqref{eq:6}]. The dashed line indicates where the renormalized Ising interaction vanishes: $K_r=0$.}
  \label{fig:2}
\end{figure}

First, in Sec.~\ref{sec:nrg-phase-diagrams}, we present the numerically obtained phase diagrams. Then, in Sec.~\ref{sec:qual-underst-phase} we perform a strong-coupling analysis that will provide us with a qualitative understanding of the underlying physics. 
\subsection{NRG phase diagrams}
\label{sec:nrg-phase-diagrams}
\begin{figure}[t]
  \centering
  \includegraphics[width=1.0\linewidth]{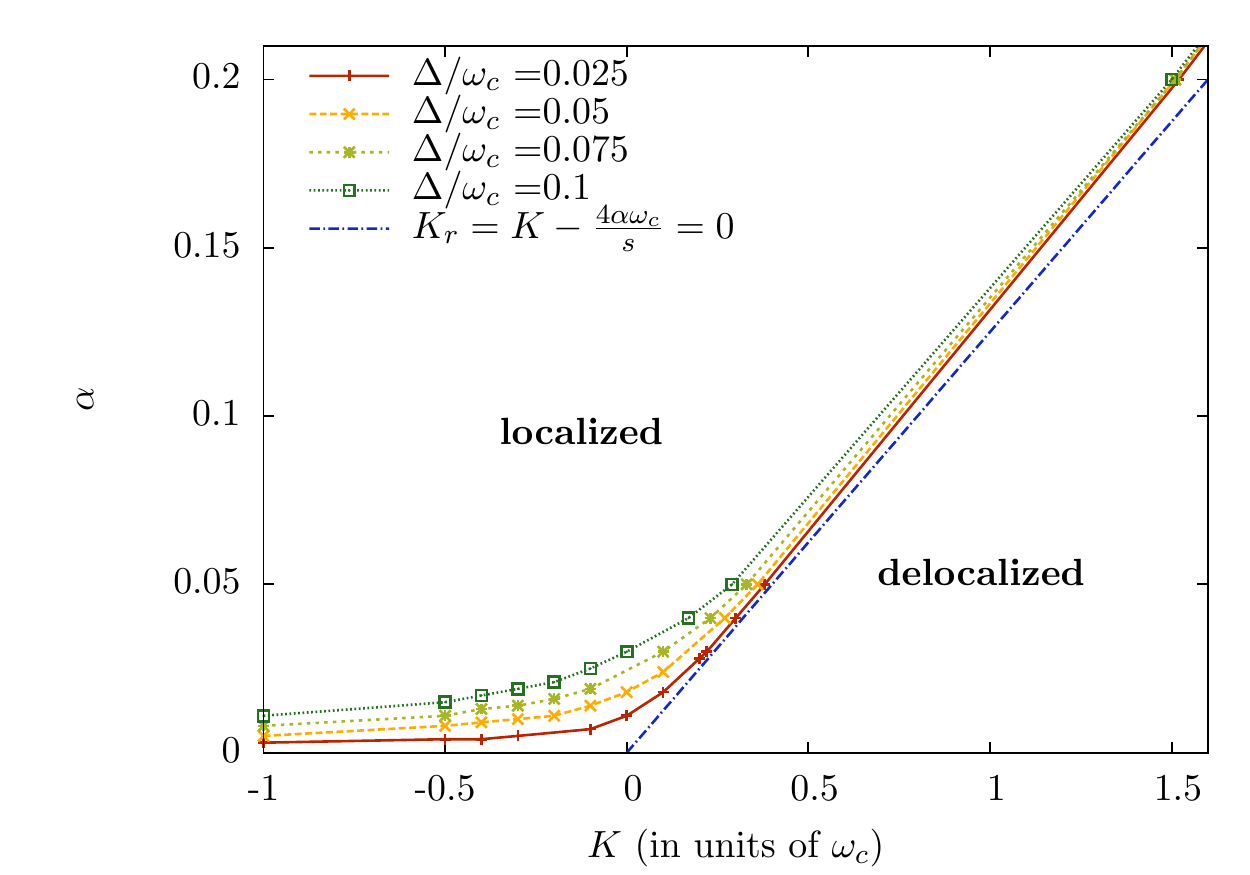}
  \caption{Phase diagram of the subohmic two-spin boson model with $s=1/2$ versus $\alpha$ and $K$, and for different values of $\Delta$. The dashed line indicates where $K_r=0$. }
  \label{fig:3}
\end{figure}

Using the NRG, we have determined the phase diagram of the two-spin boson model in Eq.~(\ref{eq:1}). We present results for an ohmic bath~\cite{tornow:035434,PhysRevB.81.235321} in Fig.~\ref{fig:2} and for a subohmic bath with $s=1/2$ in Fig.~\ref{fig:3}. Different curves correspond to different values of $\Delta/\omega_c$. Here, we assume equal tunneling amplitudes of the two spins $\Delta_1 = \Delta_2 \equiv \Delta$. 
Introducing slightly asymmetric tunneling elements $\Delta_1 \neq \Delta_2$, however, does not affect the location of the phase boundary much. Hereafter, we use units of the bath cutoff frequency, {\it i.e.}, we set $\omega_c = 1$, and we shall be mainly interested in the case where both $\Delta_{1,2} \ll \omega_c$ and $\epsilon_{1,2} \ll \omega_c$.

As shown in Figs.~\ref{fig:2} and ~\ref{fig:3}, the two-spin boson model exhibits two ground state phases: a \emph{delocalized phase}, where the spin expectation values $\av{\sigma^{z}_{1,2}}$ vanish in the ground state for $\epsilon_{1,2} \rightarrow 0$, and a \emph{localized phase}, where the spins develop a finite magnetization $\av{\sigma^{z}_{1}} = \av{\sigma^z_2} = \pm m$ ($m > 0)$ for infinitesimal bias fields $\epsilon_{1,2} = 0^\mp$. Like in the single spin-boson model, the system is delocalized for weak dissipation and enters a localized phase upon increasing $\alpha$. The phase boundary, however, now explicitly depends on the Ising interaction constant $K$. 

Let us first focus on the ohmic model in Fig.~\ref{fig:2}. For ferromagnetic $K < 0$, the phase boundary only weakly depends on $K$ and is located at $\alpha_c \approx 0.15 + \mathcal{O}(\frac{\Delta}{\omega_c})$, which is a much smaller value than in the single spin case, where the transition occurs at $\alpha_c^{\text{single}} = 1 + \mathcal{O}(\frac{\Delta}{\omega_c})$.~\cite{weissdissipation,RevModPhys.59.1} For antiferromagnetic $K > 0$, we find that the delocalized region extends up to larger values of $\alpha$ and we observe that the phase boundary occurs at the line $K = 4\alpha \omega_c/s$ for larger values of $K$. At this value of $K$ the renormalized Ising interaction $K_r$, which takes into account the bath induced ferromagnetic spin-spin interaction $(-4 \alpha \omega_c /s)$, vanishes. We defer the derivation of this formula until Sec.~\ref{sec:qual-underst-phase}.

Let us now turn to the subohmic case in Fig.~\ref{fig:3}. It shows the same qualitative features as the ohmic one, however, the system enters the localized phase for even smaller values of $\alpha$. On the ferromagnetic side $K<0$, our results suggest that $\alpha_c \approx 0 + \mathcal{O}(\frac{\Delta}{\omega_c})$, in agreement with the single spin case.~\cite{anders:210402} For antiferromagnetic $K>0$, the system again remains delocalized up to larger values of $\alpha$ and the phase transition occurs close to the line $K_r = 0$. Note that $K_r$ depends on the bath exponent $s$. 

\begin{figure}[t]
  \centering
  \includegraphics[width=1.0\linewidth]{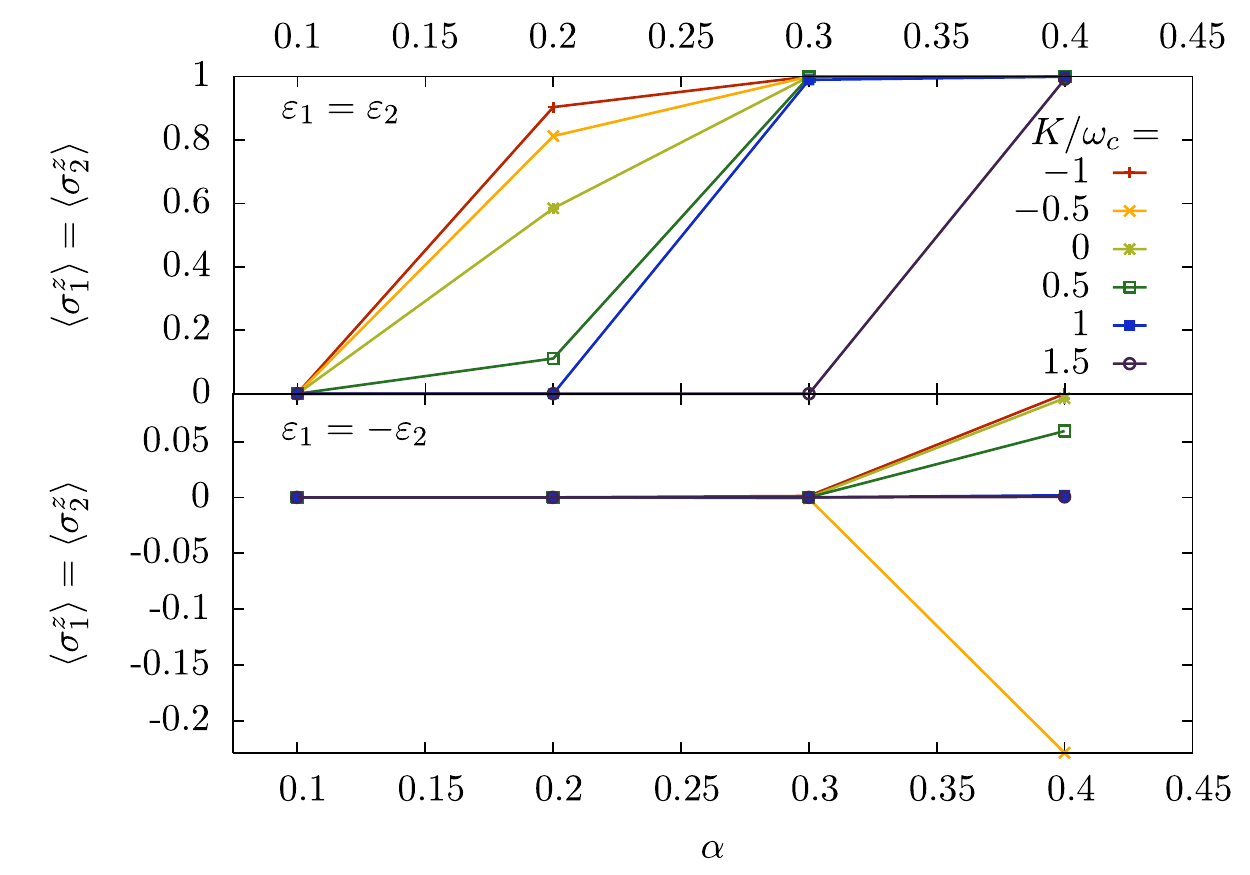}
  \caption{$\av{\sigma^z_{1,2}}$ as a function of $\alpha$ for various values of $K$ and $\Delta = 0.025 \, \omega_c$. Different bias field configurations are shown in the upper part (ferromagnetic, $\epsilon_1=\epsilon_2 = 10^{-8} \omega_c$) and lower part (antiferromagnetic, $\epsilon_1=-\epsilon_2 = 10^{-8} \omega_c$) of the figure. This plot shows that spins are always aligned in the localized phase. The expectation values $\av{\sigma^z_{1,2}}$ remain zero up to larger values of $\alpha$ simply because the antiferromagnetic bias field configuration does not lift the degeneracy of the ground states $\{\ket{\!\!\uparrow \uparrow}, \ket{\!\!\downarrow \downarrow}\}$ in the localized phase.}
  \label{fig:4}
\end{figure}

We distinguish the two phases by applying small bias fields $\epsilon_{1,2} = 10^{-8} \omega_c$ and measure $\av{\sigma^z_{1,2}}$. The latter vanishes in the delocalized region, but remains nonzero $\av{\sigma^z_1} = \av{\sigma^z_2} = -m$ ($m>0$) in the localized part of the phase diagram. We have also applied an antiferromagnetic bias field configuration $\epsilon_1 = - \epsilon_2 = 10^{-8} \omega_c$ to test whether the system can also localize in an antiferromagnetic spin configuration $\{\ket{\!\!\uparrow \downarrow}, \ket{\!\!\downarrow, \uparrow}\}$. Interestingly, however, we observe in Fig.~\ref{fig:4} that the spins always localize
in one of the \emph{ferromagnetic} spin states $\{\ket{\!\!\uparrow \uparrow}, \ket{\!\!\downarrow \downarrow}\}$. The system does not localize in any of the antiferromagnetic spin configurations. We provide a physical explanation for this phenomenon in Sec.~\ref{sec:qual-underst-phase}. Results for 
$\av{\sigma^z_{1,2}}$ as a function of $\alpha$ for both bias field configurations and different values of $K$ are shown in Fig.~\ref{fig:4}. We observe that $\av{\sigma^z_{1,2}}$ remains zero up to a larger value of $\alpha$ (for fixed $K$) simply because the antiferromagnetic bias fields $\epsilon_1 = - \epsilon_2$ do not lift the degeneracy of the two ground states $\{\ket{\!\!\uparrow \uparrow}, \ket{\!\!\downarrow \downarrow}\}$. The location of the phase boundary does of course not depend on the infinitesimal fields. 

\subsection{Qualitative understanding of the phase diagram}
\label{sec:qual-underst-phase}

From the previous considerations, immediately the questions arise why the phase diagram is not symmetric under the combined transformation of $\{K \rightarrow - K$, $\boldsymbol{\sigma}_2 \rightarrow -\boldsymbol{\sigma}_2\}$, and why the system cannot localize in one of the antiferromagnetic spin states $\{\ket{\!\!\uparrow \downarrow}, \ket{\!\!\downarrow \uparrow}\}$. 

In order to answer these questions, we perform a strong-coupling analysis which relies on the fact that the fast modes of the bath ($\omega_k \gg \Delta$) adiabatically renormalize the energy separation of different spin states.~\cite{RevModPhys.59.1} In physical terms, assuming that the bath oscillators follow the time evolution of the spins immediately (Born-Oppenheimer approximation), the spins are dressed by phonons with frequencies larger than $\Delta$. Thus, transitions between different spin states are suppressed if they involve a readjustment of the bath excitations. We will consider the ferromagnetic and antiferromagnetic cases separately.

Let us first note, however, that the bath induces a ferromagnetic interaction between the spins, which renormalizes the value of the Ising constant from its bare value of $K$ to
\begin{equation}
  \label{eq:3}
  K_r = K - \frac{4 \alpha \omega_c}{s}\,.
\end{equation}
This is most easily derived by applying the polaron unitary transformation $U = \exp[- \frac{1}{2} (\sigma^z_1 + \sigma^z_2) \sum_k \frac{\lambda_k}{\omega_k} (b^\dag_k - b_k)]$ to the Hamiltonian in Eq.~(\ref{eq:1}), which yields for $\tilde{H} = U^{-1} H U$:
\begin{align}
  \label{eq:4}
  \tilde{H} &= \sum_{j=1}^2 \Bigl[ \frac{\Delta_j}{2} ( \sigma^+_j e^{i \Omega} + \text{h.c.} ) + \frac{\epsilon_j}{2} \sigma^z_j \Bigr] + \frac{K_r}{4} \sigma^z_1 \sigma^z_2 \nonumber \\ &+ \sum_{k>0} \omega_k b^\dag_k b_k \,,
\end{align}
where the hermitian bath displacement operator reads 
\begin{equation}
  \label{eq:5}
  \Omega = - i \sum_k \frac{\lambda_k}{\omega_k} (b^\dag_k - b_k)\,.
\end{equation}
This form of the Hamiltonian makes explicit the bath induced ferromagnetic Ising interaction. In particular, if the bare Ising coupling is antiferromagnetic $K>0$, the effective interaction changes sign at a dissipation strength of $\alpha = \frac{s K}{4 \omega_c}$. For larger values of $K \geq \omega_c$, the phase transition occurs close to this critical value of $\alpha$, as shown in Figs.~\ref{fig:2} and \ref{fig:3}. From $\tilde{H}$, we can also learn immediately that a spin flip is associated with a complex excitation of the bosonic bath into a coherent state $\ket{\Omega} = e^{i \Omega} \ket{0}$, where $\ket{0}$ is the ground state of the free bath part of the Hamiltonian $H_B = \sum_k \omega_k b^\dag_k b_k$.

With this in mind, let us begin our strong-coupling analysis of the phase diagram with the \emph{ferromagnetic situation $K < 0$}, and assume that $|K| \gg \Delta_{1,2}$ and zero bias $\epsilon_{1,2} = 0$. For $\Delta_{1,2}=0$, the two lowest energy spin states are given by the two ferromagnetic states $\{\ket{\!\!\uparrow \uparrow}, \ket{\!\!\downarrow \downarrow}\}$. If we now turn on the tunneling $\Delta_1 = \Delta_2 = \Delta$, we find that the energy splitting between the two lowest states is of the order
\begin{equation}
  \label{eq:6}
  \delta E \sim \frac{ 2 \Delta^2}{|K|} \braket{\Omega}{- \Omega}\,,
\end{equation}
where the coherent state $\ket{\Omega} = e^{i \Omega} \ket{0}$ is also referred to as the displaced oscillator bath state. It occurs when all oscillators equilibrate in contact with spins that are held fixed in position $\ket{\!\!\downarrow \downarrow}$. 
In terms of the spectral density, the bath renormalized energy splitting becomes 
\begin{equation}
  \label{eq:7}
  \delta E \sim \frac{2 \Delta^2}{|K|} \exp \Bigl[ - \frac{1}{\pi} \int_{p \delta E}^\infty d\omega \frac{J(\omega)}{\omega^2} \Bigr]\,,
\end{equation}
where $p \gg 1$. To be consistent with the adiabatic renormalization scheme, the energy splitting $\delta E$ shows up as an infrared cutoff for the oscillator frequencies that are summed over. Since the bath renormalizes the energy splitting to smaller values $\delta E < \frac{2\Delta^2}{|K|}$, one can solve Eq.~\eqref{eq:7} iteratively.~\cite{RevModPhys.59.1,weissdissipation} In the case that $\delta E$ is renormalized to zero, the ground state is doubly degenerate and the system localized. This situation, where the displaced bath states $\ket{\Omega}$ and $\ket{-\Omega}$ are orthogonal to each other, is known as orthogonality catastrophe.~\cite{weissdissipation} If $\delta E$ is renormalized to a nonzero value, the ground state is unique and the system delocalized. 

For a subohmic spectral density, the iteration process yields $\delta E = 0$ for any positive value of $\alpha$, and the system is localized as soon as $\alpha > 0$. In the ohmic case, on the other hand, we find that as long as $\alpha < 1/2$, the energy splitting renormalizes to the finite value $\delta E = \delta E_0 \bigl(\frac{ \delta E_0}{\omega_c}\bigr)^{2 \alpha/(1-2\alpha)}$ where $\delta E_0 = 2 \Delta^2/|K|$. For $\alpha > 1/2$, however, one finds $\delta E = 0$ and the system is localized. The phase transition occurs at the critical value $\alpha_c = 1/2$. The same value was recently found using a variational treatment.~\cite{PhysRevB.81.235321}
Let us remark that in the case of the single spin-boson model, one has to calculate the overlap integral $\braket{\frac{\Omega}{2}}{-\frac{\Omega}{2}} = \exp[- \frac{1}{2\pi} \int_0^\infty d\omega \frac{J(\omega)}{\omega^2}]$, which leads to $\alpha_c^{\text{single}} = 1$.~\cite{RevModPhys.59.1,weissdissipation}   
This also implies that the delocalized phase in the two-spin case is characterized by a distinct
Kondo scale compared to the single spin-boson model.\cite{RevModPhys.59.1}

Our NRG calculation, which goes beyond this simple approximation and the variational approach of Ref.~\onlinecite{PhysRevB.81.235321}, indeed  shows that the critical value of $\alpha$ in the ferromagnetic regime only weakly depends on $K$. In the ohmic case, we observe, however, that $\alpha_c$ rather converges to $\alpha_c(s=1)\approx 0.15$ for large $|K|$ and $\frac{\Delta}{\omega_c} \rightarrow 0$ instead of the approximated value $\alpha_c = 1/2$. In the subohmic case, on the other hand, NRG agrees with the predicted value of $\alpha_c = 0$ as we find $\alpha_c(s=1/2) \approx 0$ for $\frac{\Delta}{\omega_c} \rightarrow 0$.


We now turn to the \emph{antiferromagnetic situation $K > 0$}. 
Since we want to investigate the antiferromagnetic regime, we thus have to assume that $K_r > 0$ (or $K \rightarrow \infty$ for any value of $\alpha$). Then, the two lowest energy states for zero tunneling ($\Delta = 0$) are degenerate in energy and given by $\{\ket{\!\!\uparrow \downarrow}, \ket{\!\! \downarrow \uparrow}\}$. If we turn on tunneling, the two states hybridize and the energy difference between the two lowest energy states reads
\begin{equation}
  \label{eq:8}
  \delta E \sim \frac{2 \Delta^2}{K} \braket{0}{0} = \frac{\Delta^2}{K}\,,
\end{equation}
where $\ket{0}$ is the unshifted bath vacuum. Hence, any nonzero value of $\Delta$ leads to a unique ground state, because the quenching of the tunneling amplitude due to the bath does not occur for a total spin zero state [compare with Eq.~(\ref{eq:6})]. [This can also be interpreted as the disappearance of
Kondo-type entanglement for a spin zero state.\cite{KarynLeHur-UnderstandingQPT-Article}] As a result, the system is always delocalized for an antiferromagnetic Ising coupling $K_r > 0$, and the phase transition to the localized state is shifted to much larger values of $\alpha$ necessary to compensate the antiferromagnetic spin-spin coupling constant $K$.

\section{Phase transitions and Scaling}
\label{sec:phase-trans-scal}

In this Section, we investigate the behavior of the system close to the localization phase transition in more detail. It is known, that the transition is in the Kosterlitz-Thouless universality class for the ohmic system,~\cite{KarynLeHur-UnderstandingQPT-Article,PhysRevB.69.214413} but it is of continuous type in the subohmic case.~\cite{vojta_philmag_2006, lehur_entanglement_spinboson} Since recent studies show that NRG is not well-suited to describe the system correctly close to the transition for $s<1/2$,~\cite{PhysRevLett.102.030601,PhysRevB.81.075122} we restrict ourselves to $s\geq 1/2$.

In Sec.~\ref{sec:entanglement-entropy}, we first study the behavior of the entanglement entropy in the ohmic and subohmic system. 
We then examine in Sec.~\ref{sec:subohm-syst-scal} the scaling of the spin expectation values $\av{\sigma^z_{1,2}}$ close to the phase transition in the subohmic system. We derive mean-field scaling relations for the critical exponents from an effective spin action functional, and compare the resulting exponents to the critical exponents that we extract from NRG.

\subsection{Static entanglement entropy}
\label{sec:entanglement-entropy}
\begin{figure}[t]
  \centering
    \includegraphics[width=1.0\linewidth]{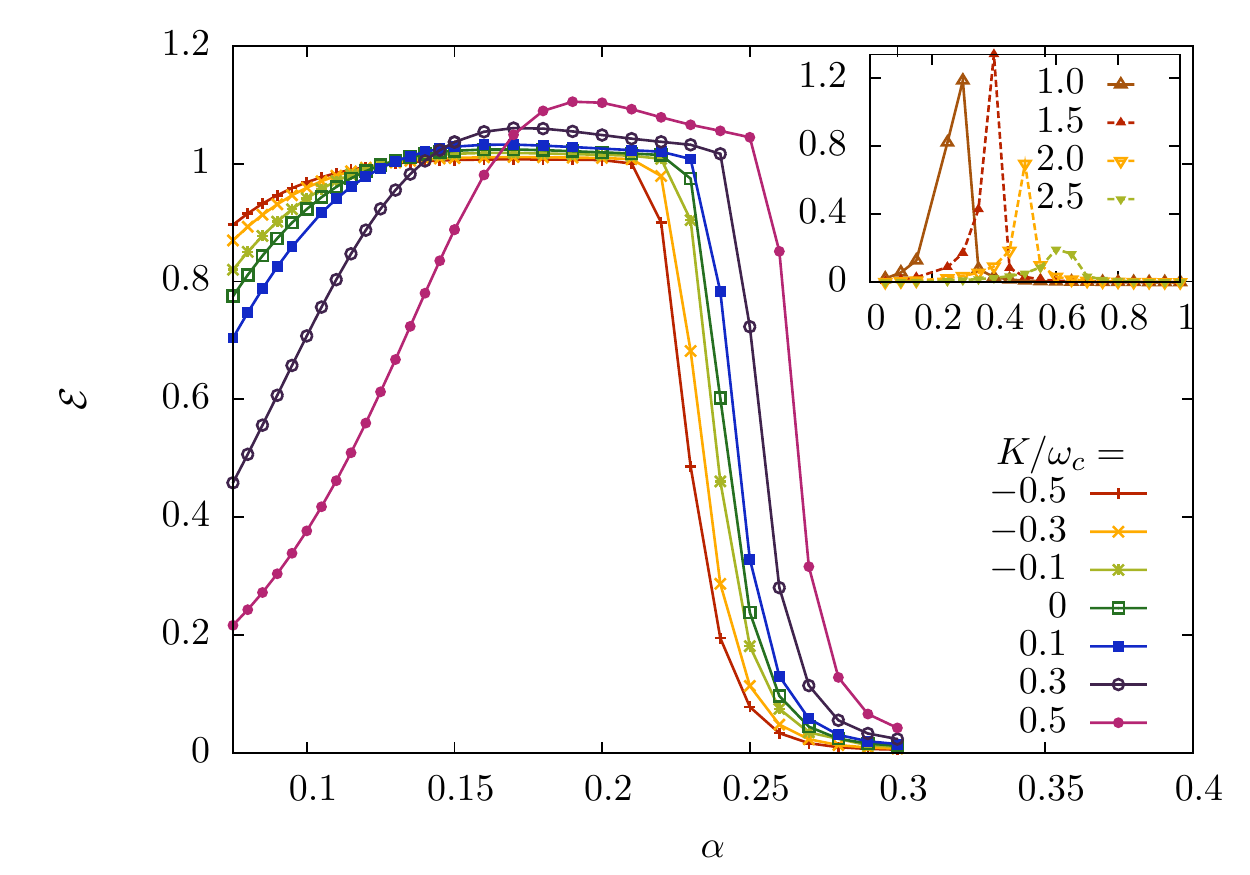}
  \caption{Entanglement entropy $\mathcal{E}$ as a function of dissipation $\alpha$ in the ohmic two-spin boson model, shown for different values of the Ising coupling $K$ and $\Delta_{1,2} = 0.1 \omega_c.$. The rapid drop to zero around $\alpha_c \approx 0.25$ signifies the transition to the localized phase. The plateau for smaller dissipation indicates the loss of phase coherence at $\alpha \approx \alpha_c/2$ similar to the single spin-boson case. The inset shows larger values of $K$ where the (incoherent) plateau shrinks to a peak-like structure, indicating that coherence is lost only right at the phase transition.  }
  \label{fig:5}
\end{figure}
The entanglement entropy $\mathcal{E}$ quantifies the degree of entanglement between the spins and the bath. It is defined as~\cite{PhysRevA.54.3824}
\begin{equation}
  \label{eq:9}
  \mathcal{E} = - \text{Tr} [ \rho_S \log_2 \rho_S]\,,
\end{equation}
where $\rho_S = \text{Tr}_B \rho$ is the reduced density matrix of the two spins. Here, $\text{Tr}_B$ denotes taking the trace over the bath degrees of freedom, and $\rho$ is the full density matrix of the spin-boson system. One finds that $0 \leq \mathcal{E}\leq \log_2 4 = 2$, where $\mathcal{E} = 0$ in the absence of entanglement between spin and bath. In Fig.~\ref{fig:5} we show results for the entanglement entropy in the ohmic system as a function of dissipation $\alpha$ for different values of Ising coupling $K$. Like in the case of the single spin-boson model, the entanglement entropy is nonzero only in the delocalized phase and rapidly falls to zero at the phase transition. It reaches a plateau for $\alpha \approx \alpha_c/2$, indicating that coherence is lost already before the system becomes localized. The plateau characterizes a region of maximal decoherence, where the spin dynamics is incoherent. This coherent-to-incoherent crossover is known from the single spin system,~\cite{kopp:220401, PhysRevA.68.034301, lehur_entanglement_spinboson} where it occurs exactly at the Toulouse point $\alpha = 1/2$. In Sec.~\ref{sec:toulouse-point-1}, we discuss the equivalent of the Toulouse point in the two-spin model where it is located at $\alpha = 1/2$ and $K = 2 \omega_c$. 

Surprisingly, as we show in the inset of Fig.~\ref{fig:5}, the plateau shrinks considerably if we go to larger positive values of $K \gtrsim \omega_c$. The plateau more and more resembles a peak-like structure. This indicates that the localization phase transition occurs much closer to the regime, where spin oscillations are coherent. Coherence is lost only right at the transition (similar to the subohmic case discussed below). This is different from the single spin case, where the incoherent regime extends between $1/2 \leq \alpha \leq 1$ and is thus much larger. 

\begin{figure}[t]
  \centering
  \includegraphics[width=.94\linewidth]{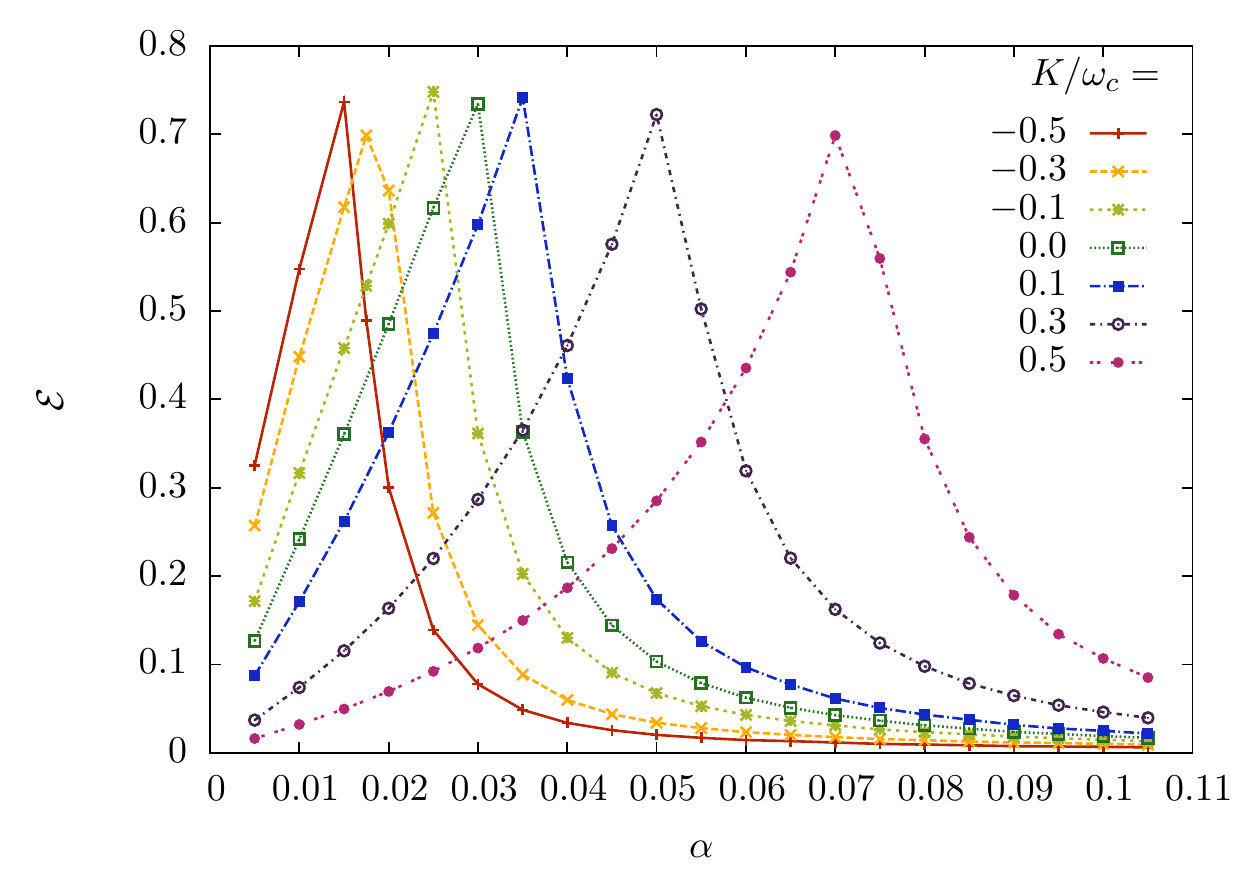}
  \caption{Entanglement entropy $\mathcal{E}$ as a function of dissipation $\alpha$ for the subohmic two-spin boson model with $s=1/2$. Different curves are for different values of the Ising coupling $K$, and $\Delta_{1,2} = 0.1 \omega_c$. The entropy $\mathcal{E}$ reaches a maximum at the phase transition (see also Fig.~\ref{fig:3}), and falls off continuously to both sides of the transition. }
  \label{fig:6}
\end{figure}

Finally, we show in Fig.~\ref{fig:6} that for a subohmic bath, the entanglement entropy rather reaches a maximum (peak) right at the localization quantum phase transition. This behavior is known from the single spin-boson system.~\cite{lehur_entanglement_spinboson} 
It signifies that the coherence of the spin oscillations (continuously) decreases toward the phase transition. There is no region where the spin transitions are completely incoherent. In fact, coherent spin oscillations of $\sigma^z(t)$ even persist into the localized phase, where they occur around a nonzero expectation value $\av{\sigma^z} \neq 0$ (see Ref.~\onlinecite{anders:210402} for the single and Sec.~\ref{sec:large-spin-bath} for the two-spin boson model). 

\subsection{Scaling of magnetization for subohmic bath}
\label{sec:subohm-syst-scal}
In this Section, we investigate the scaling of the spin expectation values $\av{\sigma^z_{1,2}}$ (magnetization) at the phase transition in the subohmic system. For the single spin-boson system, it is known that the phase transition is continuous for $s<1$, and scaling exponents have been extracted using NRG~\cite{PhysRevLett.94.070604,hur:126801,lehur_entanglement_spinboson} and Quantum Monte-Carlo calculations.~\cite{PhysRevLett.102.030601} Recently, it was realized that NRG is not well-suited to describe scaling correctly for $s<1/2$ in the single spin-boson model.~\cite{PhysRevLett.102.249904}
Therefore, we only consider exponents in the range $1/2 \leq s < 1$. 

We proceed in the following manner. First, in Sec.~\ref{sec:effect-spin-acti}, we derive an effective spin action functional by integrating over the bosonic degrees of freedom. From this action we determine, in Sec.~\ref{sec:scal-analys-comp}, the scaling dimension of the spin operators in a mean-field approximation from which follow scaling laws. We compare the resulting mean-field values for the critical exponents to those that we have extracted from the NRG calculations, and find good agreement between most of them. On the one hand this justifies our mean-field approximation, but on the other hand it also shows that the NRG analysis goes beyond this approximation. 

\subsubsection{Effective spin action functional}
\label{sec:effect-spin-acti}

An effective action functional $S_{\text{eff}}$ for the spins can be obtained by integrating over the bosonic degrees of freedom using a functional integral description.~\cite{negele_orland_book} This can be done exactly, because the Hamiltonian in Eq.~\eqref{eq:1} is quadratic in bosonic operators. 

We start with the action of the full system $S = S_S + S_B + S_{SB}$, where $S_S = \int_0^\beta d\tau \sum_{j=1}^2 [ \frac{\Delta_j}{2} \sigma^x_j(\tau) + \frac{\epsilon_j}{2} \sigma^z_j(\tau)] + \frac{K}{4} \sigma^z_1(\tau) \sigma^z_2(\tau)$ depends on spin variables only, and $S_{B} = \int_0^\beta d\tau \sum_{k} b_k^*(\tau)[ \frac{\partial}{\partial \tau} + \omega_k ] b_k(\tau)$ denotes the action of the free bath. The spin-bath interaction is described by $S_{SB} = \frac12 \int_0^\beta d\tau \sum_{k} \sum_{j=1}^2 \lambda_k \sigma^z_j(\tau) [b^*_k(\tau) + b_k(\tau)]$. Here, $\beta= 1/T$ ($k_B = 1$) is an inverse temperature, $\tau$ is an imaginary time variable and $b(\tau)$ are the usual complex boson coherent state variables. Note that in the end we will take the zero temperature limit which is well-defined in this formalism.~\cite{negele_orland_book}

Integrating over the (complex) bosonic variables~\cite{orth:051601} $\int \mathcal{D}[b^*_k(\tau),b_k(\tau)] \exp[- S_B - S_{SB}]= \exp[-S']$, leads to an effective spin action $S_{\text{eff}} = S_S + S'$. In the zero temperature limit, it takes the form
\begin{equation}
  \label{eq:10}
\begin{split}
  &S_{\text{eff}} = \int_0^\infty d\tau \Bigl\{ \sum_{j=1}^2 \Bigl[ \frac{\Delta}{2} \sigma^x_j(\tau) + \frac{\epsilon_j}{2} \sigma^z_j(\tau) \Bigr] + \frac{K_r}{4} \sigma^z_1(\tau) \sigma^z_2(\tau) \Bigr\} \\ &+ \int_0^\infty \frac{d\tau d\tau'}{16 \pi} \int d\omega J(\omega) e^{- \omega |\tau - \tau'|} \Bigl\{ \sum_{j=1}^2 \Bigl[ \sigma^z_j(\tau) - \sigma^z_j(\tau')\Bigr] \Bigr\}^2\,.
\end{split}
\end{equation}
The effect of the bosons on the spins is twofold: first, the Ising interaction constant gets renormalized to $K_r = K - 4 \alpha \omega_c/s$ by the term that is local in imaginary time. Second, the bath introduces dissipation as described by the last term in Eq.~\eqref{eq:10}, which is purely non-local in imaginary time. Integrating over frequency $\omega$, we observe that this last term describes a \emph{long-range interaction} in imaginary time
\begin{equation}
  \label{eq:11}
  \int_0^\infty d\omega J(\omega) e^{- \omega | \tau - \tau|} = \frac{2 \pi \alpha \omega_c^2 \Gamma (1+s)}{ ( 1 + \omega_c |\tau - \tau'|)^{1+s}}\,,
\end{equation}
where $\Gamma(x)$ is the Gamma function and we have used an exponential cutoff for the spectral density $J(\omega) = 2 \pi \alpha \omega_c^{1-s} \omega^s \exp[ -\omega/\omega_c]$ for convenience. 

Note that although the dissipative part still contains a term that couples the two different spins (at different times), this corresponds to a retarded Ising interaction and can thus be neglected compared to the equal-time contribution, if one is interested in ground-state properties. More specifically, the retarded term is of the form $\int d\tau d\tau' \frac{\sigma^z_1(\tau) \sigma^z_2(\tau')}{|\tau - \tau'|^{1+s}}$, which under a Fourier transformation becomes $ \sum_{\omega_n} |\omega_n|^s \sigma^z_1(\omega_n) \sigma^z_2(-\omega_n)$. Thus, if we pass to real frequencies $\omega_n \rightarrow \omega + i \delta$ and take the low-frequency limit $\omega \rightarrow 0$ these terms can be neglected compared to the static Ising interaction part $\frac{K_r}{4} \sigma^z_1(\tau) \sigma^z_2(\tau)$. This reasoning can also be justified by noting that one arrives at the same formula for the renormalized Ising constant by applying the polaron unitary transformation to the Hamiltonian in Eq.~\eqref{eq:1} as we have presented in Sec.~\ref{sec:qual-underst-phase} (see Eq.~\eqref{eq:4}).

\subsubsection{Scaling analysis: comparison between mean-field and NRG exponents}
\label{sec:scal-analys-comp}

Below, we derive mean-field critical exponents from the effective spin action in Eq.~\eqref{eq:10}, which we compare with exponents that we have extracted from our NRG calculations.

To proceed, we resort to a mean-field-like decoupling of the Ising term: $\frac{K_r}{4} \sigma^z_1(\tau) \sigma^z_2(\tau) \approx \frac{K_r}{4} [\sigma^z_1 \av{\sigma^z_2} + \av{\sigma^z_1} \sigma^z_2]$. This term then acts as a single-spin detuning, depending on the expectation value of the other spin magnetization. 

Scaling of both spins will thus be identical and we can follow the analysis for the single spin-boson model.~\cite{PhysRevLett.102.030601,PhysRevLett.102.150601,vojta_philmag_2006,lehur_entanglement_spinboson} There, one employs the quantum-to-classical mapping of the spin-boson model to the one-dimensional classical Ising model~\cite{PhysRevB.9.215,RevModPhys.59.1,sachdev_qpt_book}
\begin{equation}
  \label{eq:12}
  H_{\text{classical}} = - \sum_{i,j} J_{ij} S^z_i S^z_j + H_{\text{short-range}}
\end{equation}
with long-range interaction $J_{ij} = J/|i - j|^{1+s}$. Here, $S^z_i=\pm 1$ are classical Ising spins. There is an additional generic short-range interaction $H_{\text{short-range}}$ arising from the transverse field, but it is believed to be irrelevant for the critical behavior.~\cite{PhysRevLett.94.070604,PhysRevLett.29.917,PhysRevB.56.8945} The scaling dimensions of $\sigma^z_{1,2}$ are thus solely determined by the dissipative term, and we find from the condition that the total action is dimensionless $[S] = 1$ that 
\begin{equation}
  \label{eq:13}
  [\sigma^z_{1,2}] = T^{\frac{1-s}{2}}\,.
\end{equation}
Here, we have used units of energy (or temperature): $[\tau] = T^{-1}$. From this follows the scaling dimension of the detuning and Ising constant as 
\begin{equation}
  \label{eq:14}
  [\epsilon] = T^{\frac{1+s}{2}}, \qquad [K] = T^s\,.
\end{equation}
In order to derive scaling relations, we need to make an ansatz for the impurity part of the free energy. Since the fixed point is ``interacting'' for $s>1/2$,~\cite{vojta_philmag_2006,PhysRevLett.89.076403} we use
\begin{equation}
  \label{eq:15}
  F_{\text{imp}} = T f(|\Delta - \Delta_c|T^{-1/\nu}, \epsilon T^{-b}, |K - K_c|T^{-\kappa})\,.
\end{equation}
This ansatz can be applied for $s<1$ since the transition is continuous. Further, for a Gaussian fixed point, which occurs at $s<1/2$, the reduced free energy would also depend on dangerously irrelevant variables.

 In this ansatz we have used that in a quantum phase transition, which occurs at $T=0$, the distance to criticality is measured by the parameter deviation from the critical value of the most relevant perturbation, in this case $|\Delta - \Delta_c|$. Analogous to a classical system, where the correlation length diverges as a function of this distance, here the correlation length in imaginary time obeys $\xi \sim |\Delta - \Delta_c|^{-\nu}$ with the correlation length exponent $\nu$. The dynamic critical exponent is formally set equal to $z=1$ in this $0+1$-dimensional system. This defines a characteristic energy scale 
\begin{equation}
  \label{eq:16}
  T^* \equiv \xi^{-1} \sim |\Delta - \Delta_c|^{\nu}\,,
\end{equation}
above which critical behavior is observed.~\cite{sachdev_qpt_book}

Using the ansatz for the free energy given in Eq.~\eqref{eq:15}, we can immediately infer from $[\epsilon T^{-b}] = [|K-K_c|T^{-\kappa}] = 1$ that $b = \frac{1+s}{2}$ and $\kappa = s$. If we define the critical exponents describing the scaling of the magnetization as
\begin{align}
  \label{eq:17}
  \av{\sigma^z_{1,2}} &\sim |\Delta - \Delta_c|^\beta \\
\label{eq:18}
\av{\sigma^z_{1,2}} &\sim |\epsilon_{1,2}|^{1/\delta} \\
\label{eq:19}
\av{\sigma^z_{1,2}} &\sim |K - K_c|^\zeta\,,
\end{align}
we can derive mean-field scaling relations. For instance from Eqs.~\eqref{eq:13},~\eqref{eq:14} it immediately follows that
\begin{align}
  \label{eq:20}
  \delta_{MF} &= \frac{1 + s}{1 - s}, \qquad  \zeta_{MF} = \frac{1-s}{2s}\,.
\end{align}
We have to invoke Eq.~\eqref{eq:16} to arrive at
\begin{equation}
  \label{eq:21}
  \beta_{MF} = \nu \Bigl(\frac{1-s}{2}\Bigr)\,.
\end{equation}
If we use the result that $\nu=1/s$ for small $s$, derived in Ref.~\onlinecite{PhysRevLett.94.070604}, we find that $\zeta_{MF}=\beta_{MF}$. Close to $s=1$ it is more appropriate to use $1/\nu = \sqrt{2 (1-s)}$ as obtained in Ref.~\onlinecite{PhysRevLett.37.1577}. The resulting values for the critical exponents are shown in Table~\ref{tab:scaling_exp_NRG}.

\begin{figure}[t]
    \centering
        \includegraphics[width=\linewidth]{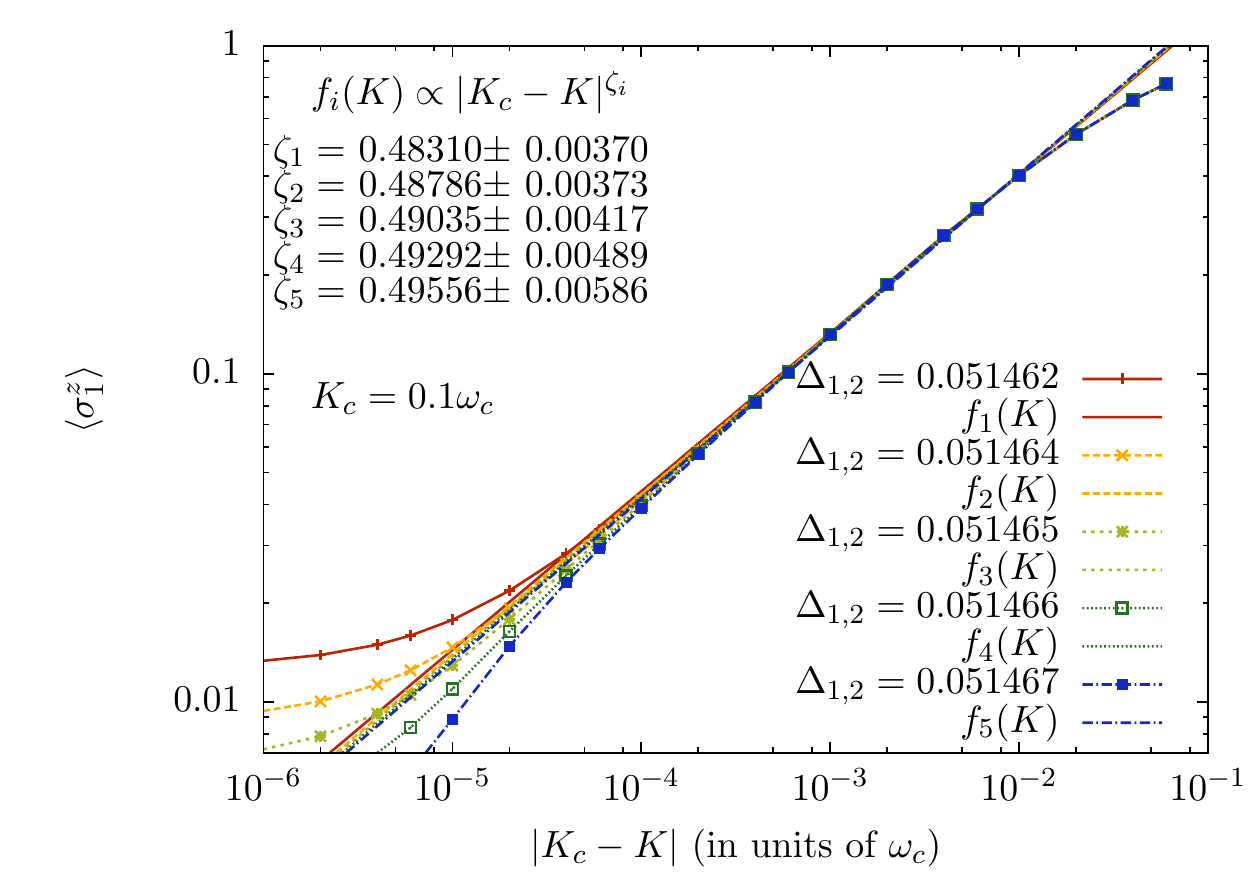}
        \caption{Scaling of magnetization at the phase transition in the subohmic system with $s=\frac12$. We fit $\av{\sigma^z_1}$ as a function of the Ising interaction against a power-law $\av{\sigma^z_1} \sim |K - K_c|^\zeta$, and find $\zeta = 0.5$. Different lines represent fits using $f_{i} \propto |K - K_c|^{\zeta_i}$. Results of the fit and error bars for $\zeta_i$, as well as the different values of $\Delta_{1,2}$ (in units of $\omega_c$) used, are shown in the plot. 
    }
    \label{fig:7}
\end{figure}

Let us now compare these mean-field predictions of the critical exponents to our NRG results. Numerically, we investigate the cases $s=\{ \frac12, \frac34, \frac{9}{10}\}$. After carefully determining the position of the phase transition, we keep all but one parameter fixed at their critical values, and study the scaling of the magnetization as a function of this remaining parameter. Typically, we find power law scaling over more than two orders of magnitude, and we find the exponents from simply fitting the slope in a log-log plot. We have checked that the extracted value of the exponent is independent of the position in the phase diagram where we cross the phase boundary. As an example, in Fig.~\ref{fig:7}, we show the scaling of $\av{\sigma^z_1}$ as a function of $|K-K_c|$. Different curves are for different values of the transverse field $\Delta$, and we extract the value of $\zeta(s=\frac12) = 0.5$, which is is perfect agreement with the mean-field prediction of $\zeta_{MF}(s=\frac12) = 1/2$.  
\begin{table}
    \centering
    \begin{tabular}[b]{c|c|c|c}
        \hline \hline
        Exponent & $s =  \frac12$ &$ s =  \frac34$ &$ s =  \frac{9}{10}$ \\
        \hline
        $\delta $&$ 4 $&$ 10 $&$ 40 $ \\
        $\delta_{\text{MF}}$ & $ 3 $&$ 7 $&$ 19 $ \\
        \hline
        $\zeta $&$ 0.5 $&$ 0.2 $&$ 0.1 $ \\
        $\zeta_{\text{MF}} $&$ 1/2 $&$ 1/6 \simeq 0.17 $&$ 1/18 \simeq 0.06 $ \\
        \hline
        $\beta $&$ 0.5 $&$ 0.2 $&$ 0.09 $ \\
        $\beta_{\text{MF}}(\nu=1/s)$ & $1/2$ & $1/6$ & $1/18$ \\
        $\beta_{\text{MF}}(\nu=1/\sqrt{2(1-s)})$ & $1/4 $ & $1/4 \sqrt{2} \simeq 0.18$ & $1/4\sqrt{5} \simeq 0.11$ \\ 
        \hline
    \end{tabular}
    \caption{Comparison of critical exponents as predicted by our mean-field analysis $\{\delta_{\text{MF}}, \zeta_{\text{MF}}, \beta_{\text{MF}}\}$ and as extracted from NRG $\{\delta, \zeta, \beta\}$.}
    \label{tab:scaling_exp_NRG}
\end{table}
In Table~\ref{tab:scaling_exp_NRG} we show a full comparison of the critical exponents derived in the mean-field approximation and extracted from NRG.  Agreement is good for the exponents $\zeta$ and $\beta$ for all values of $s$ (using the different expansions of $\nu$ as a function of $s$). For the exponent $\delta$, however, the agreement is not so good in the cases $s = \{\frac34, \frac{9}{10}\}$. Note that the trend is captured correctly and that $\delta$ diverges as $s \rightarrow 1$ which makes it increasingly hard to extract its value numerically. 

\section{Nonequilibrium spin dynamics}
\label{sec:non-equil-dynam}
Let us now turn to the dissipative nonequilibrium dynamics of the two-spin boson model of Eq.~\eqref{eq:1}. We will concentrate on the ohmic ($s=1$) as well as the subohmic case of $s=\frac12$. To access the system's rich dynamical behavior arising from the interplay of spin-spin and spin-bath interactions, we employ the time-dependent numerical renormalization group technique (TD-NRG), recently introduced by Anders and Schiller.~\cite{PhysRevLett.95.196801} Using this extension of the standard bosonic numerical renormalization 
group,~\cite{PhysRevB.71.045122,RevModPhys.80.395,PhysRevLett.95.086406} we are able to calculate the real-time evolution of an impurity observable as a reaction to a single sudden change of parameters. 
Since this method is non-perturbative as well as non-Markovian, it is capable to accurately describe the spin dynamics over the whole range of parameter values, including strong coupling. We note that a first dynamical study of the ohmic system in a very limited region of parameter space using analytical methods was given in Ref.~\onlinecite{DubeStamp-IntJModPhys-1998}. 

As common to all applications of the NRG to 
bosonic quantum impurity models, we have to restrict the 
maximal number of bosonic degrees of freedom that are added in each step of the 
iterative diagonalization procedure performed within the NRG method. We have 
checked that this cutoff does not alter our results.
We use the same NRG parameters as for the equilibrium calculations: a discretization parameter of $\Lambda = 1.4$, a total of $N_{b,0} = 599$ bosonic modes in the first iteration and $N_{b,N} = 6$ in the following ones, while keeping $N_{\text{Lev}} = 200$ low-energy levels in each NRG iteration. For the TD-NRG calculations we have additionally averaged the real-
time data using  $N_z = 8$ independent NRG runs ($z$-trick averaging). 
For more details about the method, we refer the reader to Refs.~\onlinecite{PhysRevB.74.245113,PhysRevB.41.9403}.

In the following, we discuss a number of different nonequilibrium situations. 

In Sec.~\ref{sec:exact-decoh-with}, we show that TD-NRG results perfectly agree with the exact solution that is available for zero transverse field $\Delta_{1,2}=0$, where the Hamiltonian only contains the $z$-component of the spin operators.

In Sec.~\ref{sec:beat-decoh-breakd}, we focus on the case of weak spin-bath coupling and compare TD-NRG to the commonly used perturbative Bloch-Redfield method. We provide quantitative limits at which dissipation strength this method begins to fail. 

We discuss, in Sec.~\ref{sec:synchr-spin-dynam}, the fascinating phenomenon of dynamical synchronization of the spin oscillations induced by the bath. Most importantly, this feature occurs even at weak spin-bath coupling and synchronization can thus be observed over many oscillation periods. It relies on the coherent exchange of bath excitations between the two spins, which gives rise to the bath induced part of the Ising interaction. The phenomenon cannot be observed within the Bloch-Redfield master equation approach, where the backaction of the bath on the spins is neglected. 

In Sec.~\ref{sec:case-k_r-=}, we investigate the spin dynamics for vanishing (renormalized) Ising coupling $K_r = 0$. Qualitatively, the system behaves like a single spin-boson model for $0< \alpha < 1/2$, where it exhibits damped coherent oscillations. The quality factor of the oscillations, however, is smaller in the two-spin case as the damping is stronger. 
Yet most importantly, for larger values of $\alpha$ we find that the two spins remain delocalized for $K_r=0$ up to a dissipation strength as large as $\alpha = 1.5$ in the ohmic case. The single spin-boson model, in contrast, becomes localized at $\alpha = 1$, where the spin remains frozen in its initial state.
In Sec.~\ref{sec:weak-dissipation}, we first elaborate on the region $0<\alpha<1/2$, and use an approximation that is known to be equivalent to the Non-Interacting Blip Approximation (NIBA)~\cite{RevModPhys.59.1,PhysRevA.35.1436} in the single spin case. It allows us to understand the dynamics qualitatively. In Sec.~\ref{sec:toulouse-point-1}, we then focus on the generalized Toulouse point $\alpha = 1/2$ and $K_r=0$, where $\sigma^z_{1,2}(t)$ decays purely exponentially. We show that one obtains slightly different decay rates for the single and two-spin boson model. We qualitatively explain this difference by employing a bosonization mapping to a fermionic resonant level model. In the single spin case, the fermionic model can be solved exactly. For two spins, however, the fermionic model contains an additional interaction term that stems from the Jordan-Wigner transformation of the spins and impedes an exact solution.  

We discuss the spin dynamics at large spin bath coupling in Sec.~\ref{sec:large-spin-bath}. Comparing the ohmic and subohmic cases, we find that while coherence is lost prior to localization in the ohmic system, the spins exhibit oscillations even inside the localized regime for a subohmic bath, a feature only recently discovered~\cite{anders:210402} in the single spin-boson system. 

Finally, as presented in Sec.~\ref{sec:gener-highly-entangl}, an interesting situation arises if we prepare the spins in an antiferromagnetic initial state at a location in the phase diagram which corresponds to a localized (ferromagnetic) ground state. Following the spin's dynamics over time, we observe a non-trivial steady-state, where the spins are highly entangled with the bath while developing and maintaining coherence between the two antiferromagnetic spin states. We give a simple physical explanation for this behavior.

\subsection{Decoherence without transverse field}
\label{sec:exact-decoh-with}
In this Section we discuss a specific case where one can exactly solve for the (non-trivial) dissipative spin dynamics of the two-spin boson model. We compare the exact solution to the TD-NRG results and find perfect agreement, which provides another validation of this powerful method in the strong coupling regime. 
\begin{figure}[t]
  \centering
  \includegraphics[width=1.0\linewidth]{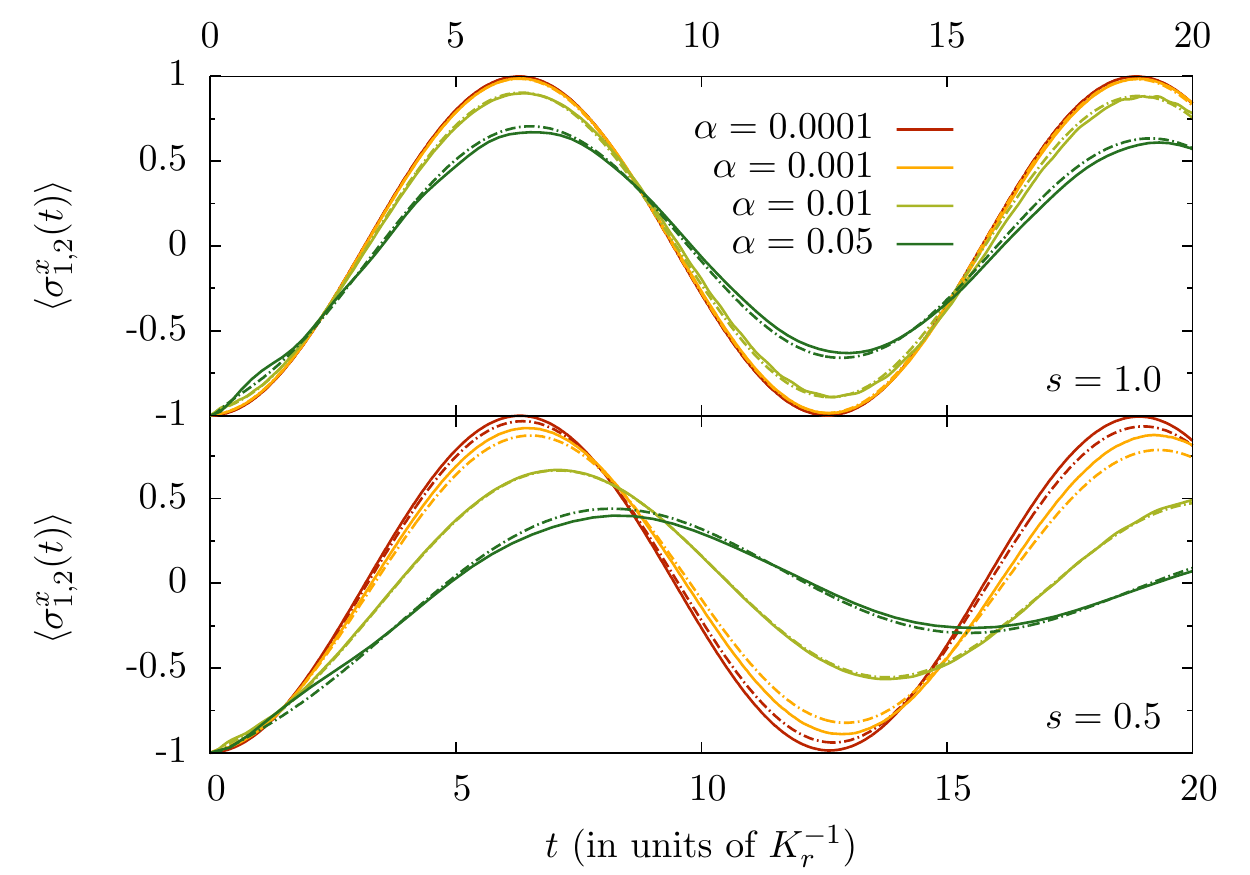}
  \caption{Comparison of $\av{\sigma^x_{1,2}(t)}$ between exact solution in Eq.~\eqref{eq:23} (dashed) and TD-NRG result (solid) for different values of $\alpha$ and bath exponents $s=1$ (upper part), $s=\frac12$ (lower part). Parameters used are $\epsilon_{1,2} = \Delta_{1,2} = 0$, $K=0$, $\omega_c = 1$, and $\alpha$ as specified in the plot. }
  \label{fig:8}
\end{figure}

For vanishing transverse fields, $\Delta_{1,2} = 0$, the Hamiltonian in Eq.~\eqref{eq:1} takes the form
\begin{align}
\label{eq:22}
    H[\Delta_{1,2}=0] &= \sum_{j=1}^2 \Bigl\{ \frac{\sigma^z_j}{2} \bigl[ \epsilon_j +  \sum_{k>0} \lambda_k (b^\dag_k + b_k)\bigr] \Bigr\} + \frac{K}{4} \sigma^z_1 \sigma^z_2  \nonumber \\ & \quad+ \sum_{k>0} \omega_k b_k^\dag b_k\,,
\end{align}
which only contains the z-component of the spin operators. Thus, the spin dynamics is non-trivial only if the initial state of the spins contains a transverse component (in the x or y-direction). For instance, spins that are initially polarized along the $x$-direction, $\av{\sigma^x_{j}(t=0)}= -1$ for $j=1,2$, will undergo damped oscillations in $\av{\sigma^x_{j}}$ as exactly described by~\cite{PhysRevA.57.737}
  \begin{equation}
    \label{eq:23}
    \av{\sigma^x_{j}(t)} = - \cos \Bigl[ \bigl( \epsilon_{j} + \frac{K_r}{2} \bigr) t \Bigr] \cos\Bigl[ \frac{Q_1(t)}{\pi} \Bigr] \exp \Bigl [\frac{-Q_2(t)}{\pi} \Bigr]\,,
  \end{equation}
with functions $Q_1(t) = \int_0^\infty d\omega J(\omega) \sin \omega t$ and $Q_2(t) = \int_0^\infty d\omega J(\omega) [1- \cos \omega t]$. For an ohmic spectral density with exponential cutoff, they read~\cite{RevModPhys.59.1}
\begin{align}
  \label{eq:24}
  Q_1(t) &= 2 \pi \alpha \tan^{-1} \omega_c t \\
\label{eq:25}
  Q_2(t) &= \pi \alpha \ln[1 + \omega_c^2 t^2]\,.
\end{align}
Note that Eq.~\eqref{eq:23} can also be derived using the polaron transformation of Sec.~\ref{sec:qual-underst-phase}. 
In Fig.~\ref{fig:8} we compare our TD-NRG results with this exact analytical prediction for bath exponents $s=\frac12$ and $s=1$, and we find perfect agreement between them.

\subsection{Breakdown of Bloch-Redfield description}
\label{sec:beat-decoh-breakd}
In this Section, we compare the results from TD-NRG with those from the commonly employed Bloch-Redfield~\cite{weissdissipation,PhysRevA.67.042319,Campagnano2010} formalism at weak spin-bath coupling. We give quantitative limits on the applicability of this perturbative and Markovian technique. 

Let us briefly outline the Bloch-Redfield approach to dissipative spin dynamics. The time-evolution of the spin reduced density matrix $\rho_S = \text{Tr}_B (\rho)$, where $\text{Tr}_B$ denotes tracing out the bath degrees of freedom and $\rho$ is the full density matrix of the spin-boson system, is given by the Bloch-Redfield equations
\begin{equation}
  \label{eq:26}
  \dot{\rho}_{S,ab}(t) = - i \omega_{ab} \rho_{S,ab}(t) - \sum_{k,l} R_{abkl} \rho_{S,kl}(t)\,.
\end{equation}
Here, $a,b,k,l\in \{1,\ldots,4\}$ label the four eigenstates (with eigenenergy $E_a$) of the free spin part of the Hamiltonian $H_S = \sum_{j=1}^2 \bigl[ \frac{\Delta_j}{2} \sigma^x_j + \frac{\epsilon_j}{2} \sigma^z_j\bigr] + \frac{K}{4} \sigma^z_1 \sigma^z_2$, and $\omega_{ab}= E_a-E_b$ are transition frequencies. For zero bias $\epsilon_{1,2} = 0$, the eigenenergies are given by $E_{1,2} = \mp \Omega_-/2$ and $E_{3,4} = \mp \Omega_+/2$ with $\Omega_\pm = \sqrt{(\Delta_1 \pm \Delta_2)^2 + K^2/4}$. The relevant transition frequencies for which $\braopket{a}{\sigma^z_{1,2}}{b} \neq 0$, read $\omega_{41} = \omega_{23} = \Omega$ and  $\omega_{42} = \omega_{13} = \delta$ with $\Omega = \frac12 (\Omega_+ + \Omega_-)$ and $\delta = \frac12 (\Omega_+ - \Omega_-)$. For nonzero bias $\epsilon_{1,2} \neq 0$, one can easily diagonalize $H_S$ numerically for a specific choice of parameters.

The Redfield tensor $R_{abkl}$ describes the effect of the bath onto the spin dynamics in the Born-Markov approximation.~\cite{CohenTannoudji-AtomPhotonInteractions} The real part of $R_{abkl}$ describes the damping induced by the bath, and the imaginary part the renormalization of the transition frequencies, up to second order in the spin-bath coupling constants $\{\lambda_k\}$.  
\begin{figure}[t]
    \begin{center}
        \includegraphics[width=1.0\linewidth]{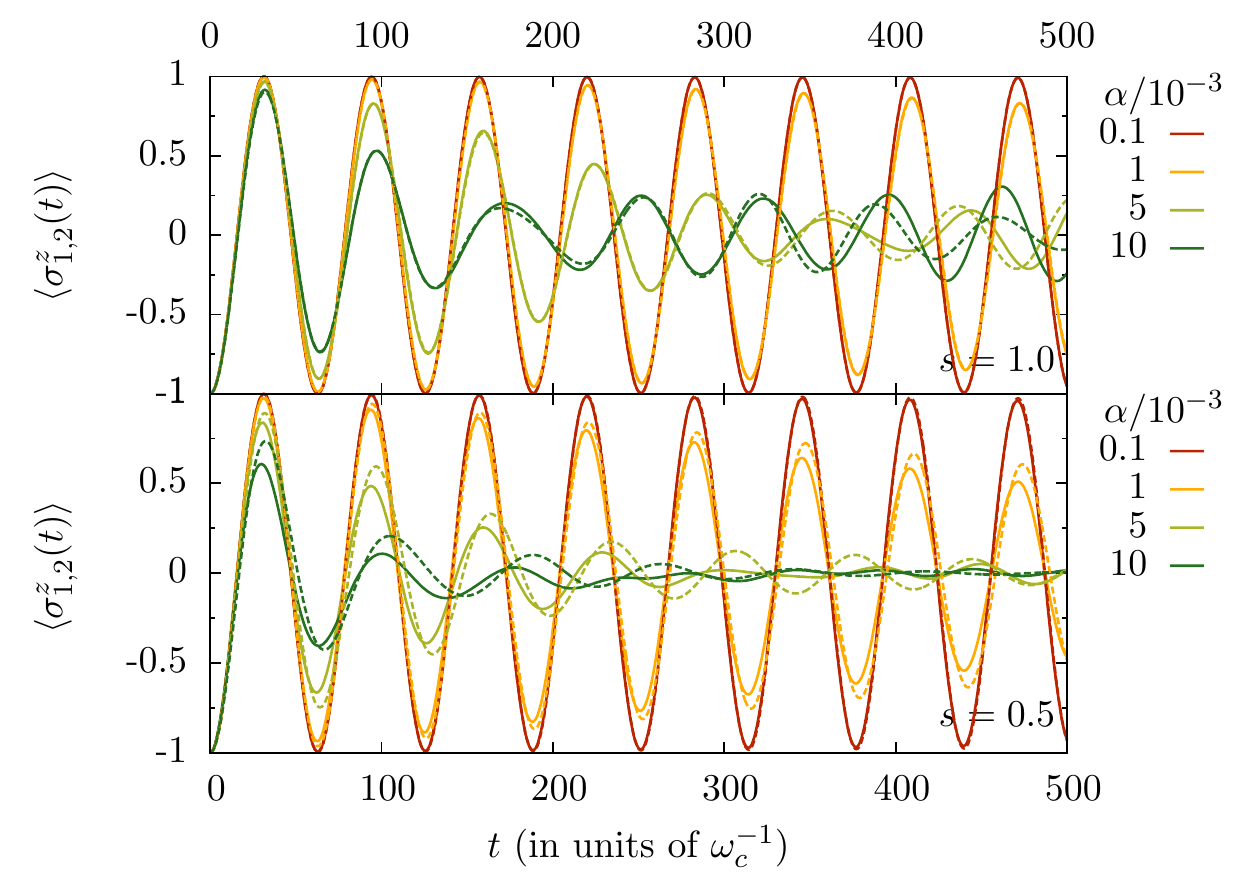}
    \end{center}
    \caption{Comparison of the results for $\av{\sigma^z_{1,2}(t)}$ from TD-NRG (solid) and Bloch-Redfield approach (dashed) for $K=\epsilon_{1,2}=0$, $\Delta_{1,2} = 0.1 \omega_c$ in the perturbative regime $\alpha \ln \frac{\omega_c}{\Delta} \ll 1$. Upper (lower) part shows the case $s=1$ ($s=\frac12$). Deviations between the two solutions are visible already for $\alpha \ln \frac{\omega_c}{\Delta} = 0.01$. Note the beatings in oscillations due to the bath induced Ising coupling.  }
    \label{fig:9}
\end{figure}

The Redfield tensor is explicitly given by golden rule transition rates and reads
\begin{equation}
  \label{eq:27}
  R_{abkl} = \delta_{bl} \sum_r \Gamma_{arrk}^{(+)} + \delta_{ak} \sum_r \Gamma_{lrrb}^{(-)} - \Gamma_{lbak}^{(+)} - \Gamma_{lbak}^{(-)}\,.
\end{equation}
The golden rule rates at temperature $T = 1/\beta$ are calculated to
\begin{equation}
  \label{eq:28}
  \begin{split}
    \Gamma^{(\pm)}_{lbak} &= \frac{\Lambda_{lbak} \tilde{J}(\omega_{ij})}{4}  [ \coth(\beta \hbar \omega_{ij}/2) \mp 1] \\ & + \frac{i \Lambda_{lbak}}{2 \pi} \mathcal{P} \int_0^\infty d\omega \frac{\tilde{J}(\omega)}{\omega^2 - \omega_{ij}^2} [ \coth(\beta \hbar \omega/2) \omega_{ij} \mp \omega ]\,,
  \end{split}
\end{equation}
where $\omega_{ij} = \omega_{lb}$ for the plus rates $\Gamma^{(+)}_{lbak}$ and $\omega_{ij} = \omega_{ak}$ for the minus rates $\Gamma^{(-)}_{lbak}$. Here, we have defined the transition matrix element
\begin{equation}
  \label{eq:29}
  \Lambda_{lbak} = \sigma^{z}_{1,lb} \sigma^{z}_{1,ak} + \sigma^{z}_{1,lb}\sigma^{z}_{2,ak} + \sigma^{z}_{2,lb}\sigma^{z}_{1,ak} + \sigma^{z}_{2,lb}\sigma^{z}_{2,ak}
\end{equation}
and a spectral density that is antisymmetrically continued to negative frequencies $\tilde{J}(\omega) = \text{sign}(\omega) \pi \alpha |\omega|^s \omega_c^{1-s} \theta(\omega_c - |\omega|)$. 
At zero temperature, the real part of the rates becomes
\begin{align}
  \label{eq:30}
  \text{Re} \Gamma_{lbak}^{(\pm)} &= \frac{\Lambda_{lbak}}{4} J(\omega_{ij})\,,
\end{align}
where again $\omega_{ij} = \omega_{ak}$ for the plus rate and $\omega_{ij} = \omega_{lb}$ for the minus rate. Note that Eq.~\eqref{eq:30} vanishes unless $\omega_{ij} > 0$. 
The principal part integral in the imaginary part of the rates can be performed analytically, and for instance in the ohmic case and for a Drude bath cutoff $J(\omega) = 2 \pi \alpha \omega/(1 + \frac{\omega^2}{\omega_c^2})$, we obtain
\begin{align}
  \label{eq:60}
  \text{Im} \Gamma^{(\pm)}_{lbak} &= \frac{\Lambda_{lbak}}{4 \pi} \frac{\pi \alpha \omega_{ij} \omega_c^2}{\omega_{ij}^2 + \omega_c^2} \Bigl[ 2 \ln \bigl| \frac{\omega_c}{\omega_{ij}} \bigr| \mp \frac{\pi \omega_c}{\omega_{ij}} \Bigr]\,.
\end{align}
In all our calculations, we use the corresponding expression for a hard bath cutoff [see Eq.~\eqref{eq:2}] which turns out to look more complicated than Eq.~\eqref{eq:60}, but leads to the same results as long as $\omega_c$ is the largest energy scale in the system. 
\begin{figure}[t]
    \begin{center}
        \includegraphics[width=1.0\linewidth]{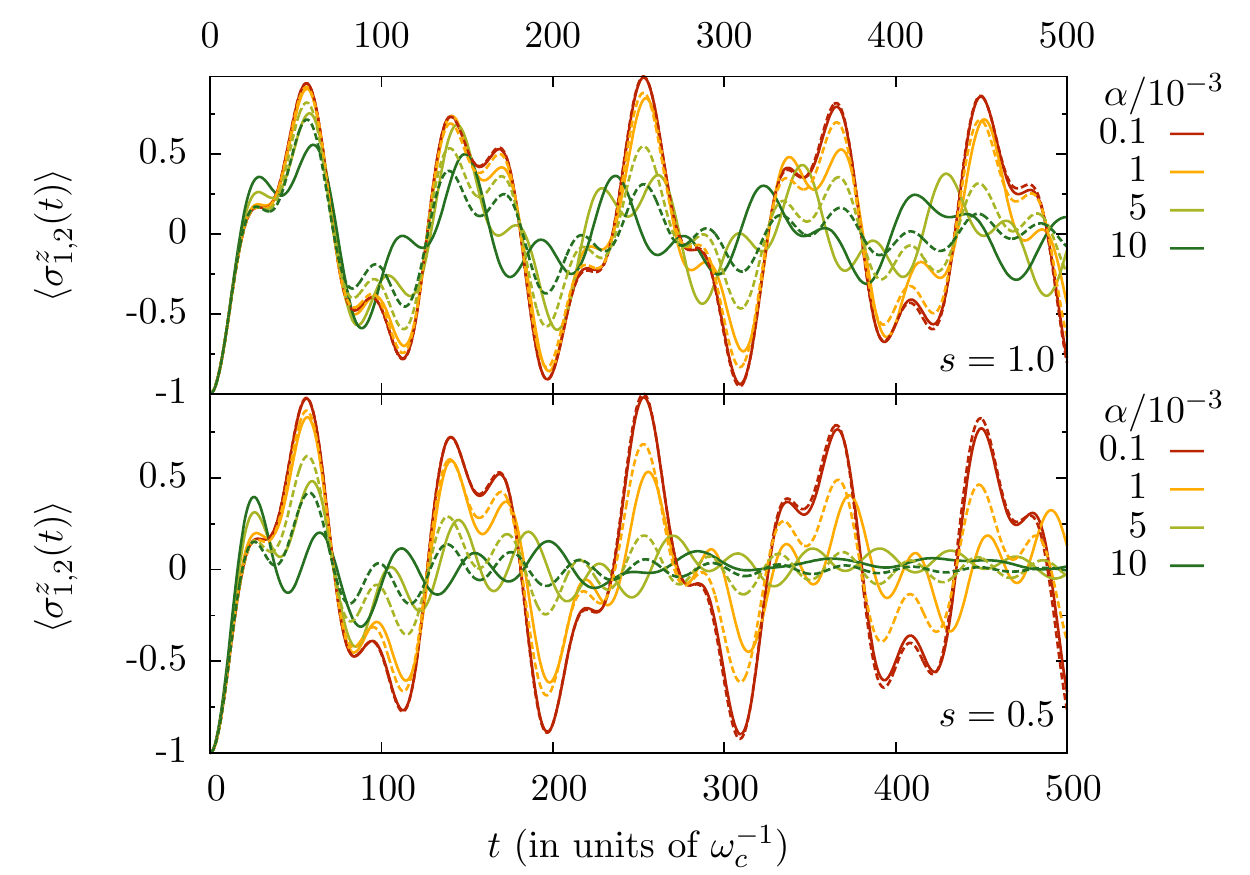}
    \end{center}
    \caption{Comparison of results for $\av{\sigma^z_{1,2}(t)}$ obtained with TD-NRG (solid) and Bloch-Redfield approach (dashed) for $K=0.2\,\omega_c$, $\epsilon_{1,2}=0$. Other parameters are as in Fig.~\ref{fig:9}. Deviations between the two solutions are visible already for $\alpha = 0.005$. }
  \label{fig:10}
\end{figure}


In the Redfield approach, the total density matrix is assumed to always factorize into a spin and a bath part. Further, by taking the long time limit in Eq.~\eqref{eq:28}, any reversible energy exchange between spins and bath, and thus any back action of the bath on the spins is neglected. Therefore, the Redfield approach does not capture the bath induced non-dissipative spin-spin interaction $- 4 \alpha \omega_c/s = K_r - K$ correctly. As we will discuss in Sec.~\ref{sec:synchr-spin-dynam} this has the important consequence that the phenomenon of a bath induced dynamical synchronization of the spin oscillations cannot be observed within the Redfield approach. 

In this Section, we want to focus on a symmetric setup of the two-spin system ($\Delta_1 = \Delta_2\equiv \Delta $) and zero bias $\epsilon_{1,2}=0$. To determine the breakdown of the Bloch-Redfield description, we compare in Fig.~\ref{fig:9} our TD-NRG results for $\av{\sigma^z_{1,2}(t)}$ with Bloch-Redfield solutions of Eq.~\eqref{eq:26} for an ohmic and a subohmic bath. Both results agree for very weak spin-bath coupling $\alpha \ln \frac{\omega_c}{\Delta} < 0.01$. However, already at $\alpha \ln \frac{\omega_c}{\Delta} = 0.01$ we find significant differences. They are more pronounced in the subohmic case, and grow with the coupling strength. Even in the absence of a direct Ising coupling term $K=0$, we observe beatings in the oscillations due to the bath induced Ising interactions $K_r$. In Fig.~\ref{fig:10} we show results for a system with a direct Ising coupling, where the beatings are stronger. Here, we find significant differences between the TD-NRG and Bloch-Redfield results to occur already for $\alpha = 0.005$ or $\alpha \ln \frac{\omega_c}{\Delta_{1,2}} \approx 0.01$. 

In summary, since the Redfield approach does not correctly account for the bath induced Ising interaction, its breakdown occurs not just when $\alpha \ln \frac{\omega_c}{\omega_{ij}} \approx 1$, but already for $\alpha \omega_c \approx \omega_{ij}$. Here $\omega_{ij}$ is a (nonzero) transition frequency of the system which is of the order $\{\Omega, \delta\}$. Since $\omega_c \gg \omega_{ij}$ the breakdown of the master equation description occurs for much smaller values of $\alpha$ compared to the single spin case, where it takes place when $\alpha \ln \frac{\omega_c}{\omega_{ij}} \approx 1$.

\subsection{Synchronization of spin dynamics}
\label{sec:synchr-spin-dynam}
In this Section, we address how the coupling of spins to a common bath can be employed to obtain a \emph{dynamical synchronization} of spin oscillations. Notably, this feature occurs already at weak spin-bath coupling, where the bath induced decoherence is small. It provides an alternative technique to synchronize the dynamics of a two-spin system, when a strong direct coupling of the spins is unavailable. 
\begin{figure}[t]
    \begin{center}
        \includegraphics[width=1.0\linewidth]{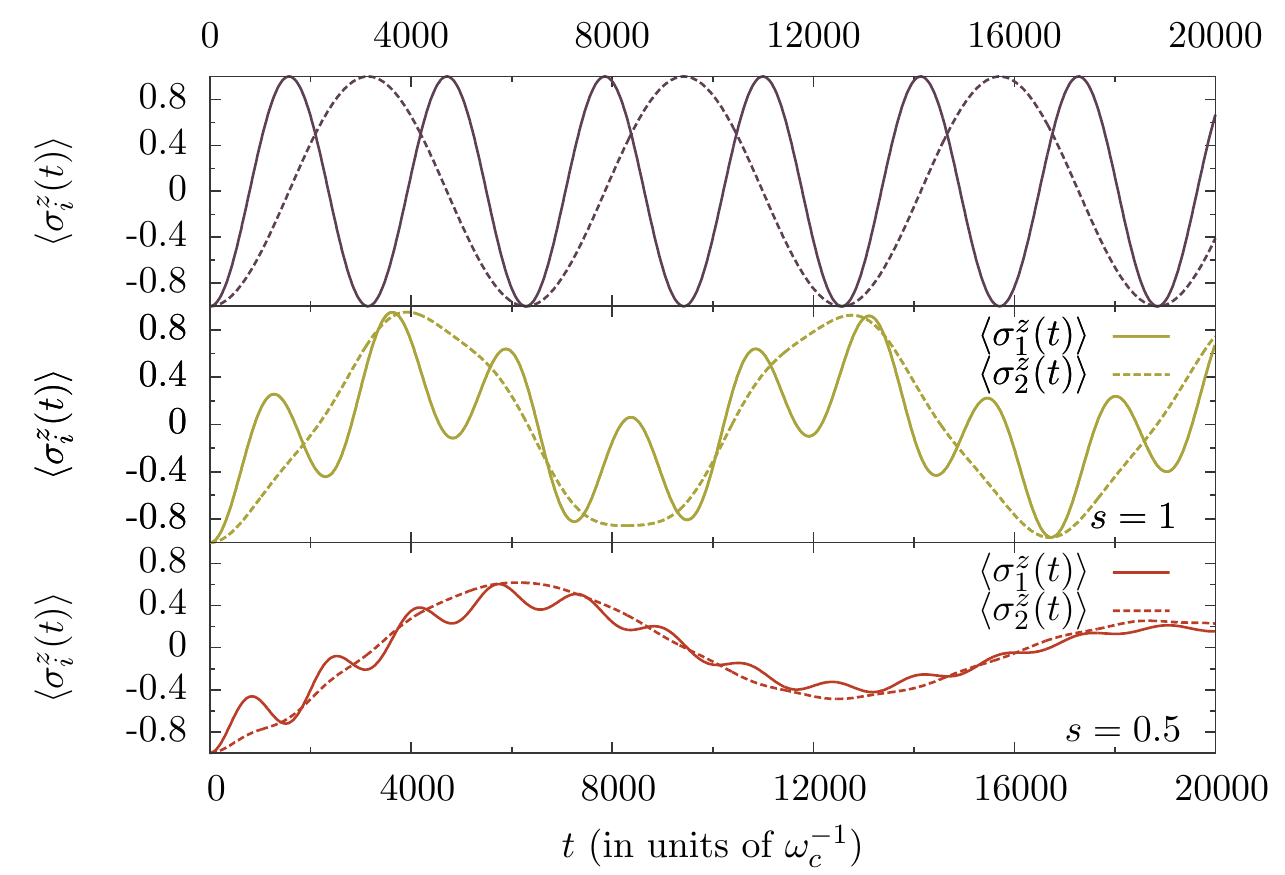}
    \end{center}
    \caption{Synchronization of two spins with different spin-flip terms
    $\Delta_{2} = \frac12 \Delta_{1} = 10^{-3} \omega_{c}$ by weak coupling to a common bath. There is no direct coupling between the spins $K=0$. The upper part of the figure shows the uncoupled case $\alpha=0$. The middle part shows the ohmic case and the lower part the subohmic one with $s=\frac12$. We use the same strength $\alpha = 8 \cdot 10^{-4}$ for the different bath dispersions, which lies in the perturbative regime: $\alpha \ln \frac{\omega_c}{\Delta_2} = 6\cdot 10^{-3} \ll 1$.}
    \label{fig:11}
\end{figure}

Let us start with two free and uncoupled spins ($K=\alpha=0$) that are driven by different tunneling amplitudes, say $\Delta_1 = 2 \Delta_2$. As shown in the upper part of Fig.~\ref{fig:11}, the spins will then undergo undamped Rabi oscillations with frequencies $\Delta_1$ and $\Delta_2$, respectively. We now consider a weak coupling to the bath in the perturbative regime where $\alpha \ln \frac{\omega_c}{\Delta_{1,2}} \ll 1$. The frequency corrections in $\Delta_{1,2}$ are small in this case. However, the bath induced Ising interaction $K_r = - 4 \alpha \omega_c/s$ can still be comparable to $\Delta_{1,2}$, because it scales with the (large) bath cutoff frequency $\omega_c$. In this case, where $K_r$ and $\Delta_{1,2}$ are of the same order of magnitude, the bath is capable of synchronizing the spin oscillations as depicted in the two lower parts of Fig.~\ref{fig:11} for an ohmic (middle) and a subohmic bath with $s=1/2$ (bottom). The synchronization is more complete for the subohmic system, since there is an increased number of slow oscillator modes present and the induced Ising interaction, which scales as $s^{-1}$, is twice as large (for the same value of $\alpha$). 

The two oscillation frequencies $\{\Omega, \delta\}$ that occur in Fig.~\ref{fig:11} can be calculated from the free spin dynamics of Eq.~\eqref{eq:1}, if we set the Ising interaction $K$ equal to its renormalized value $K = K_r = - 4 \alpha \omega_c/s$. 

\begin{figure}[t]
  \centering
  \includegraphics[width=\linewidth]{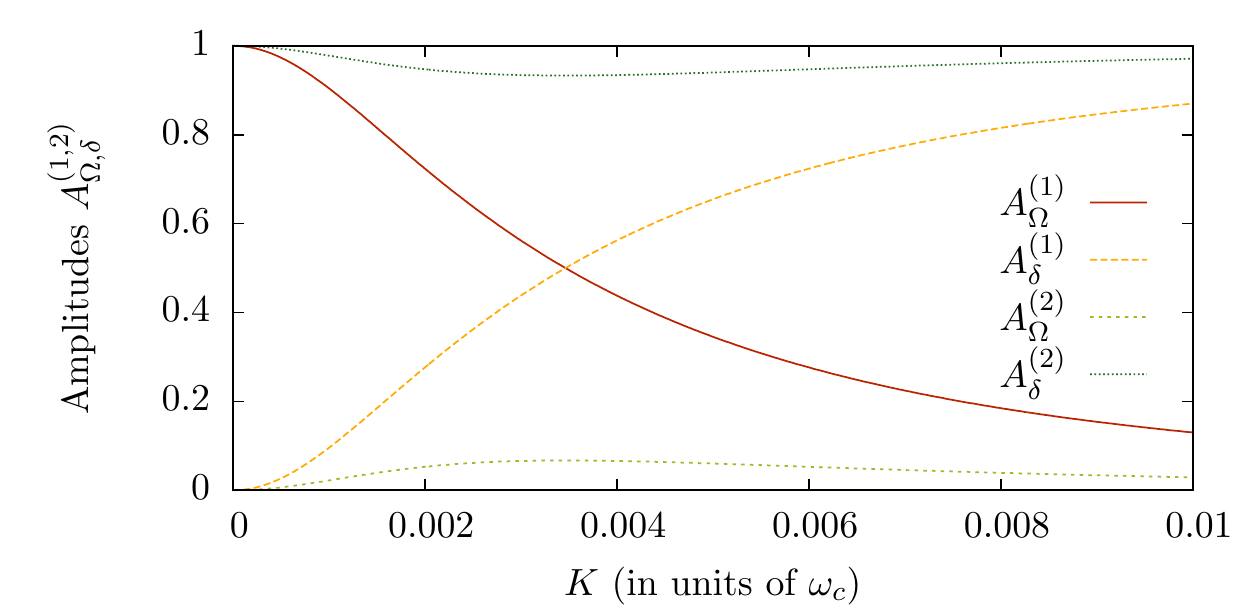}
  \caption{Oscillation amplitudes $A_{\Omega, \delta}^{(1,2)}$ as a function of Ising coupling $K$ (see Eqs.~\eqref{eq:35} and~\eqref{eq:36}). Other parameters read $\Delta_1 = 2 \Delta_2 = 2\cdot 10^{-3} \omega_c$. The two spin expectation values $\av{\sigma^z_{1,2}(t)}$ evolve according to Eq.~\eqref{eq:31}. }
    \label{fig:12}
\end{figure}

For zero bias $\epsilon_{1,2} = 0$, the free spin part in Eq.~\eqref{eq:1} reads $H_S = \sum_{j=1}^2 \frac{\Delta_j}{2} \sigma^x_j + \frac{K}{4} \sigma^z_1 \sigma^z_2$ with eigenvalues $E_{1,2} = \mp\Omega_{-}/2$ and $E_{3,4}= \mp \Omega_+/2$, where $\Omega_{\pm} = \sqrt{(\Delta_1\pm \Delta_2)^2 + K^2/4}$. 
We can find the spin dynamics from $\av{\sigma^z_{1,2}(t)} =  \text{Tr}_S [\rho_S(t) \sigma^z_{1,2}]$, where, in the absence of the bath, the spin density matrix $\rho_S(t)$ evolves in time according to the von-Neumann equation of motion $\dot{\rho}_{S} = -i [H_S, \rho_S]$. With initial condition $\rho_S(0) = \ket{\!\!\downarrow \downarrow}\bra{\downarrow \downarrow\!\!}$, we find for $j=1,2$:
\begin{align}
  \label{eq:31}
  \av{\sigma^z_j} &= \sum_{a,b} \rho_{S,ab}(0) e^{- i \omega_{ab} t} \braopket{b}{\sigma^z_j}{a} \nonumber \\ & = 2 A^{(j)}_\Omega \cos \Omega t + 2 A^{(j)}_\delta \cos \delta t \,,
\end{align}
where $a,b$ label the eigenstates of $H_S$ and $\omega_{ab} = E_a - E_b$ are the transition frequencies. They obey $ \omega_{41} = \omega_{23} = \Omega$ and $\omega_{42} = \omega_{13} = \delta$. The two oscillation frequencies that appear in Fig.~\ref{fig:11} are thus given by $\Omega = \frac12 (\Omega_+ + \Omega_-)$ and $\delta = \frac12 (\Omega_+ - \Omega_-)$. 
 For the other transitions, we find that the matrix elements $\braopket{b}{\sigma^{z}_j}{a}$ are equal to zero. 
The two oscillation amplitudes are given by ($j=1,2$)
\begin{align}
  \label{eq:32}
  A^{(j)}_\Omega &= \rho_{S,41}(0)\braopket{1}{\sigma^z_{j}}{4} + \rho_{S, 23}(0) \braopket{3}{\sigma^z_{j}}{2} \\
A^{(j)}_\delta &= \rho_{S,42}(0) \braopket{2}{\sigma^z_{j}}{4} + \rho_{S,13}(0) \braopket{3}{\sigma^z_{j}}{1}\,.
\end{align}
They are shown in Fig.~\ref{fig:12} as a function of Ising coupling $K$, and are responsible for the synchronization phenomenon. At $K=0$, the first spin oscillates with frequency $\Omega(K=0)=\Delta_1$ and $A_\Omega^{(1)}=1$. The second spin oscillates with frequency $\delta(K=0)=\Delta_2$ and $A_\delta^{(2)} = 1$. As we increase $K$, the amplitude $A_\delta^{(1)}$ increases, while $A_{\Omega}^{(1)}$ decreases, and $A_{\Omega, \delta}^{(2)}$ remain almost the same. For large $K \gg \Delta_{1,2}$, both spins oscillate with frequency $\delta \simeq 2 \Delta_1 \Delta_2/K$. 


In fact, we can derive analytical expressions for $A_{\Omega, \delta}^{(1,2)}$ if we solve the Heisenberg equation of motion for $\sigma^z_{1,2}(t) = i [H_S, \sigma^z_{1,2}(t)]$ in Laplace space.~\cite{1367-2630-10-11-115010} 
One finds that 
\begin{align}
  \label{eq:33}
  \av{\sigma^z_{1}(\lambda)} &= \frac{\lambda (\frac{K^2}{4} + \Delta_2^2 + \lambda^2)}{(\lambda^2 + \Omega^2) (\lambda^2 + \delta^2)} \\
\label{eq:34}
  \av{\sigma^z_{2}(\lambda)} &= \frac{\lambda (\frac{K^2}{4} + \Delta_1^2 + \lambda^2)}{(\lambda^2 + \Omega^2) (\lambda^2 + \delta^2)}\,,
\end{align}
which yields Eq.~\eqref{eq:31} in real space. We identify the amplitudes $A_{\Omega,\delta}^{(1,2)}$ as the respective residues of Eqs.~\eqref{eq:33} and ~\eqref{eq:34} at $\Omega$ and $\delta$. Explicitly, they read
\begin{align}
  \label{eq:35}
  A_{\Omega,\delta}^{(1)} &= \frac{\pm [-K^2 + 4 (\Delta_1^2 - \Delta_2^2)] + w}{4 w}  \\
\label{eq:36}
  A_{\Omega, \delta}^{(2)} &= \frac{\pm [-K^2 + 4(\Delta_2^2 - \Delta_1^2)] + w}{w} \,,
\end{align}
where $w = \sqrt{[K^2 + 4(\Delta_1^2 + \Delta_2^2)]^2 - 64 \Delta_1^2 \Delta_2^2}$ and the upper sign relates to $A_\Omega^{(1,2)}$. Synchronization sets in when $A_\Omega^{(1)} \simeq A_\delta^{(1)}$ which occurs for an Ising interaction strength of
\begin{equation}
  \label{eq:37}
   K = 2 \sqrt{|\Delta_1^2 - \Delta_2^2|}\,.
\end{equation}

The damping rates of the oscillation amplitudes are proportional to $J(\Omega)$ and $J(\delta)$, respectively [see Eq.~\eqref{eq:30}]. Since $J(\omega) \sim \alpha \omega^s$ they are small for the parameters in Fig.~\ref{fig:11}, and the synchronized oscillations can be seen over many periods.

We like to emphasize that the synchronization effect cannot be seen in the Bloch-Redfield master equation treatment, which does not correctly account for the bath induced Ising interaction $K_r$ (see Sec.~\ref{sec:beat-decoh-breakd}). The damping rates in Fig.~\ref{fig:11}, however, agree with the ones calculated with the perturbative Redfield approach. 


In summary, the bath induced Ising interaction scales with the bath cutoff frequency like $K_r \sim \alpha \omega_c$ whereas the bath induced damping is proportional to $\alpha \Omega^s$ and $\alpha \delta^s$, where $\{\Omega,\delta\}$ are spin transition frequencies.  A common bath can thus synchronize spin oscillations at weak coupling if the bath cutoff frequency is large: $\omega_c \gg \Omega, \delta$.

\subsection{Vanishing Ising interaction $K_r = 0$: similarities and differences with the single spin-boson model}
\label{sec:case-k_r-=}
\begin{figure}[t]
  \centering
  \includegraphics[width=\linewidth]{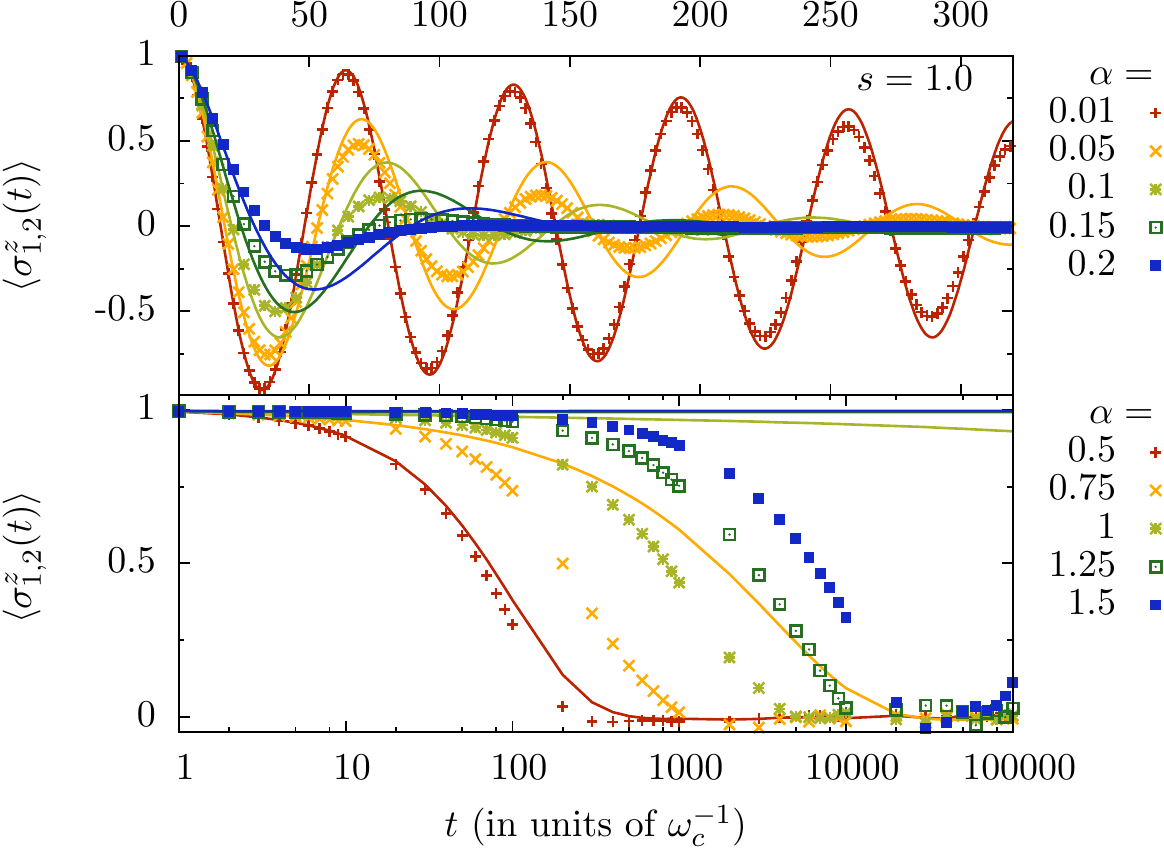}
  \caption{Comparison of spin dynamics between the ohmic single spin-boson (thin lines) and ohmic two-spin boson model with $K_r = 0$ (symbols). Different curves correspond to different values of dissipation strength $\alpha$. The Ising interaction is chosen accordingly to be $K = 4 \alpha \omega_c$. Other parameters are $\Delta_{1,2} = \Delta = 0.1 \, \omega_c$, $\epsilon_{1,2} = \epsilon = 0$. The upper panel shows the spin dynamics in the coherent regime $0 < \alpha < 1/2$. Here, the two-spin oscillations have a slightly larger frequency and are stronger damped than the single spin oscillations (see also Fig.~\ref{fig:15} ). The lower panel displays the dynamics for stronger dissipation $\alpha \geq 1/2$. For $1/2 \leq \alpha \leq 1$ both systems display incoherent decay. For even larger values of $\alpha$, we observe that, in contrast to the single spin-boson model which localizes at $\alpha = 1$, the two-spin dynamics remains incoherent at least up to $\alpha = 1.5$. This is in agreement with the phase diagram of Fig.~\ref{fig:2}.  }
    \label{fig:13}
\end{figure}

\begin{figure}[t]
  \centering
  \includegraphics[width=\linewidth]{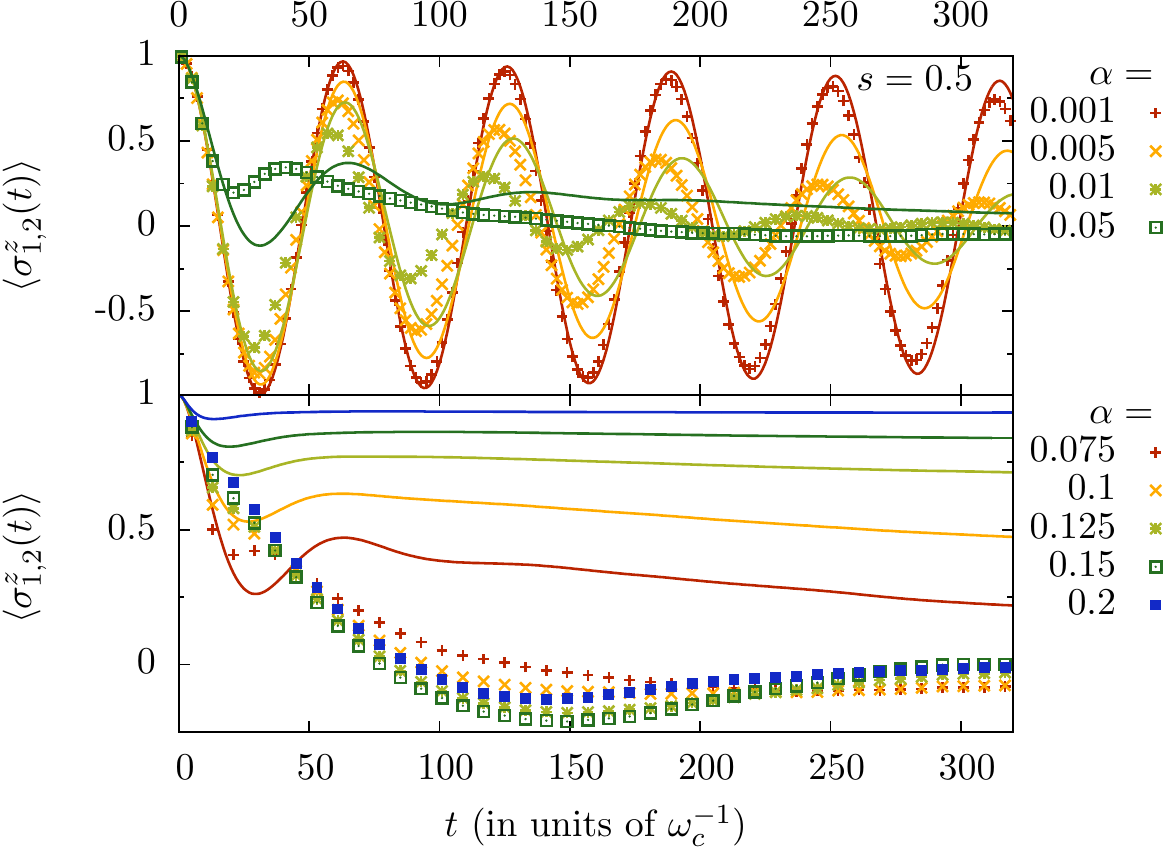}
  \caption{Comparison of spin dynamics between the subohmic single spin (thin lines) and two-spin boson model with $K_r = 0$ (symbols). Different curves correspond to different values of dissipation strength $\alpha$. Ising interaction is chosen accordingly to be $K = 8 \alpha \omega_c$. Other parameters are $\Delta_{1,2} = \Delta = 0.1 \, \omega_c$, $\epsilon_{1,2} = \epsilon = 0$. The upper panel shows the spin dynamics for weak dissipation up to $\alpha = 0.05$. The two-spin oscillations have slightly larger frequency and are stronger damped compared to the single spin results. The lower panel shows the case of strong dissipation $\alpha > 0.05$. The single spin-boson model localizes at $\alpha_c = 0.107$. In contrast, the two-spin boson model remains delocalized along $K_r =0$, which is in agreement with the phase diagram of Fig.~\ref{fig:3}.}
    \label{fig:14}
\end{figure}

In this Section, we investigate the spin dynamics along the line $K_r = 0$ in the phase diagram, i.e. for a vanishing renormalized Ising interaction (see Figs.~\ref{fig:2} and~\ref{fig:3}). At first sight, one might expect that the dynamics would be identical to that of two uncoupled spin-boson systems. However, as can be seen from the perturbative treatment in Sec.~\ref{sec:beat-decoh-breakd} already, the two spins do not decouple from each other even to linear order in $\alpha$, and the golden rule rates in Eq.~\eqref{eq:28} contain the terms $\sigma^z_{1,lb} \sigma^z_{2,ak}$ and $\sigma^z_{2,lb}\sigma^z_{1,ak}$. Qualitatively though, as displayed in Figs.~\ref{fig:13} and~\ref{fig:14}, the spin dynamics of the single and two-spin boson models agree for $K_r=0$ and small $\alpha$. 

In the ohmic case of Fig.~\ref{fig:13}, we observe a crossover from damped coherent oscillations at $0 \leq \alpha < 1/2$ to incoherent behavior at $\alpha \geq 1/2$ in both models.
In the two-spin case, however, we find stronger damping due to the terms proportional to $\sigma^z_1 \sigma^z_2$ mentioned above. This results in a smaller quality factor of the oscillations. We compare the quality factor of oscillations for the single and two-spin boson system computed with TD-NRG and Bloch-Redfield in Fig.~\ref{fig:15}. 
A detailed discussion of the dynamics at the special Toulouse point $\alpha = 1/2$ is given separately in Sec.~\ref{sec:toulouse-point-1}. 


If we further increase $\alpha$, we surprisingly observe that the two-spin boson model does not enter the localized phase (for $K_r = 0$). Unlike the single spin case, the two spins remain delocalized up to values of $\alpha > 1$. Our time-dependent numerical results in Fig.~\ref{fig:13} show that $\av{\sigma^z_{1,2}}$ relax to zero even for values as large as $\alpha = 1.5$. We note that this is in agreement with the NRG phase diagram in Fig.~\ref{fig:2}, which shows that the position of the localization phase transition converges toward the line $K_r = 0$ from the side where $K_r < 0$. 

In Fig.~\ref{fig:14}, we show the same comparison between single and two-spin boson model for a subohmic bath with $s = 1/2$. As before, for increasing dissipation the coherence of the spin oscillations is lost more rapidly in time, and a comparison of the quality factors of the single and two-spin boson system is presented in Fig.~\ref{fig:15}. Again, the system does not localize along $K_r = 0$, which is in agreement with the NRG phase diagram of Fig.~\ref{fig:3}. 

In the following, we first derive in Sec.~\ref{sec:weak-dissipation} a decoupling approximation that is equivalent to the well-known Non-Interacting Blip Approximation NIBA for the single spin-boson dynamics. In this approximation the spins decouple completely for $K_r = 0$. It allows us to understand the spin dynamics along the line $K_r=0$ qualitatively. In contrast to the case of the single spin-boson model, however, the approximation does not give quantitatively correct results for two spins, not even to linear order in $\alpha$. The reason is that dissipative second-order processes that involve both spins [see Eqs.~\eqref{eq:29} and~\eqref{eq:30}] are neglected. 

Then, in Sec.~\ref{sec:toulouse-point-1} we focus on the so-called Toulouse point $\alpha = 1/2$ of the ohmic model. For a single spin, one can solve for the dynamics exactly and $\av{\sigma^z(t)}$ exhibits pure exponential decay. The exact solution can most easily be derived by employing a bosonization mapping to a non-interacting fermionic resonant level model. In the two-spin case, we explicitly show that this mapping does not lead to a non-interacting fermionic model, which hence cannot be solved exactly. Further, our numerical results prove that the two-spin dynamics at $\alpha = 1/2$ differs slightly from the single spin case. We associate this with the influence of the retarded part of the bath induced Ising interaction which is still present even at $K_r = 0$ [see also Eq.~\eqref{eq:10}]. 

\subsubsection{Weak dissipation: Quality factors and the Non-Interacting Blip Approximation (NIBA)}
\label{sec:weak-dissipation}
In this Section, we derive a decoupling approximation that allows us to qualitatively understand the spin dynamics for $K_r=0$, but not necessarily small $\alpha$. In this approximation, the two spins decouple completely for $K_r=0$, and their Heisenberg equations of motion are identical to the ones of the single spin-boson model in the well-known NIBA.~\cite{PhysRevA.35.1436}

Our starting point to investigate the dynamics is the polaron transformed Hamiltonian in Eq.~\eqref{eq:4}. For zero detuning and at $K_r = 0$, it reduces to 
\begin{equation}
  \label{eq:38}
  \tilde{H} = \sum_{j=1}^2 \frac{\Delta_j}{2} ( \sigma^+_j e^{i \Omega} + \text{h.c.} ) + \sum_{k>0} \omega_k b^\dag_k b_k \,,
\end{equation}
where $\Omega = - i \sum_k \frac{\lambda_k}{\omega_k} (b^\dag_k - b_k)$. The Heisenberg equation of motion for $\sigma^z_j(t)$ with $j=1,2$ reads 
\begin{equation}
  \label{eq:39}
  \dot{\sigma}^z_j(t) = - i \Delta_j \sigma^+_j(t) e^{i \Omega(t)} + \text{h.c.}\,.
\end{equation}
It contains the elements $\sigma^{\pm}_j(t)$ which are given by
\begin{equation}
  \label{eq:40}
  \sigma^{+}_j(t) = - \frac{i \Delta_j}{2} \int_{-\infty}^t ds \sigma^z_j(s) e^{- i \Omega(s)}\,,
\end{equation}
and $\sigma^-_j = (\sigma^+_j)^*$. Inserting Eq.~\eqref{eq:40} in Eq.~\eqref{eq:39} yields
\begin{equation}
  \label{eq:41}
  \dot{\sigma}^z_j(t) = - \frac{\Delta_j^2}{2} \int_{-\infty}^t ds [\sigma_j^z(s) e^{i \Omega(t)} e^{- i \Omega(s)} + \text{h.c.} ]\,.
\end{equation}
Note that the two spins are still coupled to each other via the time-dependent bath operator $\Omega(t)$. This coupling describes the retarded part of the bath induced Ising interaction that we have seen already in the spin effective action of Eq.~\eqref{eq:10}. If we neglect this interaction, the two spins decouple from each other. 
\begin{figure}[t]
  \centering
  \includegraphics[width=\linewidth]{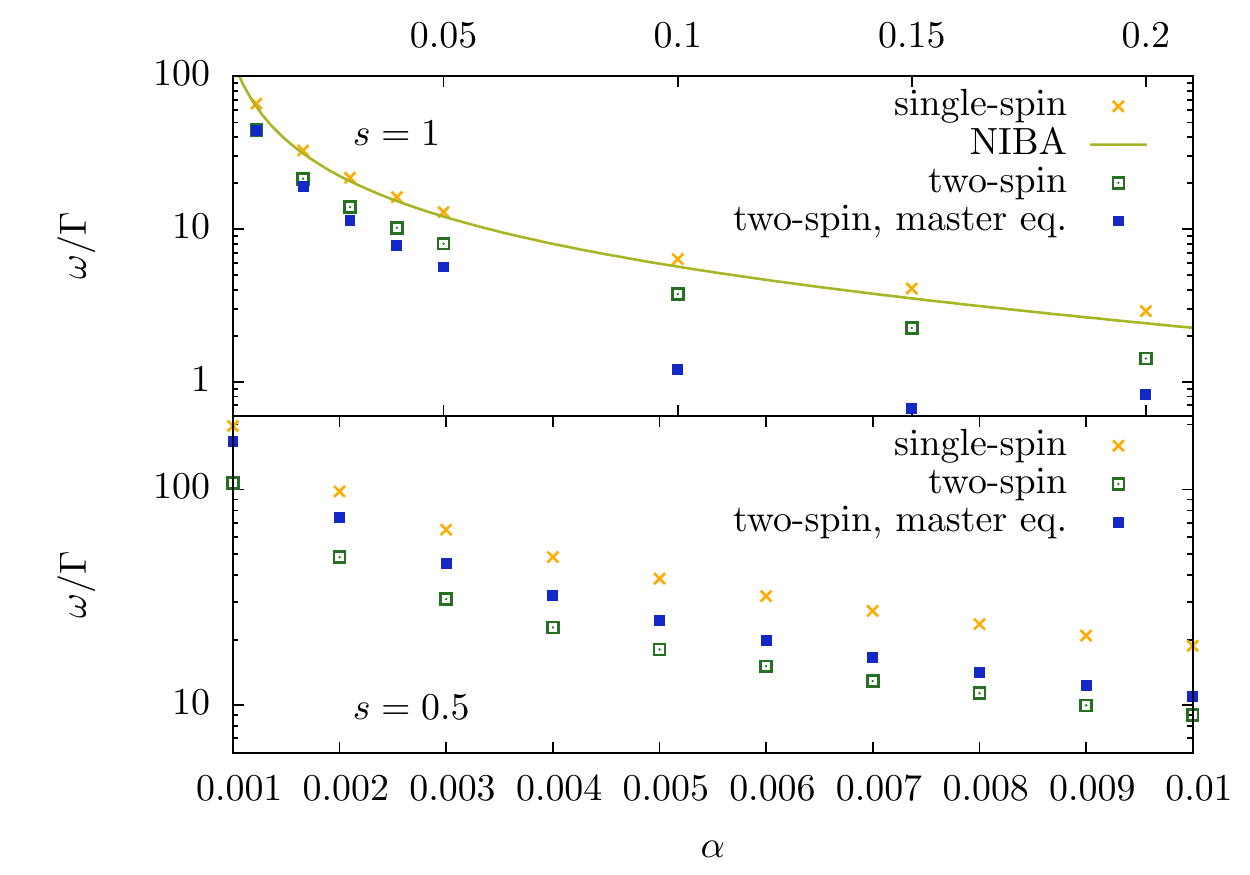}
  \caption{Quality factor of oscillations in the ohmic (upper panel) and subohmic (lower panel) single and two-spin boson models. TD-NRG results for the single (two-spin) boson model are shown as crosses (open squares). Quality factors derived from the Bloch-Redfield approach (see Sec.~\ref{sec:beat-decoh-breakd}) are displayed as filled squares. The solid line denotes the quality factor derived from the NIBA approximation of Eq.~\eqref{eq:43}.}
  \label{fig:15}
\end{figure}

More formally, we employ two approximations which are known to be equivalent to the NIBA in the single spin case.~\cite{PhysRevA.35.1436} First, we assume that the bath evolves freely $b_k(t) = b_k(0) e^{- i \omega_k t}$, neglecting any backaction of the bath on the spins. The reduced density matrix of the bath remains unaffected by the spins. Second, we trace out the bath degrees of freedom in a weak-coupling sense 
\begin{equation}
  \label{eq:42}
 \text{Tr}_B [e^{i \Omega(t)} e^{- i \Omega(s) }] = \exp\Bigl\{ \frac{1}{\pi} \bigl[ i Q_1(t-s) - Q_2(t-s) \bigr] \Bigr\}\,,
\end{equation}
which contains the bath correlation functions $Q_1(t) = \int_0^\infty d\omega J(\omega) \sin \omega t$ and $Q_2(t) = \int_0^\infty d\omega J(\omega) [1- \cos \omega t]$. As a result, the two spins are now completely decoupled from each other and their dynamics is described by
\begin{equation}
  \label{eq:43}
  \dot{\sigma}^z_j(t) = - \Delta_j^2 \int_{-\infty}^t ds \biggl\{ \sigma_j^z(s) \cos\biggl[ \frac{Q_1(t-s)}{\pi} \biggr] e^{- Q_2(t-s)/\pi} \biggl\}\,.
\end{equation}
Eq.~\eqref{eq:43} is known to describe the dynamics of the single spin-boson model in the famous NIBA.~\cite{RevModPhys.59.1,PhysRevA.35.1436} It can readily be solved by Laplace transformation. We refer to Refs.~\onlinecite{RevModPhys.59.1} and~\onlinecite{weissdissipation} for details. 

In the ohmic case, we find from Eq.~\eqref{eq:43} that the spin undergoes damped oscillations for $0 < \alpha < 1/2$. The frequency of the oscillations is given by $\omega_{\text{NIBA}} = \Delta_{\text{eff}} \cos \frac{\pi \alpha}{2(1-\alpha)}$ and the damping rate reads $\Gamma_{\text{NIBA}} = \Delta_{\text{eff}} \sin \frac{\pi \alpha}{2(1-\alpha)}$, where $\Delta_{\text{eff}} = [\Gamma(1-2\alpha) \cos \pi \alpha]^{1/2(1-\alpha)} \Delta (\Delta/\omega_c)^{\alpha/(1-\alpha)}$ is a renormalized tunneling element. The quality factor of the damped oscillations thus reads $\omega_{\text{NIBA}}/\Gamma_{\text{NIBA}} = \cot \frac{\pi \alpha}{2 (1-\alpha)}$. The NIBA also predicts an incoherent contribution to the spin dynamics, which is absent in the TD-NRG results. 
At $\alpha=1/2$, Eq.~\eqref{eq:43} predicts purely exponential relaxation $\av{\sigma^z(t)} = \exp[ - \Gamma t]$ with a decay rate given by $\Gamma = \Delta_{\text{eff}}(\alpha = 1/2) = \frac{\pi \Delta^2}{2 \omega_c}$. 
We refer to Ref.~\onlinecite{roosen_hofstetter_lehur_unpublished} for a detailed analysis of the single spin-boson dynamics within the TD-NRG.

In Fig.~\ref{fig:15}, we present a comparison of the quality factor of the oscillations $\omega/\Gamma$ for the single and two-spin boson systems as computed by TD-NRG and Bloch-Redfield. It is obtained by fitting the numerical results to the function $\sigma^z_{1,2}(t) = e^{-\Gamma t} \cos(\omega t)$. In the ohmic case, we also include the prediction from the NIBA, which agrees with the TD-NRG results of the single spin-boson model.
In general, we observe that the quality factor is smaller for the two-spin system. The Bloch-Redfield approach yields accurate results only in the ohmic case for small $\alpha$. It fails completely in the subohmic case due to the increased spectral weight of slow oscillator modes, even for weak dissipation. 





If we increase the dissipation strength further to values $\alpha > 1/2$, we observe an important difference between the single and the two-spin boson models. 
The two-spin boson model does not enter a localized phase for increasing values of $\alpha$ along the line $K_r=0$. 

For an ohmic bath, spin transitions occur even for a dissipation constant as large as $\alpha \geq 1.5$ (see Fig.~\ref{fig:13}). This is in stark contrast to the single ohmic spin-boson model, which displays a localization phase transition at a critical dissipation strength of $\alpha_c = 1 + \mathcal{O}(\frac{\Delta}{\omega_c})$. This explicitly shows that the approximations that lead to Eq.~\eqref{eq:43} even fail to give the correct qualitative dynamical behavior for stronger coupling $\alpha$.

In the subohmic case, the NIBA cannot be justified and erroneously yields localization for all $\alpha > 0$,~\cite{RevModPhys.59.1} while TD-NRG results for the single spin-boson model show that the system remains delocalized up to a finite critical value of $\alpha$.~\cite{anders:210402} Again, we find in Fig.~\ref{fig:14} that unlike the single spin-boson system, which localizes at a value of $\alpha_c = 0.107$ (for $\Delta/\omega_c = 0.1$ and $s=1/2$), the two-spin boson model always remains delocalized for $K_r = 0$.

We finally want to emphasize that the NIBA in the single spin-boson model breaks down for finite bias $\epsilon$.~\cite{weissdissipation} Thus, it cannot be applied to the two-spin problem away from the line $K_r=0$, since the Ising interaction acts as a mutual bias between the spins. One common approach is to account for the interblip correlations up to first order in the spin-bath coupling $\alpha$.~\cite{weissdissipation,Naegele2010622} This procedure, however, is equivalent to the perturbative Redfield approach that we have discussed in Sec.~\ref{sec:beat-decoh-breakd}.


\subsubsection{Toulouse point: relation to the single spin  case}
\label{sec:toulouse-point-1}
In this Section, we investigate the dynamics of the ohmic two-spin boson model at the special parameter point $K_r = 0$ and $\alpha = 1/2$.
\begin{figure}[t]
  \centering
  \includegraphics[width=\linewidth]{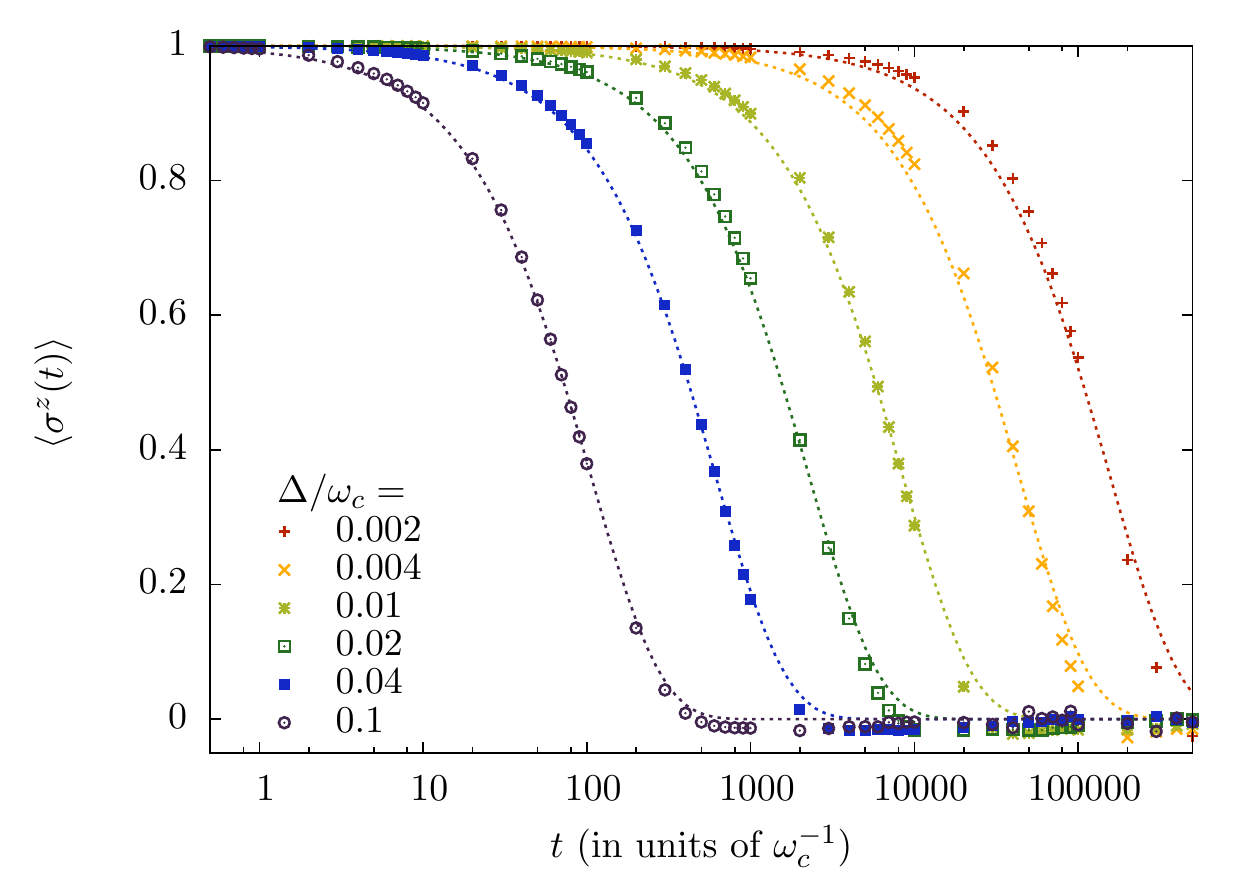}
  \caption{Exponential decay of $\av{\sigma^z(t)}$ at the Toulouse point $\alpha = 1/2$ in the single spin-boson model. Symbols are TD-NRG results for different tunneling amplitudes $\Delta/\omega_c$ which result in different decay rates $\Gamma = \frac{\pi \Delta^2}{2 \omega_c}$. Solid lines are fit with $f_i(t) = \exp[ - a_i \Gamma t]$ using fit parameters $a_i$ given in Table~\ref{tab:fit_toulouse}.}
  \label{fig:16}
\end{figure}

It is well-known that one can map the ohmic single spin-boson Hamiltonian in the scaling limit $\Delta/\omega_c \ll 1$ onto a fermionic resonant level model using bosonization and refermionization techniques.~\cite{RevModPhys.59.1} The fermionic model describes a localized level (dot) that is coupled via tunneling to a lead of free spinless fermions. In general the resulting model contains a Coulomb interaction term between the fermions on the dot and the ones in the lead. At the special (Toulouse) point of $\alpha=1/2$, however, this interaction vanishes, and the fermionic model can be solved exactly, also in nonequilibrium situations.~\cite{PhysRevB.50.5528,PhysRevB.79.115137,roosen_hofstetter_lehur_unpublished,PhysRevB.78.235110}

For a spin that is initially polarized along the $z$-direction, one finds purely exponential relaxation for $t>0$ (and $\epsilon = 0$)~\cite{roosen_hofstetter_lehur_unpublished}:
\begin{equation}
  \label{eq:44}
  \av{\sigma^z(t)} = \exp[ - \Gamma t]
\end{equation}
with decay rate $\Gamma = \pi \Delta^2/2 \omega_c$. It is worth noting that the NIBA predicts the same behavior, since it becomes exact at the Toulouse point of the single spin-boson model.~\cite{RevModPhys.59.1}

To prove the validity of the TD-NRG method in this strong coupling regime, we compare in Fig.~\ref{fig:16} our numerical results for a single spin-boson model with the exact solution of Eq.~\eqref{eq:44}. We observe that the decay is indeed purely exponential, and the decay rate is given by $\Gamma = \frac{\pi \Delta^2}{2\omega_c}$ in the scaling limit $\Delta/\omega_c \rightarrow 0$ (see Table~\ref{tab:fit_toulouse}).

\begin{figure}[t]
  \centering
  \includegraphics[width=\linewidth]{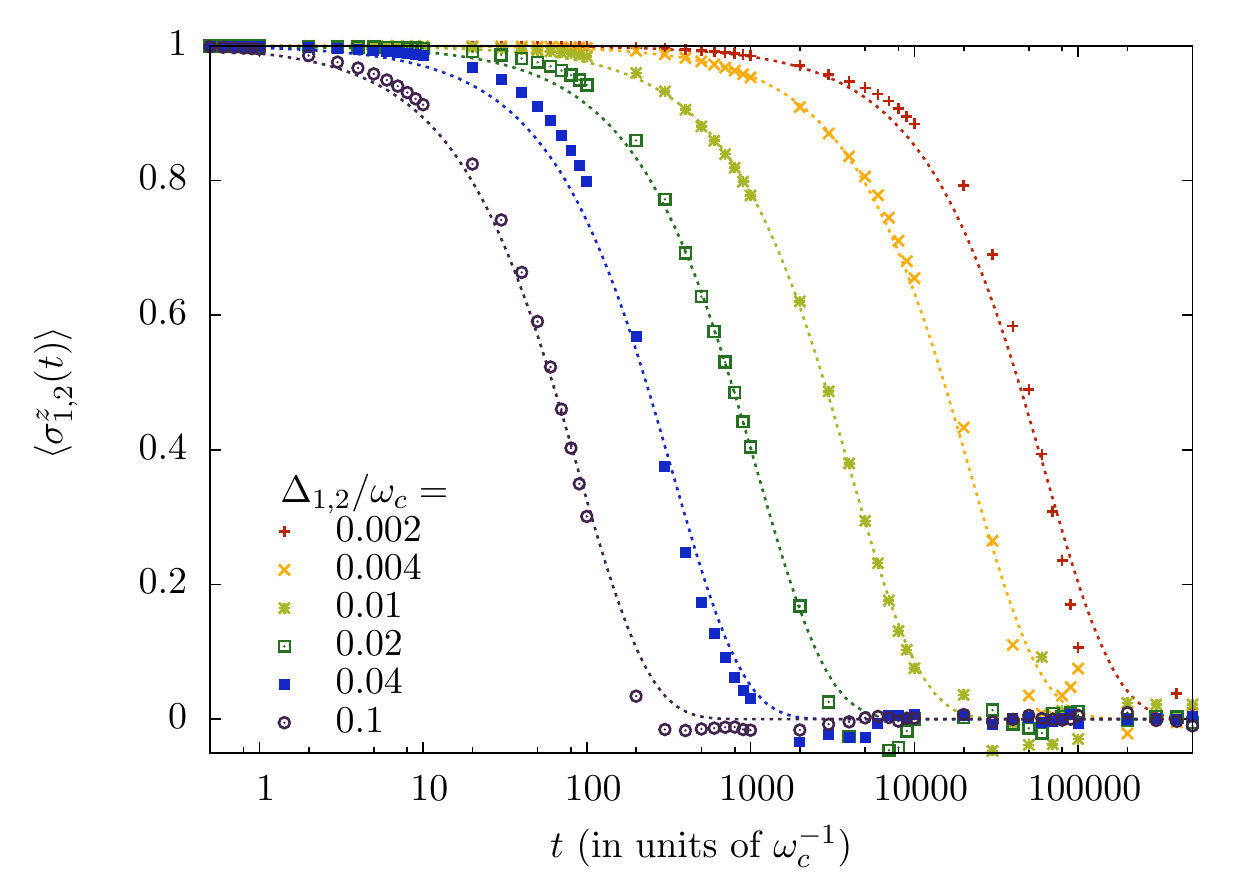}
  \caption{Exponential decay of $\av{\sigma^z_{1,2}(t)}$ at the generalized Toulouse point $K = 2\omega_c$, $\alpha = 1/2$ in the two-spin boson model. Symbols are TD-NRG results for different tunneling amplitudes $\Delta_{1,2}/\omega_c$, which result in different decay rates $\Gamma_{1,2} = \pi \Delta_{1,2}^2/(2 \omega_c)$. Since $\Delta_1 = \Delta_2$, we observe $\Gamma_1 = \Gamma_2$. Solid lines are fit with $f_i(t) = \exp[ - b_i \Gamma t]$ using fit parameters $b_i$ given in Table~\ref{tab:fit_toulouse}.}
  \label{fig:17}
\end{figure}

From Eq.~\eqref{eq:43}, we expect a similar behavior for the two-spin boson model at the (generalized) Toulouse point $K_r = 0$ and $\alpha = 1/2$. 
Indeed, as shown in Fig.~\ref{fig:17}, we observe that $\av{\sigma^z_{1,2}(t)}$ decay purely exponentially in the two-spin case as well. The decay rates of single and two-spin models, however, are slightly different. We find in Table~\ref{tab:fit_toulouse} that the decay rate of the two-spin boson model is about twice as large as the decay rate for the single spin-boson system. 
The difference of the decay rates 
is, again, due to the retarded part of the bath induced Ising interaction neglected in the derivation to Eq.~\eqref{eq:43}. We will qualitatively explain the factor two difference below, using a mapping to a fermionic resonant level model. 

One might ask whether the two-spin boson model can also be solved exactly via the bosonization mapping to a fermionic resonant level model. For two spins, however, it turns out that the fermionic model remains interacting at the Toulouse point, and thus cannot be solved exactly. As we show in detail in the Appendix, the (additional) interaction term is proportional to the tunneling elements $\Delta_{1,2}$, which describe tunneling between dot and lead in the fermionic model. Since we are interested in a solution that is non-perturbative in $\Delta_{1,2}$ we cannot treat this additional term as a weak perturbation. 

 
Specifically, the ohmic two-spin boson Hamiltonian in Eq.~\eqref{eq:1} can be mapped to a fermionic resonant level model with two energy levels on the dot. The mapping becomes exact in the scaling limit $\Delta_{1,2}/\omega_c \rightarrow 0$.~\cite{RevModPhys.59.1} As we derive in the Appendix, the resulting fermionic model is described by the Hamiltonian
\begin{align}
\label{eq:45}
    &H_{\text{RL}} = v_F \sum_{k>0} k c^\dag_k c_k + V \sum_{j=1}^2 \bigl[ d^\dag_j \psi(0) + \text{h.c.} \bigr] \nonumber \\ & - V \bigl[ (1-i) n_1 d^\dag_2 \psi(0) + (1+i) n_2 d_1^\dag \psi(0) +  \text{h.c.} \bigr]    \nonumber \\ & + \sum_{j=1}^2 \epsilon_j d^\dag_j d_j + 2 U \sum_{j=1}^2 \Bigl( d^\dag_j d_j - \frac12 \Bigr) :\!\psi^\dag(0) \psi(0)\!: \nonumber \\ & +  K_{\text{RL}} \Bigl(d^\dag_1 d_1 - \frac12 \Bigr) \Bigl( d^\dag_2 d_2 - \frac12 \Bigr) \,.
\end{align}
Here, $c_k$ annihilates a spinless fermion of momentum $k$ and energy $\omega_k = v_F k$ in the lead ($v_F$ is the Fermi velocity), and one defines $\psi(0) = L^{-1/2} \sum_k c_k$, where $L$ is the length of the lead. The colons denote normal ordering $:\!\!\psi^\dag(0) \psi(0)\!\!: \,= \psi^\dag(0) \psi(0) - \av{\psi^\dag(0) \psi(0)}$. The spin operators have been expressed in terms of fermionic operators on the dot using the renowned Jordan-Wigner transformation in a symmetric form~\cite{sachdev_qpt_book,Giamarchi-QuantumPhysIn1D,PhysRevB.48.13833}
\begin{align}
  \label{eq:46}
  \sigma_1^- &= [1- (1-i)n_2] d_1 \\
  \sigma_2^- &= [ 1 - (1+i) n_1] d_2 \\
  \sigma_j^z &= 2 n_j -1 \,, \text{for} \; j = 1,2\,,
\end{align}
where $n_j = d_j^\dag d_j$ are the dot occupation number operators. 
The parameters in $H_{RL}$ can be expressed in terms of the spin-boson parameters as
\begin{equation}
  \label{eq:47}
\begin{split}
  2 \pi \rho V^2 &\equiv \Gamma = \frac{\pi \Delta^2}{2 \omega_c}, \quad \rho U = \frac{1 - \sqrt{2\alpha}}{2}, \\ K_{RL} &= K + 2 \omega_c (1 - 2 \sqrt{2 \alpha})\,,
\end{split}
\end{equation}
where the fermionic density of states is defined as $\rho = 1/(2 \pi v_F)$. The bias field $\epsilon_j$ of the spin-boson model corresponds to the energy of the dot level $j$ with respect to the Fermi energy of the lead. 
\begin{table}[t]
  \centering
  \begin{tabular}{l|cc|cc}
\hline 
\hline
\;\;$\Delta/\omega_c$ \;\; &\;\;  $a_i$ \;\;& \;\;$\sigma(a_i)$ \;\;&\;\; $b_i$ \;\;&\;\; $\sigma(b_i)$ \;\;\\
\hline
\;\; 0.002 & 1.03& 0.03 & 2.54 & 0.26 \\
\;\; 0.004 & 0.93& 0.02 & 1.81 & 0.10 \\
\;\; 0.006 & 0.84& 0.02 & 1.64 & 0.06 \\
\;\; 0.01  & 0.76& 0.01 & 1.56 & 0.04 \\
\;\; 0.02  & 0.71& 0.01 & 1.45 & 0.04 \\
\;\; 0.04  & 0.66& 0.01 & 1.20 & 0.04 \\
\;\; 0.06  & 0.64& 0.01 & 0.94 & 0.04 \\
\;\; 0.1   & 0.61& 0.01 & 0.72 & 0.01 \\
\hline
\end{tabular}
  \caption{Fit parameters for single spin-boson model $\{a_i\}$ and two-spin boson model $\{b_i\}$ with standard error $\sigma(a_i)$ and $\sigma(b_i)$, for different values of $\Delta/\omega_c$, or $\Gamma = \pi \Delta^2/2 \omega_c$.  We fit the TD-NRG results to an exponential decay function which reads $f_i(t) = \exp[ - a_i \Gamma t]$ for the single spin-boson model, and $f_i(t) = \exp[ - b_i \Gamma t]$ for the two-spin boson model.}
\label{tab:fit_toulouse}
\end{table}

The last two interaction terms vanish at the Toulouse point: $U = K_{\text{RL}} = 0$ for $\alpha = 1/2$ and $K = 2 \omega_c$. The term, $V \bigl[ (1-i) n_1 d^\dag_2 \psi(0) + (1+i) n_2 d_1^\dag \psi(0) +  \text{h.c.} \bigr]$, 
however, is proportional to the dot-lead tunneling and thus remains. It arises due to the Jordan-Wigner string that accounts for the distinct commutation rules of fermions and spins at different sites. 

The dynamics of Eq.~\eqref{eq:45} cannot be solved exactly. Nevertheless, we can use the fermionic description to qualitatively understand that the decay rate of the two-spin boson model is about twice as large as in the single spin case.  To this end, we introduce the symmetric and antisymmetric combination of dot operators $D_s = [d_1 + d_2]/\sqrt{2}$ and $D_a = [d_1 - d_2]/\sqrt{2}$. The occupation numbers can then be expressed as $n_{1,2} = \frac12 [D_s^\dag D_s + D_a^\dag D_a \pm D_s^\dag D_a \pm D_a^\dag D_s]$, where the upper sign refers to $n_1$.  At the Toulouse point, the Hamiltonian then takes the form
\begin{align}
  \label{eq:48}
      H_{\text{RL}} &= H_0 + 
E (n_s + n_a) + \Delta E (D^\dag_s D_a + \text{h.c.})  \nonumber \\ & + \sqrt{2} V \bigl\{  \bigl( D_s^\dag - D^\dag_s n_a - i D_a^\dag n_s \bigr) \psi(0) + \text{h.c.} \bigr\} \,,
\end{align}
which contains the energy sum $E = (\epsilon_1 + \epsilon_2)/2$ and difference $\Delta E = (\epsilon_1 - \epsilon_2)/2$. We write $n_s = D^\dag_s D_s$, $n_a = D^\dag_a D_a$ and denote the free part of the lead electrons as $H_0 = v_F \sum_{k>0} k c^\dag_k c_k$. Both symmetric and antisymmetric state have the same energy $E$, and the original energy level difference translates into an effective tunneling coupling between them. 

For $\Delta E = 0$ and initially empty dots such that $n_s=n_a=0$, the antisymmetric state decouples from the system completely. The tunneling coupling between the symmetric state $D_s$ and the lead, however, is stronger than for each individual level $d_{1,2}$. It is given by $\sqrt{2} V$ instead of $V$ [see Eq.~\eqref{eq:45}], and the level will therefore fill twice as fast because $\Gamma \sim V^2$. As soon as $n_s > 0$, the antisymmetric state couples to the lead as well, and in equilibrium one finds that $\av{n_s} = \av{n_a} = 1/2$ for $E=0$. 
For $\Delta E = 0$, symmetry requires that $\av{n_1} = \av{n_2}$ and the expectation values $\av{D^\dag_s D_a}$ and $\av{D^\dag_a D_s}$ are thus purely imaginary. It then follows that $\av{n_1} = \av{n_2} = 1/2$ and $\av{\sigma^z_{1}} = \av{\sigma^z_2} = 0$ in equilibrium. For $\Delta E \neq 0$ the level correlations acquire a finite real part which gives rise to a difference in the level occupations $\av{n_{1,2}}$ in equilibrium. 


\subsection{Strong spin bath coupling}
\label{sec:large-spin-bath}
\begin{figure}[t]
    \begin{center}
        \includegraphics[width=\linewidth]{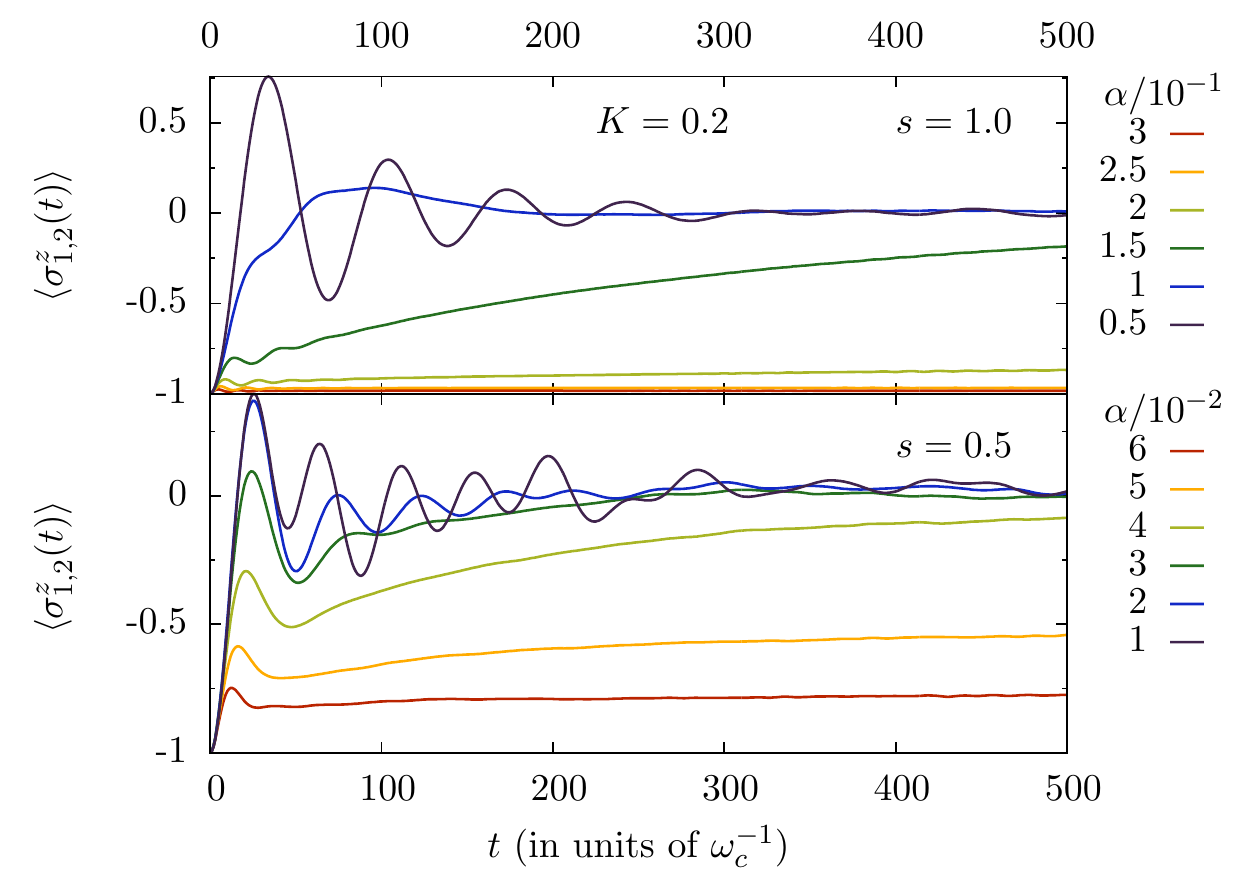}
    \end{center}
    \caption{Spin dynamics $\av{\sigma^z_{1,2}(t)}$ for different values of $\alpha$ in the regime of strong spin-bath coupling for ohmic (upper part) and subohmic bath with $s=1/2$ (lower part). Other parameters read $\Delta= 0.1 \omega_c$, $K=0.2 \, \omega_c$, $\epsilon_{1,2} = 0$. For this choice of $K$ the localization phase transition occurs at $\alpha_c \approx 0.25 (0.043)$ in the ohmic (subohmic) system. }
    \label{fig:18}
\end{figure}

In this Section, we focus on the regime of strong spin-bath coupling, where perturbative approaches are not applicable. We thus use the TD-NRG to calculate the spin dynamics, and focus on the differences between the case of an ohmic and a subohmic bath. We find that, qualitatively, the behavior in the two-spin boson systems resemble the one known from the (respective) single spin-boson model. For an ohmic bath, we observe in the upper part of Fig.~\ref{fig:18} that the coherence of oscillations is lost above a certain bath coupling strength, roughly given by $\alpha_c/2$, where $\alpha_c(K, \Delta_{1,2})$ denotes the critical value above which spin transitions are completely suppressed (localized regime of the phase diagram in Fig.~\ref{fig:2}). 

The situation is completely different for a subohmic bath as shown in the lower part of Fig.~\ref{fig:18}. Here, oscillations persist even into the localized region. This phenomenon was only recently discovered in the single spin-boson model~\cite{anders:210402,PhysRevB.81.054308} and we confirm that it also holds in the two-spin case. 

This qualitative difference between the ohmic and subohmic models at which point the coherence of the spin oscillations is lost (as a function of $\alpha$ and $K$), is also reflected in the behavior of the static entanglement entropy (see Sec.~\ref{sec:entanglement-entropy}).

\subsection{Generation of highly entangled steady states}
\label{sec:gener-highly-entangl}
In this Section, we show that the two-spin boson model displays interesting steady states for certain initial preparations. In this state, the spins are entangled with the bath while maintaining coherence between different spin configurations. 

Let us ask the question what happens if we polarize the spins initially in an antiferromagnetic configuration such as $\ket{\!\!\uparrow \downarrow}$ in a region of the phase diagram where the ground state phase is localized. At $t=0$, we then turn off the external bias fields completely, and follow the evolution of the spin reduced density matrix $\rho_S$ over time. Note that the system can only localize in one of the ferromagnetic spin states $\{\ket{\!\!\uparrow \uparrow}, \ket{\!\!\downarrow \downarrow}\}$ as discussed in Sec.~\ref{sec:qual-underst-phase}. We calculate $\rho_S(t)$ using TD-NRG and observe that after a time of the order $1/\Gamma = 2 \omega_c/(\pi \Delta^2)$ the system reaches a steady-state where the spin reduced density matrix takes the form 
\begin{equation}
  \label{eq:49}
  \rho_{S,ss} = \frac14 \begin{pmatrix} 1 & 0 & 0 & 0 \\ 0 & 1 & -1 & 0 \\ 0 & -1 & 1 & 0 \\ 0 & 0 & 0 & 1 \end{pmatrix}\,,
\end{equation}
where we use the standard basis $\{\ket{\!\!\uparrow \uparrow}, \ket{\!\! \uparrow \downarrow}, \ket{\!\!\downarrow \uparrow}, \ket{\!\!\downarrow \downarrow} \}$. With a probability of $\frac14$, the spins are thus localized in one of the ferromagnetic spin states $\{\ket{\!\!\uparrow \uparrow}, \ket{\!\!\downarrow \downarrow} \}$, and with a probability of $\frac12$ the spins are in the spin singlet state. The entanglement entropy $\mathcal{E}$, which is a measure of the entanglement between spins and bath, is nonzero in this state. Specifically, $\mathcal{E}(\rho_{S,ss}) = \frac32$ from Eq.~\eqref{eq:49}.

We can easily understand this form of the steady state by writing the initial state in terms of the singlet state $\ket{S=0, m=0} = [\ket{\!\!\uparrow \downarrow} - \ket{\!\!\downarrow \uparrow}]/\sqrt{2}$ and the triplet state $\ket{S=1, m=0} = [\ket{\!\!\uparrow \downarrow} + \ket{\!\!\downarrow \uparrow}]/\sqrt{2}$ as
\begin{equation}
  \label{eq:50}
  \ket{\!\!\uparrow \downarrow} = \frac{1}{\sqrt{2}} \bigl(\ket{S=1,m=0} + \ket{S=0, m=0}\bigr)\,.
\end{equation}
Whereas the singlet state does not couple to the bath at all, the triplet state localizes in one of the two ferromagnetic configurations. In this steady state, the spins are highly entangled with the bath modes, while developing and maintaining coherence between the two antiferromagnetic spin configurations.

\section{Conclusions}
\label{sec:conclusions}
We have presented an extensive study of a system of two Ising-coupled quantum spins in contact with a common bosonic bath. We have investigated several distinct equilibrium and nonequilibrium situations, both for the case of an ohmic as well as a subohmic bath. Employing the bosonic numerical renormalization group (NRG) and its recently developed time-dependent version (TD-NRG), we were able to describe the complete range of parameter space, from weak-to-strong coupling. We have applied a variety of different analytical approaches to comprehend, interpret and validate the numerical results. 

Using NRG we have calculated the ground state phase diagram of the model for $s=1/2$ and $s= 1$. We find a striking asymmetry in the behavior for ferromagnetic ($K<0$) and antiferromagnetic ($K>0$) Ising coupling, which we have understood as being the result of the fact that the system only localizes in a ferromagnetic spin configuration $\{\ket{\!\!\uparrow \uparrow}, \ket{\!\!\downarrow \downarrow}\}$. 

Let us briefly comment on the case of an SU($2$)-symmetric spin-spin interaction $\frac{K}{4} \boldsymbol{\sigma}_1\cdot \boldsymbol{\sigma}_2$. First, due to the fact that the spin couples to the bath via its $\sigma^z$-component, only the Ising component of the spin-spin interaction becomes renormalized by the bath $K^z_r$. This generates an anisotropic XXZ-coupling of the form $\frac{K^\perp}{4} (\sigma^x_1 \sigma^x_2 + \sigma^y_1 \sigma^y_2) + \frac{K^z_r}{4} \sigma^z_1 \sigma^z_2$, where $K^\perp = K$. To argue that the physical properties of a such a model are quite distinct from the Ising case where $K^\perp = 0$, we employ the well-established mapping to a two-impurity Kondo model.~\cite{RevModPhys.59.1,PhysRevB.69.214413} In our case, it turns out that the transverse part of the coupling is invariant under this mapping, and the Ising component renormalizes to $K^z_r = K + 4 \omega_c (1 - 2 \sqrt{\alpha})$. As shown in Refs.~\onlinecite{PhysRevB.69.214413,PhysRevB.60.R5125,PhysRevB.52.9528} for the isotropic two-impurity Kondo model, the behavior of a Kondo system with $K^{\perp} \neq 0$ greatly differs from the pure Ising case. In particular, in the absence of particle-hole symmetry the phase transition is replaced by a smooth crossover, whereas in the presence of particle-hole symmetry a phase transition occurs, but it is not of Kosterlitz-Thouless type.~\cite{PhysRevB.52.9528}
A detailed (numerical) analysis of the SU($2$)-symmetric two-spin boson system is left to further studies. 

Here, we have then investigated the behavior of the Ising two-spin boson system close to the localization phase transition, which is in different universality classes for $s=1$ and $s<1$. In the ohmic case, we find that coherence in the ground state is lost prior to localization. This is reflected in a plateau in the entanglement entropy, which describes the entanglement between spins and bath. Eventually at a critical coupling strength, the spin is localized where the entanglement entropy quickly drops to zero. We have reported that the size of the plateau shrinks considerably for larger values of the Ising coupling constant $K \gtrsim \omega_c$, indicating that, in this case, spin coherence is lost only close to the phase transition. Whereas the transition is in the Kosterlitz-Thouless universality class for the ohmic system, it is of continuous type for a subohmic bath, where we have studied the scaling of the spin magnetization $\av{\sigma^z_{1,2}}$ close to the transition. We have extracted critical exponents using NRG and compared them to analytical mean-field exponents. The agreement is reasonable though not perfect, which shows that NRG goes beyond the mean-field approximation that we have used. 

In the last part, we have discussed a number of different nonequilibrium scenarios. First, we have investigated the exactly solvable case of zero transverse fields where TD-NRG agrees perfectly with the exact analytical solution. For weak-spin bath coupling, we have provided quantitative limits on the applicability of the commonly used perturbative Bloch-Redfield method.

The coupling to the bath can be exploited to dynamically synchronize spin oscillations, which can prove useful in cases where a direct coupling between the spins is unavailable. Since the bath induced Ising coupling scales with the large bath cutoff frequency $\omega_c$, synchronization even occurs at small $\alpha$ where decoherence is weak. Nevertheless, this phenomenon could not be observed within the perturbative and Markovian Bloch-Redfield approach. 

We have then investigated the dynamics of the two-spin boson model for $K_r=0$, and have pointed out similarities and differences with the case of a single spin. We have derived the mapping of the two-spin boson model to a fermionic resonant level model, which contains two levels on the dot. In contrast to the single spin case, this model remains interacting at the Toulouse point due to an additional interaction term that arises from the Jordan-Wigner transformation of the spins. 

We have further studied $\av{\sigma^z_{1,2}(t)}$ in the crossover from weak to strong coupling where perturbative approaches cannot be applied. For strong coupling we have found that while spin transitions do not occur in the localized regime for the ohmic system, coherent spin oscillations persist into the localized regime for a subohmic bath. 

Finally, we have shown that the system features an interesting steady-state if we initially prepare it in an antiferromagnetic spin configuration within the localized regime. In this state, the spins are highly entangled with the bath degrees of freedom, and still develop and maintain coherence between different spin states. 

\section{Acknowledgments}
\label{sec:acknowlegdements}
This work is supported by NSF through Grant No. DMR-0803200, by the Yale Center for Quantum Information Physics through NSF Grant No. DMR-0653377, and by the DFG via Sonderforschungsbereich SFB/TR 49.

\appendix*
\section{Mapping of the two-spin boson model to the fermionic resonant level model}
\label{sec:toulouse-point}

In the Appendix we provide details of the mapping of the two-spin boson model to the fermionic resonant level model. Due to the Jordan-Wigner string, the corresponding fermionic model remains interacting at the Toulouse point in the case of two spins. 

Using bosonization techniques,~\cite{RevModPhys.59.1,Giamarchi-QuantumPhysIn1D,PhysRevB.69.214413} one can map the two-spin boson Hamiltonian in Eq.~\eqref{eq:1} with an ohmic spectral density $J(\omega) = 2 \pi \alpha \omega \exp[ - \omega/\omega_c]$ onto a fermionic resonant level model, which describes a central region (dot) coupled via tunneling to free spinless electrons in the lead. The number of spins in the spin-boson model is equal to the number of levels on the dot, and the number of bosonic baths is equal to the number of leads. 

Our starting point is the two-spin boson Hamiltonian in Eq.~\eqref{eq:1}
\begin{align}
  \label{eq:51}
  H_{\text{SB}} &=  \sum_{j=1}^2 \Bigl[ \frac{\Delta_j}{2} \sigma^x_j + \frac{\epsilon_j}{2} \sigma^z_j + \frac{\sigma^z_j}{2} \sum_{k>0} \lambda_k (b^\dag_k + b_k) \Bigr]  + \frac{K}{4} \sigma^z_1 \sigma^z_2 \nonumber \\ & \quad  + \sum_{k>0} \omega_k b_k^\dag b_k \,.
\end{align}
To obtain the mapping to the resonant level model (and similarly to the Kondo model), where the bath consists of free fermions, we choose the oscillator dispersion to be linear $\omega_k = v_F k$, with Fermi velocity $v_F$, and the coupling constants
\begin{equation}
  \label{eq:52}
  \lambda_k = - \sqrt{\alpha} 2 v_F \Bigl[ \frac{\pi k}{L} \Bigr]^{1/2} e^{-\omega_k/2 \omega_c}\,.
\end{equation}
The bath spectral density $J(\omega) = \pi \sum_{k>0} \lambda_k^2 \delta(\omega - \omega_k)$ is then of ohmic form $J(\omega) = 2 \pi \alpha \omega \exp(-\omega/\omega_c)$ up to an exponential cutoff at $\omega_c$. 
If we insert this into Eq.~\eqref{eq:51}, the spin-bath coupling term becomes 
\begin{equation}
  \label{eq:53}
  \sum_{j=1}^2 \pi \sigma^z_j \sum_{k>0} \Bigl[ - \sqrt{2 \alpha} v_F \Bigr] \Bigl[ \frac{k}{2 \pi L} \Bigr]^{1/2} e^{-a k/2} (b_k^\dag + b_k)\,,
\end{equation}
where we have defined the small distance cutoff $a = k_c^{-1} = v_F/\omega_c$.

We now apply a unitary (Luther-Emery) transformation to the Hamiltonian: $\tilde{H}_{\text{SB}} = U_\gamma H_{\text{SB}} U_\gamma^{-1}$ where $U_\gamma = \exp\Bigl[\frac{\gamma}{2} \sum_{j=1}^2 \sigma^z_j \xi\Bigr] $
with 
\begin{equation}
  \label{eq:54}
  \xi = \sum_{k>0} e^{- a k/2} \Bigl[ \frac{4 \pi }{k L} \Bigr]^{1/2} (b_k - b_k^\dag)\,.
\end{equation}
Evaluating this transformation term by term, and performing the sum over wavevectors $\frac{\pi v_F}{L} \sum_{k>0} e^{- a k} =  \frac{v_F}{2a} = \frac{\omega_c}{2}$, one finally obtains the Hamiltonian 
  \begin{equation}
    \label{eq:55}
\begin{split}
&    \tilde{H}_{\text{SB}} = v_F \sum_{k>0} k b^\dag_k b_k + \sum_{j=1}^2 \Bigl\{ \frac{\Delta_j}{2} \bigl( \sigma^+_j e^{\gamma \xi} + \text{h.c.} \bigr) + \frac{\epsilon_j}{2} \sigma^z_j \\ 
&\; + \pi v_F ( \sqrt{2} \gamma - \sqrt{2 \alpha}) \sigma^z_j \sum_{k>0} e^{- a k/2} \Bigl[ \frac{k}{2 \pi L} \Bigr]^{1/2} (b_k + b_k^\dag ) \Bigr\} \\
  & \;+ \bigl(K + 4 \omega_c \gamma^2 - 8 \omega_c \sqrt{\alpha} \gamma \bigr) \frac{\sigma^z_1 \sigma^z_2}{4} \,. 
\end{split}
  \end{equation}
One can show that a particular combination of the Bose operators $b_k, b_k^\dag$ can be made into an anticommuting Fermi field $\psi(x) = \frac{1}{\sqrt{ 2 \pi a}} \exp j(x) $ with~\cite{RevModPhys.59.1} 
\begin{equation}
  \label{eq:56}
  j(x) = \sum_{k>0} e^{- a k/2} \Bigl[ \frac{ 2 \pi }{k L} \Bigr]^{1/2} (b_k e^{i k x} - b_k^\dag e^{- i k x} )\,.
\end{equation}
The coefficients have been chosen such that $[j(x), j(y)] = - i \pi \text{sign}(x-y)$ for $a\rightarrow 0$ and thus $\{ \exp \pm j(x), \exp \pm j(y) \} = 0,  \;\; \text{for}\; x \neq y$. 
Choosing $\gamma = 1/\sqrt{2}$, one can thus identify the exponential $\exp[\xi/\sqrt{2}]$ which multiplies $\sigma^+_{1,2}$ in Eq.~\eqref{eq:55} as a localized spinless fermionic field $\psi(0) = (2 \pi a)^{-1/2} \exp \xi/\sqrt{2}$. The bosonic oscillator degrees of freedom are then interpreted as the density excitations $\rho(k) = \sum_p c^\dag_{p+k} c_p$, $\rho(-k) = \rho^\dag(k)$ of the fermions $\psi(x) = L^{-1/2} \sum_{k>0} c_k e^{i k x}$ via the bosonization identity $b_k = \Bigl[ \frac{2 \pi}{k L} \Bigr]^{1/2} \rho(-k)$.~\cite{RevModPhys.59.1}

Using refermionization  we can replace the free bosonic with a free fermionic Hamiltonian $v_F \sum_{k>0} k b^\dag_k b_k \rightarrow v_F \sum_{k>0} k c^\dag_k c_k$ and 
\begin{multline}
  \label{eq:57}
  \sum_{k>0} e^{- ak/2} \Bigl[ \frac{k}{2 \pi L} \Bigr]^{1/2} (b_k + b^\dag_k) \\= \sum_{k>0} \frac{e^{-ak/2}}{L} [\rho(-k) + \rho(k)] = \; :\!\psi^\dag(0) \psi(0)\!: \,,
\end{multline}
where $:\!\!\!\psi^\dag(0) \psi(0)\!\!\!: = \psi^\dag(0) \psi(0) - \av{\psi^\dag(0) \psi(0)}$ denotes normal ordering. Finally, we write the spin operators in terms of fermionic dot operators using the Jordan-Wigner transformation (in symmetric form)
\begin{align}
\label{eq:58}
\sigma_1^- &= [1- (1-i)n_2] d_1 \\
  \sigma_2^- &= [ 1 - (1+i) n_1] d_2 \\
  \sigma_j^z &= 2 n_j -1 \,, \text{for} \; j = 1,2\,.
\end{align}
We note that a less symmetric form of the transformation is equivalent. 
The Hamiltonian~\eqref{eq:51} thus reads in terms of fermionic fields as
\begin{align}
  \label{eq:59}
    &H_{\text{RL}} = v_F \sum_{k>0} k c^\dag_k c_k + \sum_{j=1}^2 V_j \bigl[ d^\dag_j \psi(0) + \text{h.c.} \bigr] \nonumber \\ & - \bigl[ V_2(1-i) n_1 d^\dag_2 \psi(0) + V_1 (1+i) n_2 d_1^\dag \psi(0) +  \text{h.c.} \bigr]    \nonumber \\ & + \sum_{j=1}^2 \epsilon_j d^\dag_j d_j + 2 U \sum_{j=1}^2 \Bigl( d^\dag_j d_j - \frac12 \Bigr) :\psi^\dag(0) \psi(0): \nonumber \\ & +  K_{\text{RL}} \Bigl(d^\dag_1 d_1 - \frac12 \Bigr) \Bigl( d^\dag_2 d_2 - \frac12 \Bigr) \,,
\end{align}
with tunneling coupling constant $V_j = \frac{\Delta_j}{2} \Bigl( \frac{1}{\rho \omega_c} \Bigr)^{1/2}$, dot-lead interaction $U = (1 - \sqrt{2 \alpha})/2 \rho$ and onsite coupling $K_{\text{RL}} = K + 2 \omega_c (1 - 2 \sqrt{2\alpha} )$. The interaction parameters $U$ and $ K_{\text{RL}}$ vanish at the Toulouse point $\alpha = 1/2$ and $K=2 \omega_c$. The additional interaction term $\bigl[ V_2 (1-i) n_1 d^\dag_2 \psi(0) + V_1 (1+i) n_2 d_1^\dag \psi(0) +  \text{h.c.} \bigr]$, 
however, is proportional to the tunneling couplings $V_j$ and remains present at the Toulouse point. As a result, the fermionic model cannot be solved exactly. 


\begin{thebibliography}{79}
\expandafter\ifx\csname natexlab\endcsname\relax\def\natexlab#1{#1}\fi
\expandafter\ifx\csname bibnamefont\endcsname\relax
  \def\bibnamefont#1{#1}\fi
\expandafter\ifx\csname bibfnamefont\endcsname\relax
  \def\bibfnamefont#1{#1}\fi
\expandafter\ifx\csname citenamefont\endcsname\relax
  \def\citenamefont#1{#1}\fi
\expandafter\ifx\csname url\endcsname\relax
  \def\url#1{\texttt{#1}}\fi
\expandafter\ifx\csname urlprefix\endcsname\relax\def\urlprefix{URL }\fi
\providecommand{\bibinfo}[2]{#2}
\providecommand{\eprint}[2][]{\url{#2}}

\bibitem[{\citenamefont{Le~Hur}(2008)}]{lehur_entanglement_spinboson}
\bibinfo{author}{\bibfnamefont{K.}~\bibnamefont{Le~Hur}},
  \bibinfo{journal}{Ann. Phys. (NY)} \textbf{\bibinfo{volume}{323}},
  \bibinfo{pages}{2208} (\bibinfo{year}{2008}).

\bibitem[{\citenamefont{Leggett et~al.}(1987)\citenamefont{Leggett,
  Chakravarty, Dorsey, Fisher, Garg, and Zwerger}}]{RevModPhys.59.1}
\bibinfo{author}{\bibfnamefont{A.~J.} \bibnamefont{Leggett}},
  \bibinfo{author}{\bibfnamefont{S.}~\bibnamefont{Chakravarty}},
  \bibinfo{author}{\bibfnamefont{A.~T.} \bibnamefont{Dorsey}},
  \bibinfo{author}{\bibfnamefont{M.~P.~A.} \bibnamefont{Fisher}},
  \bibinfo{author}{\bibfnamefont{A.}~\bibnamefont{Garg}}, \bibnamefont{and}
  \bibinfo{author}{\bibfnamefont{W.}~\bibnamefont{Zwerger}},
  \bibinfo{journal}{Rev. Mod. Phys.} \textbf{\bibinfo{volume}{59}},
  \bibinfo{pages}{1} (\bibinfo{year}{1987}).

\bibitem[{\citenamefont{Weiss}(2008)}]{weissdissipation}
\bibinfo{author}{\bibfnamefont{U.}~\bibnamefont{Weiss}},
  \emph{\bibinfo{title}{Quantum dissipative systems}},
  vol.~\bibinfo{volume}{13} of \emph{\bibinfo{series}{Series in Modern
  Condensed Matter Physics}} (\bibinfo{publisher}{World Scientific},
  \bibinfo{address}{Singapore}, \bibinfo{year}{2008}), \bibinfo{edition}{3rd}
  ed.

\bibitem[{\citenamefont{Carr}(2010)}]{KarynLeHur-UnderstandingQPT-Article}
\bibinfo{editor}{\bibfnamefont{L.~D.} \bibnamefont{Carr}}, ed.,
  \emph{\bibinfo{title}{Understanding Quantum Phase Transitions}}
  (\bibinfo{publisher}{CRC Press, Cleveland/Taylor \& Francis},
  \bibinfo{address}{Cleveland}, \bibinfo{year}{2010}), \bibinfo{edition}{1st}
  ed.

\bibitem[{\citenamefont{Zurek}(2003)}]{RevModPhys.75.715}
\bibinfo{author}{\bibfnamefont{W.~H.} \bibnamefont{Zurek}},
  \bibinfo{journal}{Rev. Mod. Phys.} \textbf{\bibinfo{volume}{75}},
  \bibinfo{pages}{715} (\bibinfo{year}{2003}).

\bibitem[{\citenamefont{Orth et~al.}(2010)\citenamefont{Orth, Imambekov, and
  Le~Hur}}]{PhysRevA.82.032118}
\bibinfo{author}{\bibfnamefont{P.~P.} \bibnamefont{Orth}},
  \bibinfo{author}{\bibfnamefont{A.}~\bibnamefont{Imambekov}},
  \bibnamefont{and} \bibinfo{author}{\bibfnamefont{K.}~\bibnamefont{Le~Hur}},
  \bibinfo{journal}{Phys. Rev. A} \textbf{\bibinfo{volume}{82}},
  \bibinfo{pages}{032118} (\bibinfo{year}{2010}).

\bibitem[{\citenamefont{Guo et~al.}(2009)\citenamefont{Guo, Weichselbaum,
  Kehrein, Xiang, and von Delft}}]{PhysRevB.79.115137}
\bibinfo{author}{\bibfnamefont{C.}~\bibnamefont{Guo}},
  \bibinfo{author}{\bibfnamefont{A.}~\bibnamefont{Weichselbaum}},
  \bibinfo{author}{\bibfnamefont{S.}~\bibnamefont{Kehrein}},
  \bibinfo{author}{\bibfnamefont{T.}~\bibnamefont{Xiang}}, \bibnamefont{and}
  \bibinfo{author}{\bibfnamefont{J.}~\bibnamefont{von Delft}},
  \bibinfo{journal}{Phys. Rev. B} \textbf{\bibinfo{volume}{79}},
  \bibinfo{pages}{115137} (\bibinfo{year}{2009}).

\bibitem[{\citenamefont{Nalbach and Thorwart}(2009)}]{nalbach:220401}
\bibinfo{author}{\bibfnamefont{P.}~\bibnamefont{Nalbach}} \bibnamefont{and}
  \bibinfo{author}{\bibfnamefont{M.}~\bibnamefont{Thorwart}},
  \bibinfo{journal}{Phys. Rev. Lett.} \textbf{\bibinfo{volume}{103}},
  \bibinfo{eid}{220401} (\bibinfo{year}{2009}).

\bibitem[{\citenamefont{Roosen et~al.}(2010)\citenamefont{Roosen, Le~Hur, and
  Hofstetter}}]{roosen_hofstetter_lehur_unpublished}
\bibinfo{author}{\bibfnamefont{D.}~\bibnamefont{Roosen}},
  \bibinfo{author}{\bibfnamefont{K.}~\bibnamefont{Le~Hur}}, \bibnamefont{and}
  \bibinfo{author}{\bibfnamefont{W.}~\bibnamefont{Hofstetter}}
  (\bibinfo{year}{2010}), \bibinfo{note}{in preparation}.

\bibitem[{\citenamefont{Uhrig}(2007)}]{PhysRevLett.98.100504}
\bibinfo{author}{\bibfnamefont{G.~S.} \bibnamefont{Uhrig}},
  \bibinfo{journal}{Phys. Rev. Lett.} \textbf{\bibinfo{volume}{98}},
  \bibinfo{pages}{100504} (\bibinfo{year}{2007}).

\bibitem[{\citenamefont{Grabert and
  Wipf}(1990)}]{GrabertWipf-SpinBosonTunnelingOfDefects1990}
\bibinfo{author}{\bibfnamefont{H.}~\bibnamefont{Grabert}} \bibnamefont{and}
  \bibinfo{author}{\bibfnamefont{H.}~\bibnamefont{Wipf}},
  \bibinfo{journal}{Adv. Solid State Phys.} \textbf{\bibinfo{volume}{30}},
  \bibinfo{pages}{1} (\bibinfo{year}{1990}).

\bibitem[{\citenamefont{Marcus}(1956)}]{Marcus1956}
\bibinfo{author}{\bibfnamefont{R.}~\bibnamefont{Marcus}}, \bibinfo{journal}{J.
  Chem. Phys.} \textbf{\bibinfo{volume}{24}}, \bibinfo{pages}{966}
  (\bibinfo{year}{1956}).

\bibitem[{\citenamefont{Marcus and Sutin}(1985)}]{Marcus1985}
\bibinfo{author}{\bibfnamefont{R.~A.} \bibnamefont{Marcus}} \bibnamefont{and}
  \bibinfo{author}{\bibfnamefont{N.}~\bibnamefont{Sutin}},
  \bibinfo{journal}{Biochim. Biophys. Acta} \textbf{\bibinfo{volume}{811}},
  \bibinfo{pages}{265} (\bibinfo{year}{1985}).

\bibitem[{\citenamefont{Makhlin et~al.}(2001)\citenamefont{Makhlin, Sch\"on,
  and Shnirman}}]{RevModPhys.73.357}
\bibinfo{author}{\bibfnamefont{Y.}~\bibnamefont{Makhlin}},
  \bibinfo{author}{\bibfnamefont{G.}~\bibnamefont{Sch\"on}}, \bibnamefont{and}
  \bibinfo{author}{\bibfnamefont{A.}~\bibnamefont{Shnirman}},
  \bibinfo{journal}{Rev. Mod. Phys.} \textbf{\bibinfo{volume}{73}},
  \bibinfo{pages}{357} (\bibinfo{year}{2001}).

\bibitem[{\citenamefont{Mooij et~al.}(1999)\citenamefont{Mooij, Orlando,
  Levitov, Tian, van~der Wal, and Lloyd}}]{J.E.Mooij08131999}
\bibinfo{author}{\bibfnamefont{J.~E.} \bibnamefont{Mooij}},
  \bibinfo{author}{\bibfnamefont{T.~P.} \bibnamefont{Orlando}},
  \bibinfo{author}{\bibfnamefont{L.}~\bibnamefont{Levitov}},
  \bibinfo{author}{\bibfnamefont{L.}~\bibnamefont{Tian}},
  \bibinfo{author}{\bibfnamefont{C.~H.} \bibnamefont{van~der Wal}},
  \bibnamefont{and} \bibinfo{author}{\bibfnamefont{S.}~\bibnamefont{Lloyd}},
  \bibinfo{journal}{Science} \textbf{\bibinfo{volume}{285}},
  \bibinfo{pages}{1036} (\bibinfo{year}{1999}).

\bibitem[{\citenamefont{Manucharyan et~al.}(2009)\citenamefont{Manucharyan,
  Koch, Brink, Glazman, and Devoret}}]{Manucharyan_coherent_osc_in_sc_loop}
\bibinfo{author}{\bibfnamefont{V.~E.} \bibnamefont{Manucharyan}},
  \bibinfo{author}{\bibfnamefont{J.}~\bibnamefont{Koch}},
  \bibinfo{author}{\bibfnamefont{M.}~\bibnamefont{Brink}},
  \bibinfo{author}{\bibfnamefont{L.~I.} \bibnamefont{Glazman}},
  \bibnamefont{and} \bibinfo{author}{\bibfnamefont{M.~H.}
  \bibnamefont{Devoret}}, \bibinfo{journal}{arXiv:0910.3039v1}
  (\bibinfo{year}{2009}).

\bibitem[{\citenamefont{Porras et~al.}(2008)\citenamefont{Porras, Marquardt,
  von Delft, and Cirac}}]{porras:010101}
\bibinfo{author}{\bibfnamefont{D.}~\bibnamefont{Porras}},
  \bibinfo{author}{\bibfnamefont{F.}~\bibnamefont{Marquardt}},
  \bibinfo{author}{\bibfnamefont{J.}~\bibnamefont{von Delft}},
  \bibnamefont{and} \bibinfo{author}{\bibfnamefont{J.~I.} \bibnamefont{Cirac}},
  \bibinfo{journal}{Phys. Rev. A} \textbf{\bibinfo{volume}{78}},
  \bibinfo{eid}{010101(R)} (\bibinfo{year}{2008}).

\bibitem[{\citenamefont{Dzsotjan et~al.}(2010)\citenamefont{Dzsotjan,
  S\o{}rensen, and Fleischhauer}}]{PhysRevB.82.075427}
\bibinfo{author}{\bibfnamefont{D.}~\bibnamefont{Dzsotjan}},
  \bibinfo{author}{\bibfnamefont{A.~S.} \bibnamefont{S\o{}rensen}},
  \bibnamefont{and}
  \bibinfo{author}{\bibfnamefont{M.}~\bibnamefont{Fleischhauer}},
  \bibinfo{journal}{Phys. Rev. B} \textbf{\bibinfo{volume}{82}},
  \bibinfo{pages}{075427} (\bibinfo{year}{2010}).

\bibitem[{\citenamefont{Recati et~al.}(2005)\citenamefont{Recati, Fedichev,
  Zwerger, von Delft, and Zoller}}]{recati:040404}
\bibinfo{author}{\bibfnamefont{A.}~\bibnamefont{Recati}},
  \bibinfo{author}{\bibfnamefont{P.~O.} \bibnamefont{Fedichev}},
  \bibinfo{author}{\bibfnamefont{W.}~\bibnamefont{Zwerger}},
  \bibinfo{author}{\bibfnamefont{J.}~\bibnamefont{von Delft}},
  \bibnamefont{and} \bibinfo{author}{\bibfnamefont{P.}~\bibnamefont{Zoller}},
  \bibinfo{journal}{Phys. Rev. Lett.} \textbf{\bibinfo{volume}{94}},
  \bibinfo{eid}{040404} (\bibinfo{year}{2005}).

\bibitem[{\citenamefont{Orth et~al.}(2008)\citenamefont{Orth, Stanic, and
  Le~Hur}}]{orth:051601}
\bibinfo{author}{\bibfnamefont{P.~P.} \bibnamefont{Orth}},
  \bibinfo{author}{\bibfnamefont{I.}~\bibnamefont{Stanic}}, \bibnamefont{and}
  \bibinfo{author}{\bibfnamefont{K.}~\bibnamefont{Le~Hur}},
  \bibinfo{journal}{Phys. Rev. A} \textbf{\bibinfo{volume}{77}},
  \bibinfo{eid}{051601(R)} (\bibinfo{year}{2008}).

\bibitem[{\citenamefont{Pertot et~al.}(2010)\citenamefont{Pertot, Gadway, and
  Schneble}}]{PhysRevLett.104.200402}
\bibinfo{author}{\bibfnamefont{D.}~\bibnamefont{Pertot}},
  \bibinfo{author}{\bibfnamefont{B.}~\bibnamefont{Gadway}}, \bibnamefont{and}
  \bibinfo{author}{\bibfnamefont{D.}~\bibnamefont{Schneble}},
  \bibinfo{journal}{Phys. Rev. Lett.} \textbf{\bibinfo{volume}{104}},
  \bibinfo{pages}{200402} (\bibinfo{year}{2010}).

\bibitem[{\citenamefont{Gadway et~al.}(2010)\citenamefont{Gadway, Pertot,
  Reimann, and Schneble}}]{PhysRevLett.105.045303}
\bibinfo{author}{\bibfnamefont{B.}~\bibnamefont{Gadway}},
  \bibinfo{author}{\bibfnamefont{D.}~\bibnamefont{Pertot}},
  \bibinfo{author}{\bibfnamefont{R.}~\bibnamefont{Reimann}}, \bibnamefont{and}
  \bibinfo{author}{\bibfnamefont{D.}~\bibnamefont{Schneble}},
  \bibinfo{journal}{Phys. Rev. Lett.} \textbf{\bibinfo{volume}{105}},
  \bibinfo{pages}{045303} (\bibinfo{year}{2010}).

\bibitem[{\citenamefont{Raimond et~al.}(2001)\citenamefont{Raimond, Brune, and
  Haroche}}]{RevModPhys.73.565}
\bibinfo{author}{\bibfnamefont{J.~M.} \bibnamefont{Raimond}},
  \bibinfo{author}{\bibfnamefont{M.}~\bibnamefont{Brune}}, \bibnamefont{and}
  \bibinfo{author}{\bibfnamefont{S.}~\bibnamefont{Haroche}},
  \bibinfo{journal}{Rev. Mod. Phys.} \textbf{\bibinfo{volume}{73}},
  \bibinfo{pages}{565} (\bibinfo{year}{2001}).

\bibitem[{\citenamefont{Schoelkopf and Girvin}(2008)}]{Schoelkopf_Nature_2008}
\bibinfo{author}{\bibfnamefont{R.~J.} \bibnamefont{Schoelkopf}}
  \bibnamefont{and} \bibinfo{author}{\bibfnamefont{S.~M.}
  \bibnamefont{Girvin}}, \bibinfo{journal}{Nature (London)}
  \textbf{\bibinfo{volume}{451}}, \bibinfo{pages}{664} (\bibinfo{year}{2008}).

\bibitem[{\citenamefont{Koch and Le~Hur}(2009)}]{koch:023811}
\bibinfo{author}{\bibfnamefont{J.}~\bibnamefont{Koch}} \bibnamefont{and}
  \bibinfo{author}{\bibfnamefont{K.}~\bibnamefont{Le~Hur}},
  \bibinfo{journal}{Phys. Rev. A} \textbf{\bibinfo{volume}{80}},
  \bibinfo{eid}{023811} (\bibinfo{year}{2009}).

\bibitem[{\citenamefont{Hartmann et~al.}(2008)\citenamefont{Hartmann, Brandao,
  and Plenio}}]{Hartmann_Plenio_LPRev_2008}
\bibinfo{author}{\bibfnamefont{M.}~\bibnamefont{Hartmann}},
  \bibinfo{author}{\bibfnamefont{G.}~\bibnamefont{Brandao}}, \bibnamefont{and}
  \bibinfo{author}{\bibfnamefont{M.}~\bibnamefont{Plenio}},
  \bibinfo{journal}{Laser \& Photon. Rev.} \textbf{\bibinfo{volume}{2}},
  \bibinfo{pages}{527} (\bibinfo{year}{2008}).

\bibitem[{\citenamefont{Zell et~al.}(2009)\citenamefont{Zell, Queisser, and
  Klesse}}]{PhysRevLett.102.160501}
\bibinfo{author}{\bibfnamefont{T.}~\bibnamefont{Zell}},
  \bibinfo{author}{\bibfnamefont{F.}~\bibnamefont{Queisser}}, \bibnamefont{and}
  \bibinfo{author}{\bibfnamefont{R.}~\bibnamefont{Klesse}},
  \bibinfo{journal}{Phys. Rev. Lett.} \textbf{\bibinfo{volume}{102}},
  \bibinfo{pages}{160501} (\bibinfo{year}{2009}).

\bibitem[{\citenamefont{Waugh et~al.}(1995)\citenamefont{Waugh, Berry, Mar,
  Westervelt, Campman, and Gossard}}]{PhysRevLett.75.705}
\bibinfo{author}{\bibfnamefont{F.~R.} \bibnamefont{Waugh}},
  \bibinfo{author}{\bibfnamefont{M.~J.} \bibnamefont{Berry}},
  \bibinfo{author}{\bibfnamefont{D.~J.} \bibnamefont{Mar}},
  \bibinfo{author}{\bibfnamefont{R.~M.} \bibnamefont{Westervelt}},
  \bibinfo{author}{\bibfnamefont{K.~L.} \bibnamefont{Campman}},
  \bibnamefont{and} \bibinfo{author}{\bibfnamefont{A.~C.}
  \bibnamefont{Gossard}}, \bibinfo{journal}{Phys. Rev. Lett.}
  \textbf{\bibinfo{volume}{75}}, \bibinfo{pages}{705} (\bibinfo{year}{1995}).

\bibitem[{\citenamefont{Matveev et~al.}(1996)\citenamefont{Matveev, Glazman,
  and Baranger}}]{PhysRevB.53.1034}
\bibinfo{author}{\bibfnamefont{K.~A.} \bibnamefont{Matveev}},
  \bibinfo{author}{\bibfnamefont{L.~I.} \bibnamefont{Glazman}},
  \bibnamefont{and} \bibinfo{author}{\bibfnamefont{H.~U.}
  \bibnamefont{Baranger}}, \bibinfo{journal}{Phys. Rev. B}
  \textbf{\bibinfo{volume}{53}}, \bibinfo{pages}{1034} (\bibinfo{year}{1996}).

\bibitem[{\citenamefont{Golden and Halperin}(1996)}]{PhysRevB.53.3893}
\bibinfo{author}{\bibfnamefont{J.~M.} \bibnamefont{Golden}} \bibnamefont{and}
  \bibinfo{author}{\bibfnamefont{B.~I.} \bibnamefont{Halperin}},
  \bibinfo{journal}{Phys. Rev. B} \textbf{\bibinfo{volume}{53}},
  \bibinfo{pages}{3893} (\bibinfo{year}{1996}).

\bibitem[{\citenamefont{Contreras-Pulido and
  Aguado}(2008)}]{PhysRevB.77.155420}
\bibinfo{author}{\bibfnamefont{L.~D.} \bibnamefont{Contreras-Pulido}}
  \bibnamefont{and} \bibinfo{author}{\bibfnamefont{R.}~\bibnamefont{Aguado}},
  \bibinfo{journal}{Phys. Rev. B} \textbf{\bibinfo{volume}{77}},
  \bibinfo{pages}{155420} (\bibinfo{year}{2008}).

\bibitem[{\citenamefont{Campagnano et~al.}(2010)\citenamefont{Campagnano,
  Hamma, and Weiss}}]{Campagnano2010}
\bibinfo{author}{\bibfnamefont{G.}~\bibnamefont{Campagnano}},
  \bibinfo{author}{\bibfnamefont{A.}~\bibnamefont{Hamma}}, \bibnamefont{and}
  \bibinfo{author}{\bibfnamefont{U.}~\bibnamefont{Weiss}},
  \bibinfo{journal}{Phys. Lett. A} \textbf{\bibinfo{volume}{374}},
  \bibinfo{pages}{416} (\bibinfo{year}{2010}).

\bibitem[{\citenamefont{Solenov et~al.}(2007)\citenamefont{Solenov, Tolkunov,
  and Privman}}]{PhysRevB.75.035134}
\bibinfo{author}{\bibfnamefont{D.}~\bibnamefont{Solenov}},
  \bibinfo{author}{\bibfnamefont{D.}~\bibnamefont{Tolkunov}}, \bibnamefont{and}
  \bibinfo{author}{\bibfnamefont{V.}~\bibnamefont{Privman}},
  \bibinfo{journal}{Phys. Rev. B} \textbf{\bibinfo{volume}{75}},
  \bibinfo{pages}{035134} (\bibinfo{year}{2007}).

\bibitem[{\citenamefont{Braun}(2002)}]{PhysRevLett.89.277901}
\bibinfo{author}{\bibfnamefont{D.}~\bibnamefont{Braun}},
  \bibinfo{journal}{Phys. Rev. Lett.} \textbf{\bibinfo{volume}{89}},
  \bibinfo{pages}{277901} (\bibinfo{year}{2002}).

\bibitem[{\citenamefont{Benatti et~al.}(2003)\citenamefont{Benatti, Floreanini,
  and Piani}}]{PhysRevLett.91.070402}
\bibinfo{author}{\bibfnamefont{F.}~\bibnamefont{Benatti}},
  \bibinfo{author}{\bibfnamefont{R.}~\bibnamefont{Floreanini}},
  \bibnamefont{and} \bibinfo{author}{\bibfnamefont{M.}~\bibnamefont{Piani}},
  \bibinfo{journal}{Phys. Rev. Lett.} \textbf{\bibinfo{volume}{91}},
  \bibinfo{pages}{070402} (\bibinfo{year}{2003}).

\bibitem[{\citenamefont{Garst et~al.}(2004)\citenamefont{Garst, Kehrein,
  Pruschke, Rosch, and Vojta}}]{PhysRevB.69.214413}
\bibinfo{author}{\bibfnamefont{M.}~\bibnamefont{Garst}},
  \bibinfo{author}{\bibfnamefont{S.}~\bibnamefont{Kehrein}},
  \bibinfo{author}{\bibfnamefont{T.}~\bibnamefont{Pruschke}},
  \bibinfo{author}{\bibfnamefont{A.}~\bibnamefont{Rosch}}, \bibnamefont{and}
  \bibinfo{author}{\bibfnamefont{M.}~\bibnamefont{Vojta}},
  \bibinfo{journal}{Phys. Rev. B} \textbf{\bibinfo{volume}{69}},
  \bibinfo{pages}{214413} (\bibinfo{year}{2004}).

\bibitem[{\citenamefont{N\"agele and Weiss}(2010)}]{Naegele2010622}
\bibinfo{author}{\bibfnamefont{P.}~\bibnamefont{N\"agele}} \bibnamefont{and}
  \bibinfo{author}{\bibfnamefont{U.}~\bibnamefont{Weiss}},
  \bibinfo{journal}{Physica E} \textbf{\bibinfo{volume}{42}},
  \bibinfo{pages}{622} (\bibinfo{year}{2010}).

\bibitem[{\citenamefont{N\"agele et~al.}(2008)\citenamefont{N\"agele,
  Campagnano, and Weiss}}]{1367-2630-10-11-115010}
\bibinfo{author}{\bibfnamefont{P.}~\bibnamefont{N\"agele}},
  \bibinfo{author}{\bibfnamefont{G.}~\bibnamefont{Campagnano}},
  \bibnamefont{and} \bibinfo{author}{\bibfnamefont{U.}~\bibnamefont{Weiss}},
  \bibinfo{journal}{New J. Phys.} \textbf{\bibinfo{volume}{10}},
  \bibinfo{pages}{115010} (\bibinfo{year}{2008}).

\bibitem[{\citenamefont{Tong and Vojta}(2006)}]{PhysRevLett.97.016802}
\bibinfo{author}{\bibfnamefont{N.-H.} \bibnamefont{Tong}} \bibnamefont{and}
  \bibinfo{author}{\bibfnamefont{M.}~\bibnamefont{Vojta}},
  \bibinfo{journal}{Phys. Rev. Lett.} \textbf{\bibinfo{volume}{97}},
  \bibinfo{pages}{016802} (\bibinfo{year}{2006}).

\bibitem[{\citenamefont{Kirchner and Si}(2009)}]{PhysRevLett.103.206401}
\bibinfo{author}{\bibfnamefont{S.}~\bibnamefont{Kirchner}} \bibnamefont{and}
  \bibinfo{author}{\bibfnamefont{Q.}~\bibnamefont{Si}}, \bibinfo{journal}{Phys.
  Rev. Lett.} \textbf{\bibinfo{volume}{103}}, \bibinfo{pages}{206401}
  (\bibinfo{year}{2009}).

\bibitem[{\citenamefont{Wilson}(1975)}]{Wilson1975}
\bibinfo{author}{\bibfnamefont{K.~G.} \bibnamefont{Wilson}},
  \bibinfo{journal}{Rev. Mod. Phys.} \textbf{\bibinfo{volume}{47}},
  \bibinfo{pages}{773} (\bibinfo{year}{1975}).

\bibitem[{\citenamefont{Bulla et~al.}(2005)\citenamefont{Bulla, Lee, Tong, and
  Vojta}}]{PhysRevB.71.045122}
\bibinfo{author}{\bibfnamefont{R.}~\bibnamefont{Bulla}},
  \bibinfo{author}{\bibfnamefont{H.-J.} \bibnamefont{Lee}},
  \bibinfo{author}{\bibfnamefont{N.-H.} \bibnamefont{Tong}}, \bibnamefont{and}
  \bibinfo{author}{\bibfnamefont{M.}~\bibnamefont{Vojta}},
  \bibinfo{journal}{Phys. Rev. B} \textbf{\bibinfo{volume}{71}},
  \bibinfo{pages}{045122} (\bibinfo{year}{2005}).

\bibitem[{\citenamefont{Bulla et~al.}(2008)\citenamefont{Bulla, Costi, and
  Pruschke}}]{RevModPhys.80.395}
\bibinfo{author}{\bibfnamefont{R.}~\bibnamefont{Bulla}},
  \bibinfo{author}{\bibfnamefont{T.~A.} \bibnamefont{Costi}}, \bibnamefont{and}
  \bibinfo{author}{\bibfnamefont{T.}~\bibnamefont{Pruschke}},
  \bibinfo{journal}{Rev. Mod. Phys.} \textbf{\bibinfo{volume}{80}},
  \bibinfo{pages}{395} (\bibinfo{year}{2008}).

\bibitem[{\citenamefont{Li et~al.}(2005)\citenamefont{Li, Le~Hur, and
  Hofstetter}}]{PhysRevLett.95.086406}
\bibinfo{author}{\bibfnamefont{M.-R.} \bibnamefont{Li}},
  \bibinfo{author}{\bibfnamefont{K.}~\bibnamefont{Le~Hur}}, \bibnamefont{and}
  \bibinfo{author}{\bibfnamefont{W.}~\bibnamefont{Hofstetter}},
  \bibinfo{journal}{Phys. Rev. Lett.} \textbf{\bibinfo{volume}{95}},
  \bibinfo{pages}{086406} (\bibinfo{year}{2005}).

\bibitem[{\citenamefont{Anders and Schiller}(2005)}]{PhysRevLett.95.196801}
\bibinfo{author}{\bibfnamefont{F.~B.} \bibnamefont{Anders}} \bibnamefont{and}
  \bibinfo{author}{\bibfnamefont{A.}~\bibnamefont{Schiller}},
  \bibinfo{journal}{Phys. Rev. Lett.} \textbf{\bibinfo{volume}{95}},
  \bibinfo{pages}{196801} (\bibinfo{year}{2005}).

\bibitem[{\citenamefont{Anders and Schiller}(2006)}]{PhysRevB.74.245113}
\bibinfo{author}{\bibfnamefont{F.~B.} \bibnamefont{Anders}} \bibnamefont{and}
  \bibinfo{author}{\bibfnamefont{A.}~\bibnamefont{Schiller}},
  \bibinfo{journal}{Phys. Rev. B} \textbf{\bibinfo{volume}{74}},
  \bibinfo{pages}{245113} (\bibinfo{year}{2006}).

\bibitem[{\citenamefont{Storcz and Wilhelm}(2003)}]{PhysRevA.67.042319}
\bibinfo{author}{\bibfnamefont{M.~J.} \bibnamefont{Storcz}} \bibnamefont{and}
  \bibinfo{author}{\bibfnamefont{F.~K.} \bibnamefont{Wilhelm}},
  \bibinfo{journal}{Phys. Rev. A} \textbf{\bibinfo{volume}{67}},
  \bibinfo{pages}{042319} (\bibinfo{year}{2003}).

\bibitem[{\citenamefont{Tornow et~al.}(2008)\citenamefont{Tornow, Bulla,
  Anders, and Nitzan}}]{tornow:035434}
\bibinfo{author}{\bibfnamefont{S.}~\bibnamefont{Tornow}},
  \bibinfo{author}{\bibfnamefont{R.}~\bibnamefont{Bulla}},
  \bibinfo{author}{\bibfnamefont{F.~B.} \bibnamefont{Anders}},
  \bibnamefont{and} \bibinfo{author}{\bibfnamefont{A.}~\bibnamefont{Nitzan}},
  \bibinfo{journal}{Phys. Rev. B} \textbf{\bibinfo{volume}{78}},
  \bibinfo{eid}{035434} (\bibinfo{year}{2008}).

\bibitem[{\citenamefont{McCutcheon et~al.}(2010)\citenamefont{McCutcheon,
  Nazir, Bose, and Fisher}}]{PhysRevB.81.235321}
\bibinfo{author}{\bibfnamefont{D.~P.~S.} \bibnamefont{McCutcheon}},
  \bibinfo{author}{\bibfnamefont{A.}~\bibnamefont{Nazir}},
  \bibinfo{author}{\bibfnamefont{S.}~\bibnamefont{Bose}}, \bibnamefont{and}
  \bibinfo{author}{\bibfnamefont{A.~J.} \bibnamefont{Fisher}},
  \bibinfo{journal}{Phys. Rev. B} \textbf{\bibinfo{volume}{81}},
  \bibinfo{pages}{235321} (\bibinfo{year}{2010}).

\bibitem[{\citenamefont{Anders et~al.}(2007)\citenamefont{Anders, Bulla, and
  Vojta}}]{anders:210402}
\bibinfo{author}{\bibfnamefont{F.~B.} \bibnamefont{Anders}},
  \bibinfo{author}{\bibfnamefont{R.}~\bibnamefont{Bulla}}, \bibnamefont{and}
  \bibinfo{author}{\bibfnamefont{M.}~\bibnamefont{Vojta}},
  \bibinfo{journal}{Phys. Rev. Lett.} \textbf{\bibinfo{volume}{98}},
  \bibinfo{eid}{210402} (\bibinfo{year}{2007}).

\bibitem[{\citenamefont{Vojta}(2006)}]{vojta_philmag_2006}
\bibinfo{author}{\bibfnamefont{M.}~\bibnamefont{Vojta}},
  \bibinfo{journal}{Phil. Mag.} \textbf{\bibinfo{volume}{86}},
  \bibinfo{pages}{1807} (\bibinfo{year}{2006}).

\bibitem[{\citenamefont{Winter et~al.}(2009)\citenamefont{Winter, Rieger,
  Vojta, and Bulla}}]{PhysRevLett.102.030601}
\bibinfo{author}{\bibfnamefont{A.}~\bibnamefont{Winter}},
  \bibinfo{author}{\bibfnamefont{H.}~\bibnamefont{Rieger}},
  \bibinfo{author}{\bibfnamefont{M.}~\bibnamefont{Vojta}}, \bibnamefont{and}
  \bibinfo{author}{\bibfnamefont{R.}~\bibnamefont{Bulla}},
  \bibinfo{journal}{Phys. Rev. Lett.} \textbf{\bibinfo{volume}{102}},
  \bibinfo{pages}{030601} (\bibinfo{year}{2009}).

\bibitem[{\citenamefont{Vojta et~al.}(2010)\citenamefont{Vojta, Bulla,
  G\"uttge, and Anders}}]{PhysRevB.81.075122}
\bibinfo{author}{\bibfnamefont{M.}~\bibnamefont{Vojta}},
  \bibinfo{author}{\bibfnamefont{R.}~\bibnamefont{Bulla}},
  \bibinfo{author}{\bibfnamefont{F.}~\bibnamefont{G\"uttge}}, \bibnamefont{and}
  \bibinfo{author}{\bibfnamefont{F.}~\bibnamefont{Anders}},
  \bibinfo{journal}{Phys. Rev. B} \textbf{\bibinfo{volume}{81}},
  \bibinfo{pages}{075122} (\bibinfo{year}{2010}).

\bibitem[{\citenamefont{Bennett et~al.}(1996)\citenamefont{Bennett, DiVincenzo,
  Smolin, and Wootters}}]{PhysRevA.54.3824}
\bibinfo{author}{\bibfnamefont{C.~H.} \bibnamefont{Bennett}},
  \bibinfo{author}{\bibfnamefont{D.~P.} \bibnamefont{DiVincenzo}},
  \bibinfo{author}{\bibfnamefont{J.~A.} \bibnamefont{Smolin}},
  \bibnamefont{and} \bibinfo{author}{\bibfnamefont{W.~K.}
  \bibnamefont{Wootters}}, \bibinfo{journal}{Phys. Rev. A}
  \textbf{\bibinfo{volume}{54}}, \bibinfo{pages}{3824} (\bibinfo{year}{1996}).

\bibitem[{\citenamefont{Kopp and Le~Hur}(2007)}]{kopp:220401}
\bibinfo{author}{\bibfnamefont{A.}~\bibnamefont{Kopp}} \bibnamefont{and}
  \bibinfo{author}{\bibfnamefont{K.}~\bibnamefont{Le~Hur}},
  \bibinfo{journal}{Phys. Rev. Lett.} \textbf{\bibinfo{volume}{98}},
  \bibinfo{eid}{220401} (\bibinfo{year}{2007}).

\bibitem[{\citenamefont{Costi and McKenzie}(2003)}]{PhysRevA.68.034301}
\bibinfo{author}{\bibfnamefont{T.~A.} \bibnamefont{Costi}} \bibnamefont{and}
  \bibinfo{author}{\bibfnamefont{R.~H.} \bibnamefont{McKenzie}},
  \bibinfo{journal}{Phys. Rev. A} \textbf{\bibinfo{volume}{68}},
  \bibinfo{pages}{034301} (\bibinfo{year}{2003}).

\bibitem[{\citenamefont{Vojta et~al.}(2005)\citenamefont{Vojta, Tong, and
  Bulla}}]{PhysRevLett.94.070604}
\bibinfo{author}{\bibfnamefont{M.}~\bibnamefont{Vojta}},
  \bibinfo{author}{\bibfnamefont{N.-H.} \bibnamefont{Tong}}, \bibnamefont{and}
  \bibinfo{author}{\bibfnamefont{R.}~\bibnamefont{Bulla}},
  \bibinfo{journal}{Phys. Rev. Lett.} \textbf{\bibinfo{volume}{94}},
  \bibinfo{pages}{070604} (\bibinfo{year}{2005}).

\bibitem[{\citenamefont{Le~Hur et~al.}(2007)\citenamefont{Le~Hur,
  Doucet-Beaupr\'{e}, and Hofstetter}}]{hur:126801}
\bibinfo{author}{\bibfnamefont{K.}~\bibnamefont{Le~Hur}},
  \bibinfo{author}{\bibfnamefont{P.}~\bibnamefont{Doucet-Beaupr\'{e}}},
  \bibnamefont{and}
  \bibinfo{author}{\bibfnamefont{W.}~\bibnamefont{Hofstetter}},
  \bibinfo{journal}{Phys. Rev. Lett.} \textbf{\bibinfo{volume}{99}},
  \bibinfo{eid}{126801} (\bibinfo{year}{2007}).

\bibitem[{\citenamefont{Vojta et~al.}(2009)\citenamefont{Vojta, Tong, and
  Bulla}}]{PhysRevLett.102.249904}
\bibinfo{author}{\bibfnamefont{M.}~\bibnamefont{Vojta}},
  \bibinfo{author}{\bibfnamefont{N.-H.} \bibnamefont{Tong}}, \bibnamefont{and}
  \bibinfo{author}{\bibfnamefont{R.}~\bibnamefont{Bulla}},
  \bibinfo{journal}{Phys. Rev. Lett.} \textbf{\bibinfo{volume}{102}},
  \bibinfo{pages}{249904(E)} (\bibinfo{year}{2009}).

\bibitem[{\citenamefont{Negele and Orland}(1998)}]{negele_orland_book}
\bibinfo{author}{\bibfnamefont{J.~W.} \bibnamefont{Negele}} \bibnamefont{and}
  \bibinfo{author}{\bibfnamefont{H.}~\bibnamefont{Orland}},
  \emph{\bibinfo{title}{Quantum Many-Particle Systems}}
  (\bibinfo{publisher}{Westview Press}, \bibinfo{address}{Boulder, CO, USA},
  \bibinfo{year}{1998}).

\bibitem[{\citenamefont{Alvermann and Fehske}(2009)}]{PhysRevLett.102.150601}
\bibinfo{author}{\bibfnamefont{A.}~\bibnamefont{Alvermann}} \bibnamefont{and}
  \bibinfo{author}{\bibfnamefont{H.}~\bibnamefont{Fehske}},
  \bibinfo{journal}{Phys. Rev. Lett.} \textbf{\bibinfo{volume}{102}},
  \bibinfo{pages}{150601} (\bibinfo{year}{2009}).

\bibitem[{\citenamefont{Emery and Luther}(1974)}]{PhysRevB.9.215}
\bibinfo{author}{\bibfnamefont{V.~J.} \bibnamefont{Emery}} \bibnamefont{and}
  \bibinfo{author}{\bibfnamefont{A.}~\bibnamefont{Luther}},
  \bibinfo{journal}{Phys. Rev. B} \textbf{\bibinfo{volume}{9}},
  \bibinfo{pages}{215} (\bibinfo{year}{1974}).

\bibitem[{\citenamefont{Sachdev}(1999)}]{sachdev_qpt_book}
\bibinfo{author}{\bibfnamefont{S.}~\bibnamefont{Sachdev}},
  \emph{\bibinfo{title}{Quantum Phase Transitions}}
  (\bibinfo{publisher}{Cambridge University Press},
  \bibinfo{address}{Cambridge, U.K.}, \bibinfo{year}{1999}).

\bibitem[{\citenamefont{Fisher et~al.}(1972)\citenamefont{Fisher, Ma, and
  Nickel}}]{PhysRevLett.29.917}
\bibinfo{author}{\bibfnamefont{M.~E.} \bibnamefont{Fisher}},
  \bibinfo{author}{\bibfnamefont{S.-k.} \bibnamefont{Ma}}, \bibnamefont{and}
  \bibinfo{author}{\bibfnamefont{B.~G.} \bibnamefont{Nickel}},
  \bibinfo{journal}{Phys. Rev. Lett.} \textbf{\bibinfo{volume}{29}},
  \bibinfo{pages}{917} (\bibinfo{year}{1972}).

\bibitem[{\citenamefont{Luijten and Bl\"ote}(1997)}]{PhysRevB.56.8945}
\bibinfo{author}{\bibfnamefont{E.}~\bibnamefont{Luijten}} \bibnamefont{and}
  \bibinfo{author}{\bibfnamefont{H.~W.~J.} \bibnamefont{Bl\"ote}},
  \bibinfo{journal}{Phys. Rev. B} \textbf{\bibinfo{volume}{56}},
  \bibinfo{pages}{8945} (\bibinfo{year}{1997}).

\bibitem[{\citenamefont{Ingersent and Si}(2002)}]{PhysRevLett.89.076403}
\bibinfo{author}{\bibfnamefont{K.}~\bibnamefont{Ingersent}} \bibnamefont{and}
  \bibinfo{author}{\bibfnamefont{Q.}~\bibnamefont{Si}}, \bibinfo{journal}{Phys.
  Rev. Lett.} \textbf{\bibinfo{volume}{89}}, \bibinfo{pages}{076403}
  (\bibinfo{year}{2002}).

\bibitem[{\citenamefont{Kosterlitz}(1976)}]{PhysRevLett.37.1577}
\bibinfo{author}{\bibfnamefont{J.~M.} \bibnamefont{Kosterlitz}},
  \bibinfo{journal}{Phys. Rev. Lett.} \textbf{\bibinfo{volume}{37}},
  \bibinfo{pages}{1577} (\bibinfo{year}{1976}).

\bibitem[{\citenamefont{Dub\'{e} and Stamp}(1998)}]{DubeStamp-IntJModPhys-1998}
\bibinfo{author}{\bibfnamefont{M.}~\bibnamefont{Dub\'{e}}} \bibnamefont{and}
  \bibinfo{author}{\bibfnamefont{P.~C.~E.} \bibnamefont{Stamp}},
  \bibinfo{journal}{Int. J. Mod. Phys.} \textbf{\bibinfo{volume}{12}},
  \bibinfo{pages}{1191} (\bibinfo{year}{1998}).

\bibitem[{\citenamefont{Yoshida et~al.}(1990)\citenamefont{Yoshida, Whitaker,
  and Oliveira}}]{PhysRevB.41.9403}
\bibinfo{author}{\bibfnamefont{M.}~\bibnamefont{Yoshida}},
  \bibinfo{author}{\bibfnamefont{M.~A.} \bibnamefont{Whitaker}},
  \bibnamefont{and} \bibinfo{author}{\bibfnamefont{L.~N.}
  \bibnamefont{Oliveira}}, \bibinfo{journal}{Phys. Rev. B}
  \textbf{\bibinfo{volume}{41}}, \bibinfo{pages}{9403} (\bibinfo{year}{1990}).

\bibitem[{\citenamefont{Dekker}(1987)}]{PhysRevA.35.1436}
\bibinfo{author}{\bibfnamefont{H.}~\bibnamefont{Dekker}},
  \bibinfo{journal}{Phys. Rev. A} \textbf{\bibinfo{volume}{35}},
  \bibinfo{pages}{1436} (\bibinfo{year}{1987}).

\bibitem[{\citenamefont{Duan and Guo}(1998)}]{PhysRevA.57.737}
\bibinfo{author}{\bibfnamefont{L.-M.} \bibnamefont{Duan}} \bibnamefont{and}
  \bibinfo{author}{\bibfnamefont{G.-C.} \bibnamefont{Guo}},
  \bibinfo{journal}{Phys. Rev. A} \textbf{\bibinfo{volume}{57}},
  \bibinfo{pages}{737} (\bibinfo{year}{1998}).

\bibitem[{\citenamefont{Cohen-Tannoudji
  et~al.}(2004)\citenamefont{Cohen-Tannoudji, Dupont-Roc, and
  Grynberg}}]{CohenTannoudji-AtomPhotonInteractions}
\bibinfo{author}{\bibfnamefont{C.}~\bibnamefont{Cohen-Tannoudji}},
  \bibinfo{author}{\bibfnamefont{J.}~\bibnamefont{Dupont-Roc}},
  \bibnamefont{and} \bibinfo{author}{\bibfnamefont{G.}~\bibnamefont{Grynberg}},
  \emph{\bibinfo{title}{Atom-Photon Interactions}} (\bibinfo{publisher}{JW},
  \bibinfo{year}{2004}).

\bibitem[{\citenamefont{Jauho et~al.}(1994)\citenamefont{Jauho, Wingreen, and
  Meir}}]{PhysRevB.50.5528}
\bibinfo{author}{\bibfnamefont{A.-P.} \bibnamefont{Jauho}},
  \bibinfo{author}{\bibfnamefont{N.~S.} \bibnamefont{Wingreen}},
  \bibnamefont{and} \bibinfo{author}{\bibfnamefont{Y.}~\bibnamefont{Meir}},
  \bibinfo{journal}{Phys. Rev. B} \textbf{\bibinfo{volume}{50}},
  \bibinfo{pages}{5528} (\bibinfo{year}{1994}).

\bibitem[{\citenamefont{Schmidt et~al.}(2008)\citenamefont{Schmidt, Werner,
  M\"uhlbacher, and Komnik}}]{PhysRevB.78.235110}
\bibinfo{author}{\bibfnamefont{T.~L.} \bibnamefont{Schmidt}},
  \bibinfo{author}{\bibfnamefont{P.}~\bibnamefont{Werner}},
  \bibinfo{author}{\bibfnamefont{L.}~\bibnamefont{M\"uhlbacher}},
  \bibnamefont{and} \bibinfo{author}{\bibfnamefont{A.}~\bibnamefont{Komnik}},
  \bibinfo{journal}{Phys. Rev. B} \textbf{\bibinfo{volume}{78}},
  \bibinfo{pages}{235110} (\bibinfo{year}{2008}).

\bibitem[{\citenamefont{Giamarchi}(2004)}]{Giamarchi-QuantumPhysIn1D}
\bibinfo{author}{\bibfnamefont{T.}~\bibnamefont{Giamarchi}},
  \emph{\bibinfo{title}{Quantum Physics in One Dimension}}
  (\bibinfo{publisher}{Oxford Univ. Press}, \bibinfo{address}{USA},
  \bibinfo{year}{2004}).

\bibitem[{\citenamefont{Sire et~al.}(1993)\citenamefont{Sire, Varma, and
  Krishnamurthy}}]{PhysRevB.48.13833}
\bibinfo{author}{\bibfnamefont{C.}~\bibnamefont{Sire}},
  \bibinfo{author}{\bibfnamefont{C.~M.} \bibnamefont{Varma}}, \bibnamefont{and}
  \bibinfo{author}{\bibfnamefont{H.~R.} \bibnamefont{Krishnamurthy}},
  \bibinfo{journal}{Phys. Rev. B} \textbf{\bibinfo{volume}{48}},
  \bibinfo{pages}{13833} (\bibinfo{year}{1993}).

\bibitem[{\citenamefont{Nalbach and Thorwart}(2010)}]{PhysRevB.81.054308}
\bibinfo{author}{\bibfnamefont{P.}~\bibnamefont{Nalbach}} \bibnamefont{and}
  \bibinfo{author}{\bibfnamefont{M.}~\bibnamefont{Thorwart}},
  \bibinfo{journal}{Phys. Rev. B} \textbf{\bibinfo{volume}{81}},
  \bibinfo{pages}{054308} (\bibinfo{year}{2010}).

\bibitem[{\citenamefont{Andrei et~al.}(1999)\citenamefont{Andrei, Zim\'anyi,
  and Sch\"on}}]{PhysRevB.60.R5125}
\bibinfo{author}{\bibfnamefont{N.}~\bibnamefont{Andrei}},
  \bibinfo{author}{\bibfnamefont{G.~T.} \bibnamefont{Zim\'anyi}},
  \bibnamefont{and} \bibinfo{author}{\bibfnamefont{G.}~\bibnamefont{Sch\"on}},
  \bibinfo{journal}{Phys. Rev. B} \textbf{\bibinfo{volume}{60}},
  \bibinfo{pages}{R5125} (\bibinfo{year}{1999}).

\bibitem[{\citenamefont{Affleck et~al.}(1995)\citenamefont{Affleck, Ludwig, and
  Jones}}]{PhysRevB.52.9528}
\bibinfo{author}{\bibfnamefont{I.}~\bibnamefont{Affleck}},
  \bibinfo{author}{\bibfnamefont{A.~W.~W.} \bibnamefont{Ludwig}},
  \bibnamefont{and} \bibinfo{author}{\bibfnamefont{B.~A.} \bibnamefont{Jones}},
  \bibinfo{journal}{Phys. Rev. B} \textbf{\bibinfo{volume}{52}},
  \bibinfo{pages}{9528} (\bibinfo{year}{1995}).

\end{thebibliography}

\end{document}